\documentclass[nolinenumbers, twocolumn]{aastex701}

\usepackage{tabularx}
\usepackage[T1]{fontenc}
\DeclareUnicodeCharacter{0627}{ }
\DeclareUnicodeCharacter{0644}{ }
\DeclareUnicodeCharacter{0645}{ }
\DeclareUnicodeCharacter{0632}{ }
\DeclareUnicodeCharacter{064A}{ }
\usepackage{tabularx}
\usepackage{booktabs}
\usepackage{amsmath}
\usepackage{multirow} 
\usepackage{caption}
\usepackage{subcaption}
\usepackage{placeins}
\usepackage{graphicx}
\usepackage{makecell} 

\graphicspath{ {./figures/} }

\begin{document}

\title{Surface Properties, Orbital Dynamics, and Thermophysical Modeling of the Primitive Asteroid (269) Justitia}

\author[orcid=0009-0008-5757-3121,sname='Braga']{L. Braga}
\affiliation{São Paulo State University (UNESP), School of Engineering and Sciences, Guaratinguetá, São Paulo, Brazil}
\email{lb.braga@unesp.br}  

\author[orcid=0000-0002-9448-141X, sname='Amarante']{A. Amarante}
\affiliation{São Paulo State University (UNESP), School of Engineering and Sciences, Guaratinguetá, São Paulo, Brazil}
\email{andre.amarante@unesp.br}

\author[orcid=0000-0002-6162-9078, sname='Ferreira']{A. Ferreira}
\affiliation{São Paulo State University (UNESP), School of Engineering and Sciences, Guaratinguetá, São Paulo, Brazil}
\email{alessandra.ferraz@unesp.br}

\author[orcid=0000-0003-1549-4587, sname='Monteiro']{F. Monteiro}
\affiliation{São Paulo State University (UNESP), School of Engineering and Sciences, Guaratinguetá, São Paulo, Brazil}
\email{filipe.monteiro@unesp.br}

\author[orcid=0009-0004-1941-7209, sname='Martins']{M. Martins}
\affiliation{São Paulo State University (UNESP), School of Engineering and Sciences, Guaratinguetá, São Paulo, Brazil}
\email{maria.m.goncalves@unesp.br}

\begin{abstract}

The Emirates Mission to the Asteroid Belt (EMA) will study the ultra-red asteroid (269)~Justitia. In this work, we present the first detailed investigation of Justitia's surface dynamics using the newly developed 3-D polyhedral shape model with 574 vertices and 1,144 faces. We analyze its geopotential, surface acceleration, escape speeds, slopes, and equilibrium points, and we also search for planar symmetric periodic orbits using an equivalent ellipsoidal approximation.
Our results indicate that the lowest geopotential values occur at the poles, which also correspond to regions of maximum surface acceleration. The global slope distribution suggests preferred zones of material accumulation or migration, offering clues to Justitia's long-term morphological evolution. Most slopes remain below $40^\circ$, implying that loose particles may settle stably across large portions of the surface.
We identify five equilibrium points consistent with Justitia's estimated density and slow rotation period. Two external points (E$_2$ and E$_4$) exhibit linear stability, and all equilibrium locations lie relatively far from the surface due to the body's slow spin. Additionally, we discover 28 new families of planar symmetric periodic orbits, then classify their topologies and determine their linear stability, providing a dynamical framework relevant to future spacecraft operations near Justitia.
Finally, thermal modeling reveals how thermal inertia and heliocentric distance shape Justitia's temperature distribution. The south pole receives more insolation than the north pole, reaching minimum temperatures of about 102~K and 87~K, respectively. These combined dynamical and thermal results offer valuable insights for the EMA mission and for understanding slowly rotating small bodies.

\end{abstract}



\section{Introduction} 
\label{introduction}
Primitive asteroids can transport water and complex organic molecules, which may have reached the primitive Earth through impacts, contributing to the emergence of life on the planet \citep{morbideli2000,izidoro2013}. In this sense, understanding their composition, internal structures, and evolution is crucial for comprehending the formation of the Solar System. In addition, asteroids arouse the interest of the mining industry due to their potential for economic exploration \citep{lewis1997}. Both the mining industry and space agencies, such as NASA, ESA, and JAXA, have been investigating these celestial bodies and the changes they have undergone during the evolution of the Solar System.

In 2004, the ESA Rosetta mission \citep{Glassmeieretal2007, Rolletal2016} captured images of the asteroids (2867) Steins and (21) Lutetia during its trajectory to comet 67P/Churyumov-Gerasimenko in 2008 and 2010, respectively.
The NASA New-Horizons mission in 2015 carried out a detailed study of the dwarf planet Pluto, its moon Charon, and the small satellites Styx, Nix, Kerberos, and Hydra, as well as the Kuiper Belt Object (KBO) (486958) Arrokoth in January 2019 \citep{stern2007, Sternetal2019}.
In 2014, JAXA launched the Hayabusa2 mission to collect samples from the asteroid (162173) Ryugu and return to Earth in 2020 with the samples \citep{Hirabayashietal2021}.
The NASA OSIRIS-REx mission, launched in September 2016, arrived at the asteroid (101955) Bennu in 2018 and returned to Earth with carbon-rich samples in September 2023. Bennu was chosen because it is a potentially hazardous asteroid. In addition, Bennu was the first asteroid to be seen ejecting particles from its surface \citep{laureta2017, lauretta2019}.

In October 2021, NASA launched the Lucy mission to explore seven Trojan asteroids orbiting Jupiter, which are considered fossils from the formation of the Solar System. In November 2023, the probe arrived at the main-belt asteroid (152830) Dinkinesh, confirming that this asteroid is a binary system \citep{levison2024}. In April 2025, the spacecraft made a close approach to the asteroid (52246) Donaldjohanson \citep{marchi2025}.
Finally, the NASA Psyche mission, launched in October 2023, aims to investigate the asteroid (16) Psyche, which is believed to be an exposed metallic core of an unformed planet or a composition of primordial material rich in metal that has not undergone complete fusion \citep{eikens}.

In this context, several studies have explored the surface dynamics of small bodies, such as Bennu, Ryugu, Psyche, and Arrokoth, revealing key insights into their geophysical environments.

\citet{scheeresetal2016} investigated the dynamic processes on the surface of the asteroid Bennu and identified slopes smaller than 45$^\circ$, with the highest values in the intermediate areas between the equator and the poles of the northern and southern hemispheres of its surface, and lower values in the central regions along the equator. \citet{moura2020} analyzed the gravitational field, topography, and dynamics around the asteroid Psyche in detail, revealing that it has a low-tilt surface. They also identified four external equilibrium points, of which two are stable, which could assist the mission.

\citet{amaranteandwinter2020} explored in detail the surface dynamics of KBO Arrokoth. They identified seven equilibrium points of Arrokoth with different topological stabilities. They found that Arrokoth's equator is an unstable region due to its high rotational period, while its polar regions are stable for surface particles. \citet{amarante2021} used a spherical cloud of particles to determine the reaggregation density on Bennu's surface and identified that 30\% of the initial particles that are in retrograde orbits collide with its surface, proving to be more stable than the particles that are in prograde orbits, where half of the particles collided with the asteroid's surface during the entire integration.

\citet{fuetal2024} identified 25 families of bifurcated initial periodic orbits associated with Ryugu's outer equilibrium points. The same bifurcation points connect these different bifurcated families. Furthermore, comparing Bennu and Ryugu, it is observed that in both asteroids, the bifurcation of equilibrium-point branches occurs from a degenerate point as their rotation rates increase.

Therefore, studying the surface dynamics of small bodies enables us to understand how the gravitational environment affects the surface of irregularly shaped minor bodies, which can provide important support for future missions to these small worlds of the Solar System, such as the Emirates Mission to explore the Asteroid Belt (EMA) \citep{Almazi2024, Sorli2024, scheeres2025, ElMaarry2025}.

The EMA mission will study six asteroids before reaching its final destination: the asteroid (269) Justitia, which exhibits an intensely reddish coloration due to the possible presence of organic compounds on its surface, a characteristic unusual for asteroids located in this region of the main asteroid belt. Recently, a new taxonomic type (Z-type) was introduced to classify these asteroids, which, like Justitia, exhibit very red colors and spectra and are considered ultra-red asteroids \citep{Mahlke2022}.

In this sense, the analyses and comparisons indicate that Justitia's properties resemble Centaurs and Trans-Neptunian Objects (TNOs) \citep{hasegawa2021, hume}. Thus, it is possible that these asteroids were thrown into the main belt, as predicted by migration models of Solar System formation \citep{tsiganis, morbidelli2005, morbidelli2010}. Therefore, it is essential to study this class of small bodies to understand the dynamic evolution of the inner Solar System.

For example, \citet{parkeretal2024} presented the proximity operations strategy of the EMA mission, with an emphasis on mapping and landing on (269) Justitia using carefully planned orbits and correction maneuvers. Moreover, \citet{marciniak2025} constrained the diameter of asteroid (269) Justitia to 55-60 km and derived its rotational period of 33.12962 h, shape, thermal inertia, and surface roughness. \citet{buie2025} found an irregular shape for Justitia with a circular-equivalent radius of 28.5-28.9 km and a visual albedo of 0.072$\pm$0.007 from a multi-station occultation campaign. 

Furthermore, \citet{harish2025} analyzed the composition of the surface of the asteroid Justitia and concluded that its surface is composed of a combination of silicates altered by space weathering and carbonaceous material, which explains some of its physical characteristics, such as albedo and the features observed in the thermal infrared, which indicates that Justitia possibly underwent complex evolutionary processes, thus representing an object that formed in the region of the giant planets and later implanted into the main belt.

In this work, we provide a detailed dynamical analysis of the surface of asteroid (269) Justitia, aiming to support the EMA Mission. 
We explored the surface characteristics of asteroid Justitia, including the geopotential surface, surface acceleration, escape speed, and surface slope distribution. Additionally, we analyzed the number, location, and stability of equilibrium points.

We also analyze the global behavior of the slopes on Justitia's surface, considering the third-body (3B) perturbation in the Main Belt and the solar radiation pressure (SRP) in both the Main Belt and the Kuiper Belt, assuming that Justitia originated in the latter region.

We investigate the evolution of the equilibrium points, assuming that Justitia's past spin was faster than its current one, and verify how these points behave as the rotational period increases.

Furthermore, planar symmetric periodic orbits around asteroid Justitia were explored by employing an equivalent ellipsoidal approximation for its irregular shape. These periodic solutions help to understand the system's global behavior.

Finally, we evaluate the surface temperature of Justitia under different thermal inertia values and heliocentric distances using the Thermophysical Model.

\section{Polyhedron Method} 
\label{polyhedronmethod}
Many numerical models have been developed to approximate the gravitational potential of irregularly shaped bodies, including asteroids. They include triaxial ellipsoids \citep{wernner1994}, massive straight-line segments \citep{Riaguasetal}, and solid circular rings \citep{broucke}. 
Methods such as Legendre polynomials also generate inaccuracies at specific points \citep{scheeres1994}.

A numerical approach that reduces computational effort is to model the gravitational potential of small bodies with irregular shapes using a uniform grid of point masses (mascons) that occupy the volume. Then, the total gravitational potential is obtained by summing the gravitational potential contributions of each mass \citep{Glassmeieretal2007}. However, this approach suffers from inaccuracies near the surface of the irregular body \citep{wernerl1996}. Then, more robust numerical methods are required to compute the gravitational potential near an asteroid's surface, enabling the study of surface characteristics, such as the geopotential.

In this scenario, the polyhedron method is widely used to accurately estimate the gravitational field generated by an irregularly shaped minor body, particularly near its surface. \citet{wernerl1996} refined this technique by employing a harmonic expansion based on a polyhedron with constant density. They demonstrated that errors in the expressions for the gravitational potential were more significant near the surface of an asteroid than with spherical harmonics or the mascon method. \citet{tsoulis2001,tsoulis2012} developed a numerical approach to address these singularities by deriving specific terms to manage the computation points associated with the gravitational potential from polyhedral sources near the surface. 

For this purpose, we investigate the surface characteristics of asteroid Justitia in detail, using the newly 2025 convex 3-D polyhedral shape model of asteroid Justitia consisting of 574 vertices and 1,144 faces \citep{marciniak2025}, constructed through the lightcurve inversion method \citet{2001Icar..153...24K, 2001Icar..153...37K}.
The 3-D polyhedral shape model of asteroid Justitia has an equivalent diameter of 59.0 km \citep{marciniak2025}. Moreover, a rotational period of 33.12962 h was derived by \cite{marciniak2025, buie2025}. This rotational period is consistent with those derived by \citet{Pilcher2016}.

The density of a small body is one of its essential properties and can vary considerably depending on its composition or internal structure. We applied the following bulk density ranges corresponding to the possible taxonomic classes of asteroids D-type 1.0\,g\,cm$^{-3}$ and analogous primitive meteorites $\sim1.6-2.3\,$g\,cm$^{-3}$ \citep{marchis, Mcmahon2023, parkeretal2024}.
Then, we investigate the effects of surface dynamics in a range of densities from 1.0\,g\,cm$^{-3}$ to 2.0\,g\,cm$^{-3}$.

We adopted the \textsc{Minor-Gravity} package\footnote{\url{https://github.com/a-amarante/minor-gravity}} \citep{amaranteandwinter2020, minorgravity} to compute the surface characteristics of asteroid Justitia, calculating the gravitational potential by the polyhedron method that avoids singularities over the surface of the asteroid \citep{tsoulis2001,tsoulis2012}, including its first and second-order derivatives (Appendix \ref{appendixA}, Eqs. \eqref{eq:potencial_split}-\eqref{eq:anpq}).

In addition, we also determined the number, location, and stability of equilibrium points around Justitia using the \textsc{Minor-Equilibria} package\footnote{\url{https://github.com/a-amarante/minor-equilibria-nr}} \citep{amaranteandwinter2020, minorequilibria}.

\section{Shape Model and Geopotential Surface}
\label{shapemodel}
\begin{figure*}
	\plotone{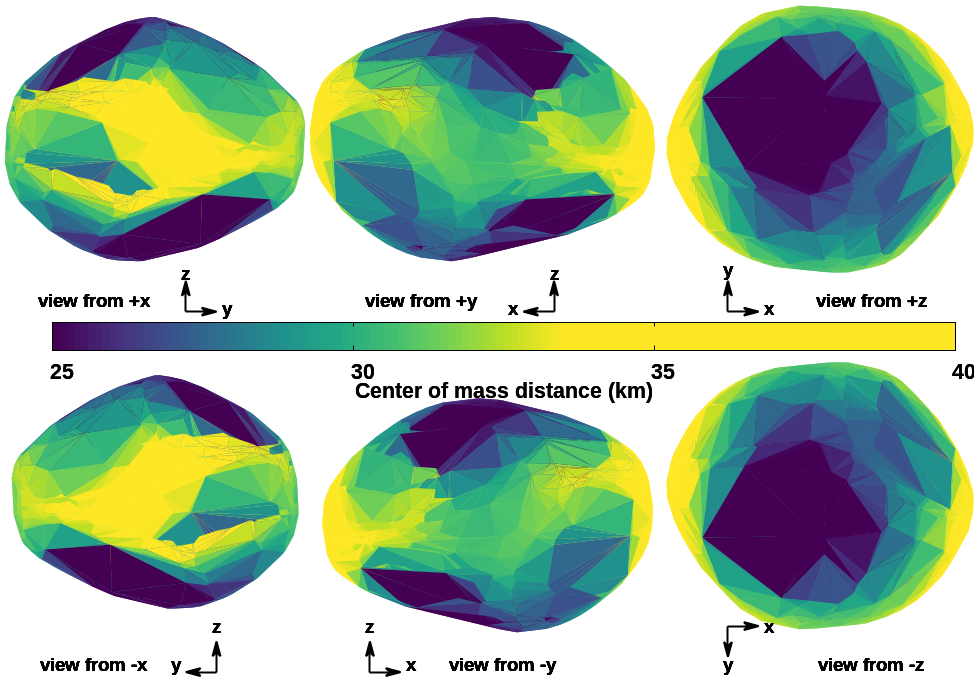}
    \caption{3-D polyhedral shape model of asteroid Justitia presented in six perspective views ($\pm$x, $\pm$y, and $\pm$z). The shape model consisted of 574 vertices and 1,144 triangular faces. The color code indicates the distance from the centroid of each face to the asteroid's center of mass, in km.}
    \label{fig:1}
\end{figure*}

Figure \ref{fig:1} illustrates the 3-D polyhedral shape of asteroid Justitia, composed of 574 vertices and 1,144 faces in six perspectives: $\pm$x, $\pm$y, and $\pm$z. The color box code indicates the distance between each triangular face and the asteroid's center of mass. These distances provide complementary information for investigating the geopotential surface and surface acceleration (sec. \ref{sec:2.3}).

The sum of gravitational and centrifugal potentials defines the geopotential of an asteroid. If the reference frame of the asteroid is on principal axes of inertia and centered on its barycenter, the geopotential could be computed as \citep{scheeres2012}:
\begin{equation}
V(\mathbf{r}) = -\frac{1}{2} \omega^2 \left(x^2 + y^2\right) + U(\mathbf{r}),
\label{eq:1}
\end{equation}
\noindent where $\mathbf{r}$ denotes the position of the particle in the reference frame attached to the asteroid about the center of mass, $\omega$ represents the angular speed, and $U(\mathbf{r})$ corresponds to the gravitational potential obtained numerically (Appendix \ref{appendixA}, Eq. \eqref{eq:potencial_split}).
\begin{figure*}
\plotone{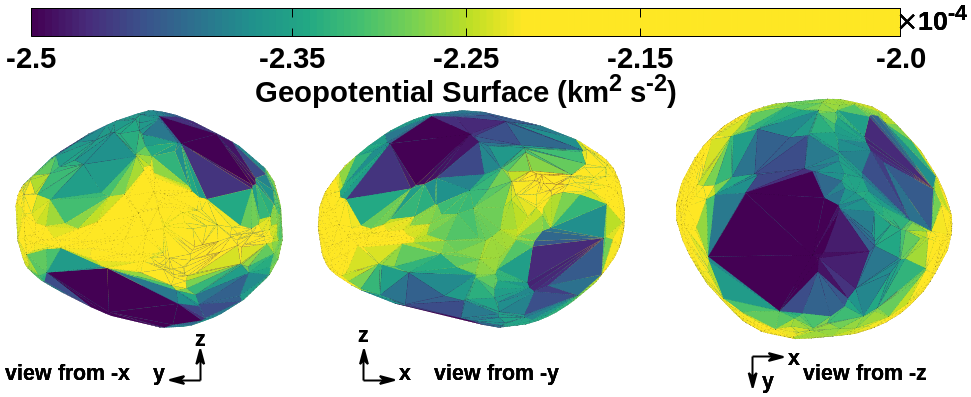}
    \caption{Geopotential distribution over the surface of asteroid Justitia, assuming a bulk density of 1.0\,g\,cm$^{-3}$ and a rotational period of 33.12962\,h, shown from views $-$x, $-$y, and $-$z. The color scale indicates the computed values from Eq. \eqref{eq:1}, expressed in km$^{2}$\,s$^{-2}$.}
    \label{fig:2}
\end{figure*}
Fig. \ref{fig:2} shows the geopotential computed across the surface of asteroid Justitia for a chosen density of 1.0\,g\,cm$^{-3}$ and a rotational period of 33.12962\,h. The views from -x, -y enhance the maximum geopotential values, while the \textbf{View from -z} represents the minimum ones.

Figures \ref{fig:1} and \ref{fig:2} show that regions far away from the center of mass have maximum values of the geopotential surface, while those close to the center of mass have low values of the geopotential surface. 

Comparing the geopotential of asteroid Justitia with those found on asteroid (25143) Itokawa and KBO Arrokoth, as it is currently located in the main asteroid belt and may have similar properties to TNOs \citep{hasegawa2021}, we have that the minimum geopotential of Justitia is $\sim35.5\times$ smaller than that of Itokawa \citep{scheeres2012}. Furthermore, it is $\sim4.5\times$ smaller than that of Arrokoth \citep{amaranteandwinter2020}.
Table \ref{tab:geopotential} shows that the global behavior of the geopotential distribution (Fig. \ref{fig:2}) over the surface of Justitia does not change significantly, keeping the same order of magnitude ($10^{-4}$), in the interval of densities 1.0-2.0\,g\,cm$^{-3}$.

\begin{table}
\centering
\caption{Global maximum and minimum geopotential, in km$^{2}$\,s$^{-2}$, computed over the surface of asteroid Justitia for different densities.}
\label{tab:geopotential}
\begin{tabular}{c c c}
\hline\hline
Density (g\,cm$^{-3}$) & Min. value ($\times10^{-4}$) & Max. value ($\times10^{-4}$) \\
\hline
1.0 & $-2.5$ & $-2.0$ \\
1.2 & $-3.1$ & $-2.5$ \\
1.4 & $-3.6$ & $-2.9$ \\
1.6 & $-4.1$ & $-3.3$ \\
1.8 & $-4.6$ & $-3.7$ \\
2.0 & $-5.1$ & $-4.1$ \\
\hline
\end{tabular}
\end{table}

\section{Equipotential Curves}
\label{2.8}
\begin{figure*}
	\plotone{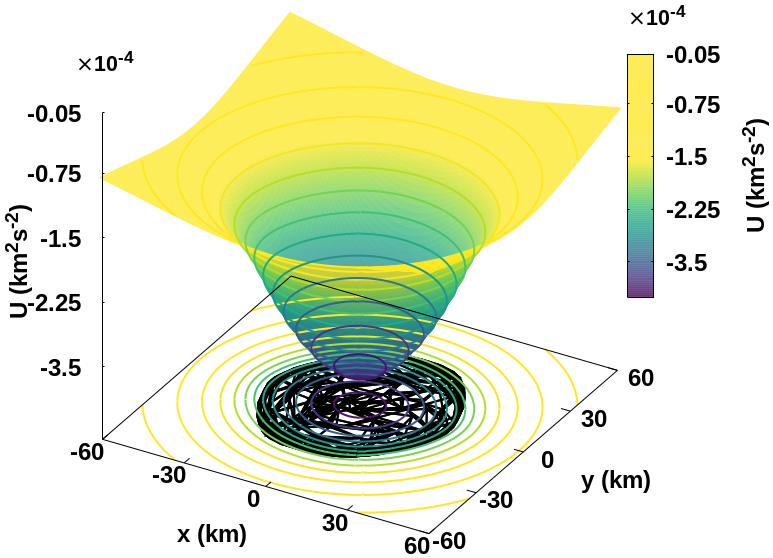}
    \caption{Equipotential surface of Justitia and its contour lines in the xOy plane. The color box denotes the gravitational potential, $U(\mathbf{r})$, in km$^{^2}$\,s$^{-2}$ around Justitia (black).}
    \label{fig:13}
\end{figure*}

As the gravity field of Justitia is dominated by the gravitational potential, in Sections \ref{2.8} and \ref{2.9} we computed the equipotential curves and gravitational acceleration around the asteroid, respectively.
Figure \ref{fig:13} shows the equipotential surface of Justitia and its contour lines in the xOy plane. The colors indicate the gravitational potential $U(\mathbf{r})$. The projection lines of asteroid Justitia are highlighted in black. Figure \ref{fig:13} shows that $U(\mathbf{r})$ near the surface of Justitia ranges from, approximately, -2.25 $\times$ 10$^{-4}$ km$^{2}$\,s$^{-2}$ to -1.5 $\times$ 10$^{-4}$ km$^{2}$\,s$^{-2}$, consistent with geopotential surface (Fig. \ref{fig:2}). This behavior suggests that the centrifugal potential does not significantly affect Justitia's gravity field.

\section{Surface Acceleration}
\label{sec:2.3}
\begin{figure*}
\plotone{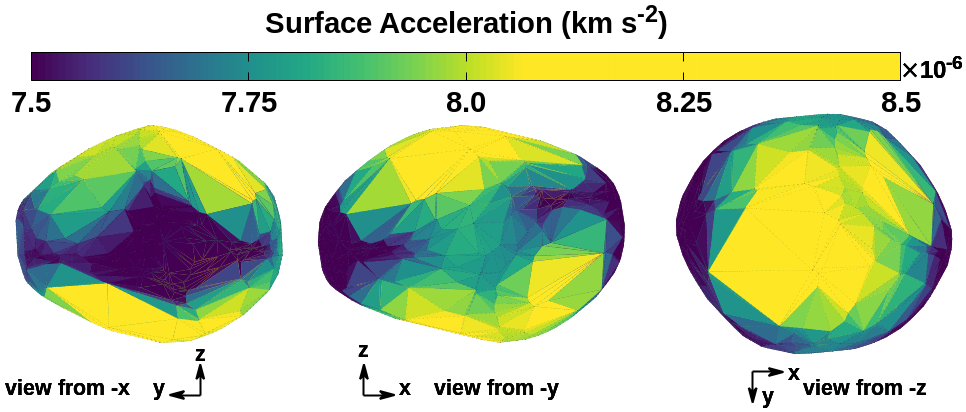}
    \caption{The surface acceleration distribution across asteroid Justitia is presented from three orthogonal perspectives ($-$x, $-$y, and $-$z). The color bar represents the magnitude of the vector sum of the gravitational and centrifugal acceleration vectors $\mid\mid-\nabla V(\mathbf{r})\mid\mid$ at the barycenter of the facet, in km\,s$^ {-2}$.}
    \label{fig:5}
\end{figure*}

By taking the gradient of geopotential at any $\mathbf{r}$ location on the surface of Justitia, we obtain its surface acceleration from Eq. \eqref{eq:2}:
\begin{align}
  \mid\mid-\nabla V(\mathbf{r})\mid\mid.
 \label{eq:2}
\end{align}
The surface acceleration is given as the magnitude of the vector sum of the gravitational and centrifugal acceleration vectors. Figure \ref{fig:5} shows the surface acceleration calculated at the barycenter of each triangular face of the polyhedral model of the asteroid Justitia. Figures \ref{fig:2} and \ref{fig:5} show that the regions of maximum surface acceleration View from -z correspond to the regions of minimum surface geopotential View from -z. In contrast, the regions of minimum surface acceleration correspond to the regions of maximum geopotential (views from -x, -y). This phenomenon occurs because the ratio between the order of magnitude of gravity and the potential of the centrifugal force is $\sim$100 ($\mathcal{K}$ parameter from Eq.\eqref{paramk}). 
\section{Gravitational Acceleration}
\label{2.9}
\begin{figure*}
    \includegraphics[width=0.5\textwidth]{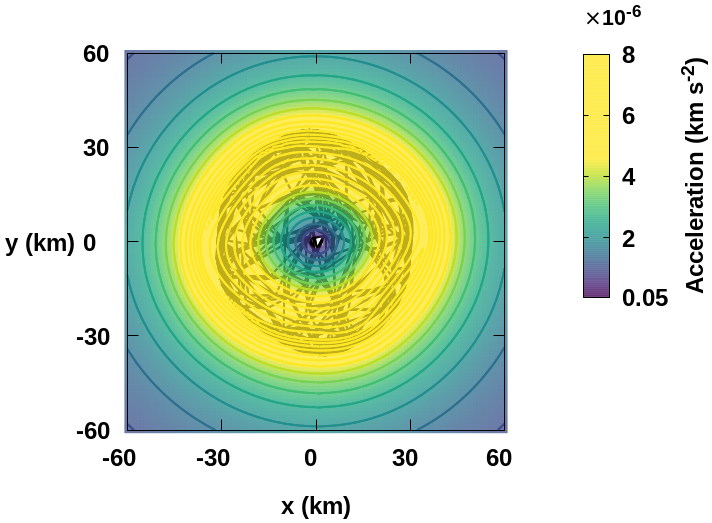}
    \includegraphics[width=0.5\textwidth]{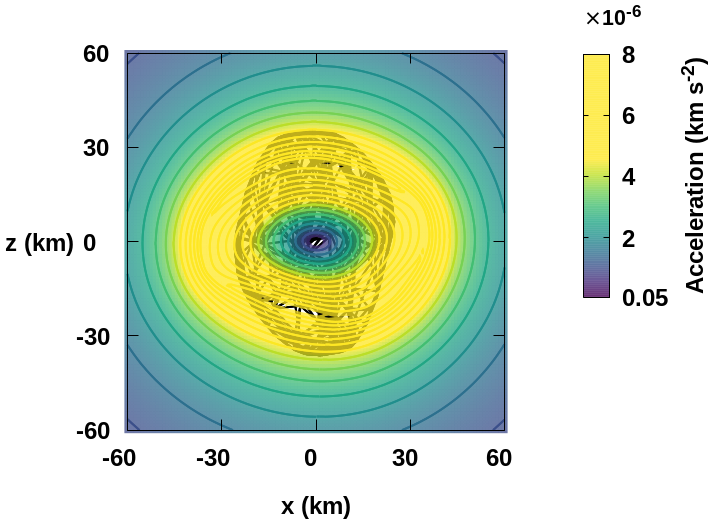}
    \includegraphics[width=0.5\textwidth]{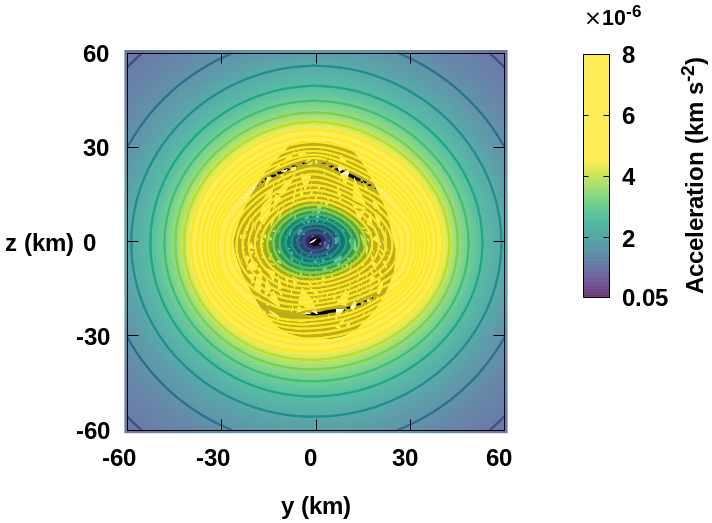}
    \caption{Gravitational acceleration around and inside Justitia in three orthogonal projections: xOy, xOz, and yOz planes. The contour lines represent lines of force. The color bar indicates the magnitude of the gravitational acceleration vector $\mid\mid-\nabla U(\mathbf{r}) \mid\mid$, measured in km s$^{-2}$.}
    \label{fig:14}
\end{figure*}

Figure \ref{fig:14} shows the distribution of the magnitude of the gravitational acceleration vector around asteroid Justitia. The contour lines are visualized through projections: xOy, xOz, and yOz, with color mapping indicating the magnitude of the gravitational acceleration vector $\mid\mid-\nabla U(\mathbf{r}) \mid\mid$. We observe that the gravitational acceleration at the surface of asteroid Justitia is stronger than in regions near its center of mass, due to its mass distribution. The higher values of gravitational acceleration are found near the asteroid's surface. They can reach values of $\sim8\times10^{-6}$\,km\,s$^{-2}$, similar to those found for the surface acceleration (Fig. \ref{fig:5}). 
Comparing the surface acceleration of Justitia with those found in the asteroids Bennu and Psyche, we have that the minimum acceleration of Justitia is $\sim310\times$ larger than that of Bennu \citep{scheeresetal2016} and $\sim18\times$ smaller than that of Psyche \citep{moura2020}. 

Table \ref{tab:surfaceacc} shows the global behavior of the results for the surface acceleration of asteroid Justitia for different densities. We observe that if the density is doubled, the surface acceleration approximately doubles. 

\begin{table}
\centering
\caption{Global maximum and minimum surface acceleration, in km\,s$^{-2}$, computed over the surface of asteroid Justitia for different densities.}
\label{tab:surfaceacc}
\begin{tabular}{c c c}
\hline\hline
Density (g\,cm$^{-3}$) & Min. value ($\times10^{-6}$) & Max. value ($\times10^{-6}$) \\
\hline
1.0 & $7.5$ & $8.5$ \\
1.2 & $8.6$ & $9.8$ \\
1.4 & $10.1$ & $11.5$ \\
1.6 & $11.5$ & $13.1$ \\
1.8 & $13.0$ & $15.0$ \\
2.0 & $14.5$ & $16.4$ \\
\hline
\end{tabular}
\end{table}

\section{Surface Slopes}
\label{sec:2.4}
\begin{figure*}	\plotone{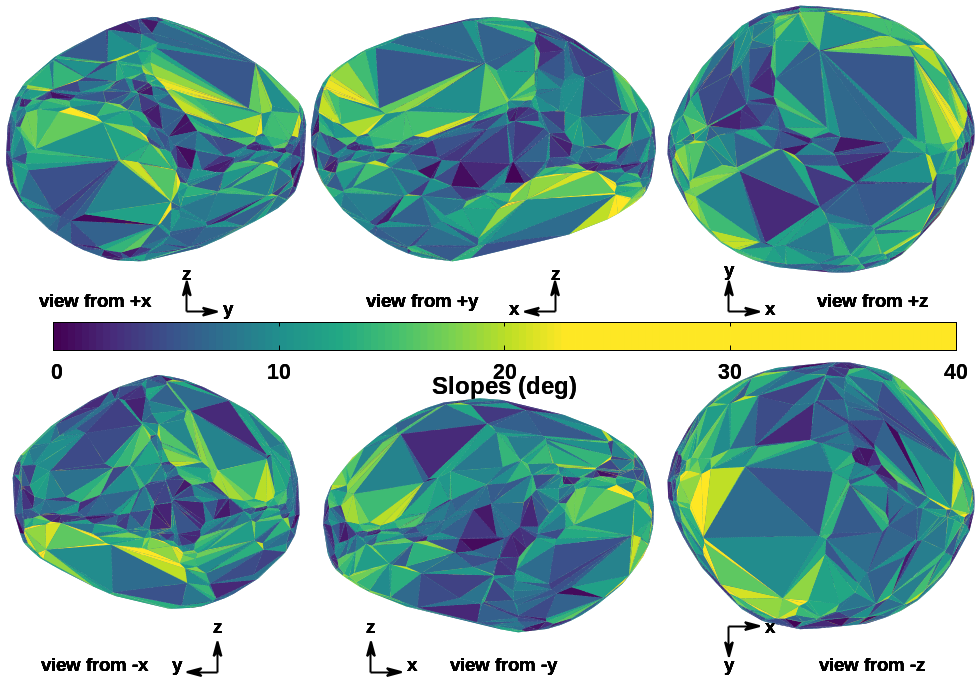}
    \caption{Surface slopes mapped across the surface of Justitia are viewed in six perspectives ($\pm$x, $\pm$y, and $\pm$z). The color bar indicates the slope $\theta$ (degrees) computed numerically from Eq. \eqref{eq:3}.}
    \label{fig:7}
\end{figure*}

Considering a location $\mathbf{r}$ on the surface of Justitia, the relative orientation between the normal unit vector to this surface $\mathbf{\hat{n}}$ and the gradient of geopotential calculated at point $\mathbf{r}$ will establish the surface slope \citep{scheeresetal2016}. The slope $\theta$ is defined as the supplement of the angle formed by the normal vector to the surface and the surface acceleration vector at point $\mathbf{r}$ (see eq. \eqref{eq:3}). This dynamic characteristic provides insight into the motion of particles on the surface of the asteroid Justitia. If $\theta$ > 90$^{\circ}$, the particles are possibly being ejected from the asteroid's surface, while if $\theta$ < 90$^{\circ}$, the particles may accumulate in some regions of the asteroid's surface.
\begin{equation}
\theta = 180^\circ - \arccos\left(\frac{-\nabla V(\mathbf{r}) \cdot \mathbf{\hat{n}}}{\mid\mid-\nabla V(\mathbf{r})\mid\mid}\right)\dfrac{180^{\circ}}{\pi}.
\label{eq:3}
\end{equation}
The analysis of the slopes of Justitia shows that they do not exceed 40$^{\circ}$ (see Fig. \ref{fig:7}). The distribution of maximum slopes reaches values around 40$\mathbf{^\circ}$, and the minimum slopes values close to 0$\mathbf{^\circ}$. Our results show that 97.5\% of the surface of asteroid Justitia has slopes below 30$^{\circ}$.

Some bodies with surface slopes less than 30$^{\circ}$, such as Beta and Gamma of the triple asteroid system (153591) 2001 SN$_{263}$ \citep{winter2020} and (16) Psyche \citep{moura2020}, suggest that globally there is a low possibility of ejecting particles \citep{scheeres2012}.

\subsection{Disturbed Slopes}
\label{DS}
In these sections, we analyze how other perturbing forces influence the slopes on Justitia's surface, for a bulk density of 1.0 g\,cm$^{-3}$. These are the solar radiation pressure (SRP), where we assume Justitia is located in the main belt (aphelion and perihelion) and that the asteroid was previously in the Kuiper belt at $\sim\,40\,$\,\textbf{au}. We also include the perturbation of a third body, Mars and Jupiter, respectively, due to Justitia's location in the central part of the main belt. We emphasize that, even considering different values of our density range for the asteroid Justitia, the overall behavior of the surface perturbations does not change significantly.
\subsection{Solar Radiation Pressure Pertubation}
\label{SRPP}
For the SRP analysis, we consider this perturbation in a particle located at the centroid of each triangular plate of the polyhedral model.
The equation of motion under the influence of SRP is given by Eq. \eqref{SRPEQ} \citep{Burns1979, Mingard1984, amarante2021}.

{
\begin{equation}
\mathbf{a}_{\mathrm{SRP}}(\mathbf{r}) = - \frac{Q_{\mathrm{pr}} S_{0} R_{0}^{2}}{c \, \|\mathbf{r}_{s} - \mathbf{r} \|^{3}} \, \frac{A}{m} \, (\mathbf{r}_{s} - \mathbf{r}),
\label{SRPEQ}
\end{equation}
}

\noindent where $\mathbf{r}_S$ is the position vector of the Sun relative to Justitia, $\|\mathbf{r}_S - \mathbf{r}\|$ is the distance from the particle's located at the barycenter of each face to the Sun, $S_0$ is a solar constant or radiation flux density at the distance of the astronomical unit $R_0$ (au), $c$ is the speed of light, and $Q_{pr}$ is the dimensionless efficiency factor for radiation pressure, which depends on the properties (e.g., density, shape, size) of the particle. In the mathematical model, 
this value is assumed to be $1$ to represent the value of an ideal material. The adopted SRP parameters are presented in Tab.~\ref{tab:srp_params}.

\begin{table}
\centering
{
\caption{The Solar Radiation Pressure parameters used in the mathematical model.}
\begin{tabular}{l l l }
\toprule
Parameter & Value & Units  \\
\midrule
$Q_{pr}$ & $1$ & $--$  \\
$S_0$ & $1.36 \times 10^3$ & kg\,s$^{-3}$  \\
$R_0$ & $1.495978707 \times 10^8$ & km  \\
$c$ & $2.99792 \times 10^5$ & km\,s$^{-1}$  \\
$\rho$ & $1.0$ & g\,cm$^{-3}$ \\
$r_{pe}$ & $1-10^{-1}$ & $\mu m$  \\
$r_{ap}$ & $3.0\times10^{-1}-9.0\times10^{-2}$ & $\mu m$  \\
$r_{Kb}$ & $5.0\times10^{-3}-5.0\times10^{-4}$ & $\mu m$  \\
\bottomrule
\end{tabular}
\label{tab:srp_params}
}
\end{table}

Our objective was to determine the acceleration vector caused by solar radiation pressure (SRP) on a particle at the centroid of each face
of the convex polyhedron and then add to this vector the total acceleration vector calculated in section \ref{sec:2.4}.
After this, we obtain the new total acceleration vector due to SRP, given by Eq. \eqref{totalsrpacc}.
\begin{equation}
a_{total} = a_{SRP}(\mathbf{r}) {-}{\nabla \mathbf{V}(\mathbf{r})}. \label{totalsrpacc}
\end{equation}
\noindent Thus, our analyses show the global behavior of the slopes, varying the particle's radius at the perihelion and aphelion of its orbit in the main belt and also varying the particle's radius in the Kuiper belt, assuming that Justitia may have originated there in the past.

\begin{figure}
    \centering
    \includegraphics[width=1.0\linewidth]{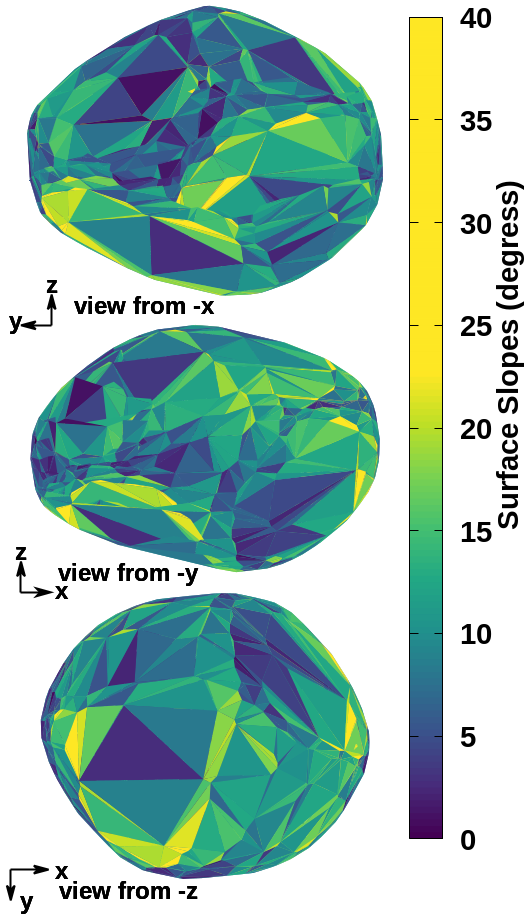}
    \caption{Global behavior of the slopes as a function of particle size of $1\,\mu m$, at approximately 2.06 au. The color scale shows slopes from 0$^{\circ}$ to 40$^{\circ}$}.
    \label{mbpa}
\end{figure}

\begin{figure}
    \centering
    \includegraphics[width=1.0\linewidth]{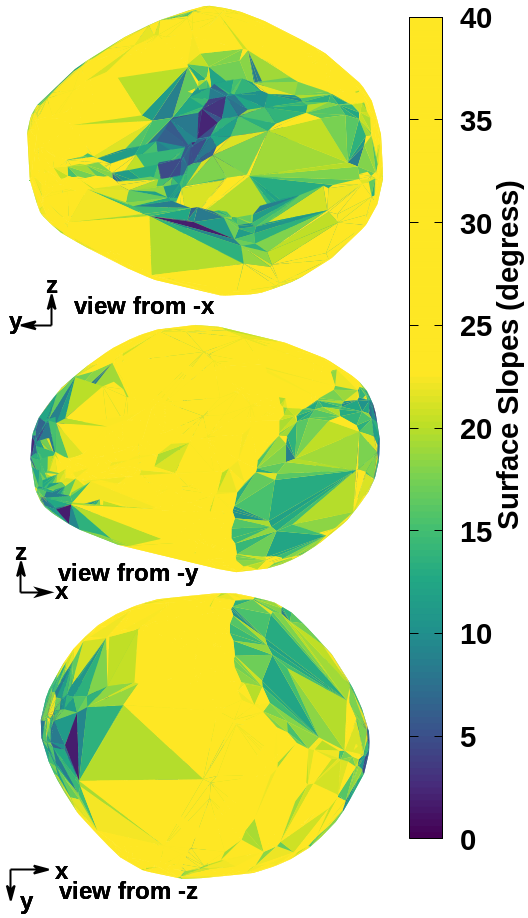}
    \caption{Global behavior of the slopes as a function of particle size of $4.0\times10^{-1}\,\mu m$, at approximately 2.06 au. The color scale shows slopes from 0$^{\circ}$ to 40$^{\circ}$}.
    \label{mbpb}
\end{figure}

\begin{figure}
    \centering
    \includegraphics[width=1.0\linewidth]{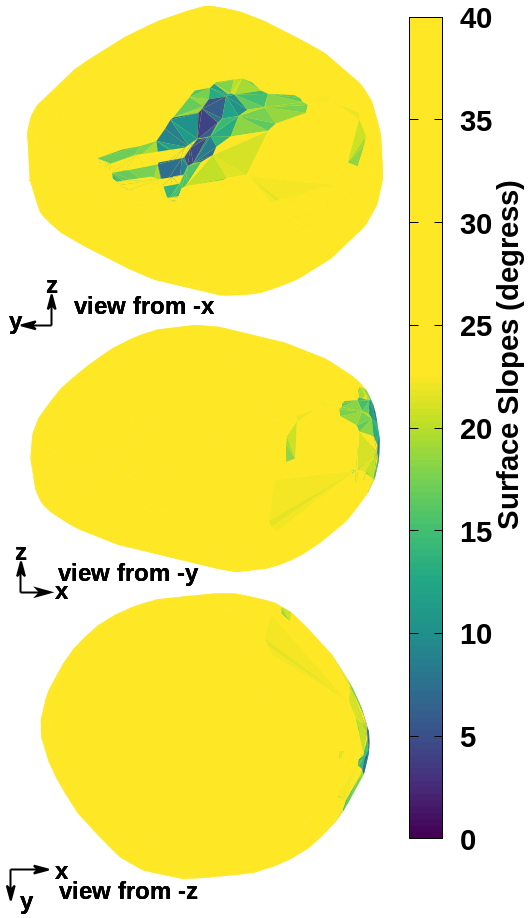}
    \caption{Global behavior of the slopes as a function of particle size of $10^{-1}\,\mu m$, at approximately 2.06 au. The color scale shows slopes from 0$^{\circ}$ to 40$^{\circ}$}.
    \label{mbpc}
\end{figure}

Figures \ref{mbpa}, \ref{mbpb}, and \ref{mbpc}, we fix the pericenter of the asteroid Justitia's orbit, and we vary the particle sizes from $1 \mu m$ to $10^{-1} \mu m$, at the centroid of each face of the convex model.
We find that particles $<\sim4\times 10^{-1} \mu m$ on the surface of asteroid Justitia are significantly influenced by the SRP at the pericenter of the asteroid's orbit, while particles on the order of $1 \mu m$ are not significantly influenced by the SRP.

\begin{figure}
    \centering
    \includegraphics[width=1.0\linewidth]{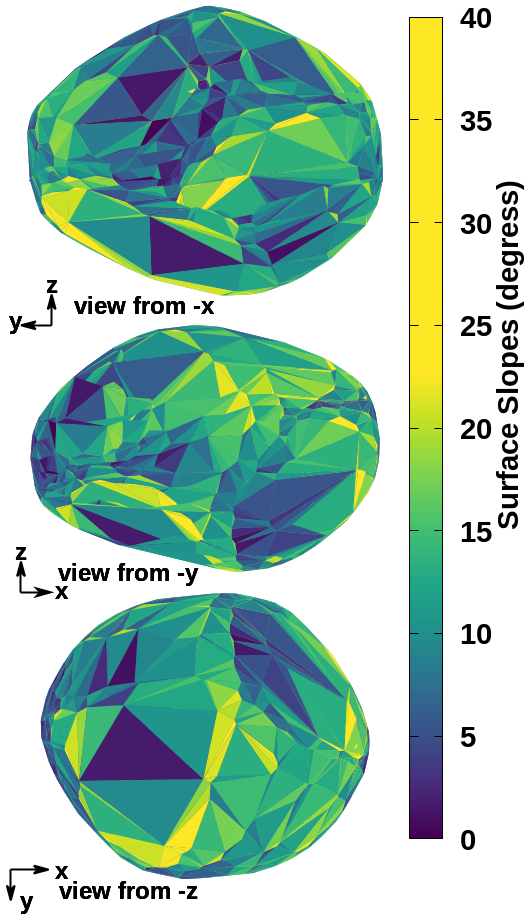}
    \caption{Global behavior of the slopes as a function of particle size of $3.0\times10^{-1}\,\mu m$, at approximately 3.17 au. The color scale shows slopes from 0$^{\circ}$ to 40$^{\circ}$.}
    \label{mbaa}
\end{figure}

\begin{figure}
    \centering
    \includegraphics[width=1.0\linewidth]{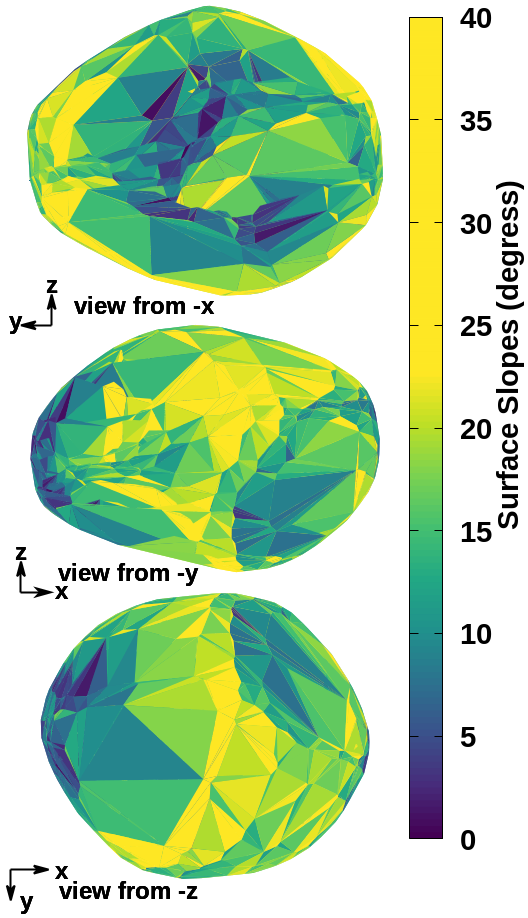}
    \caption{Global behavior of the slopes as a function of particle size of $1.5\times10^{-1}\,\mu m$, at approximately 3.17 au. The color scale shows slopes from 0$\mathbf{^{\circ}}$ to 40$^{\circ}$.}
    \label{mbab}
\end{figure}

\begin{figure}
    \centering
    \includegraphics[width=1.0\linewidth]{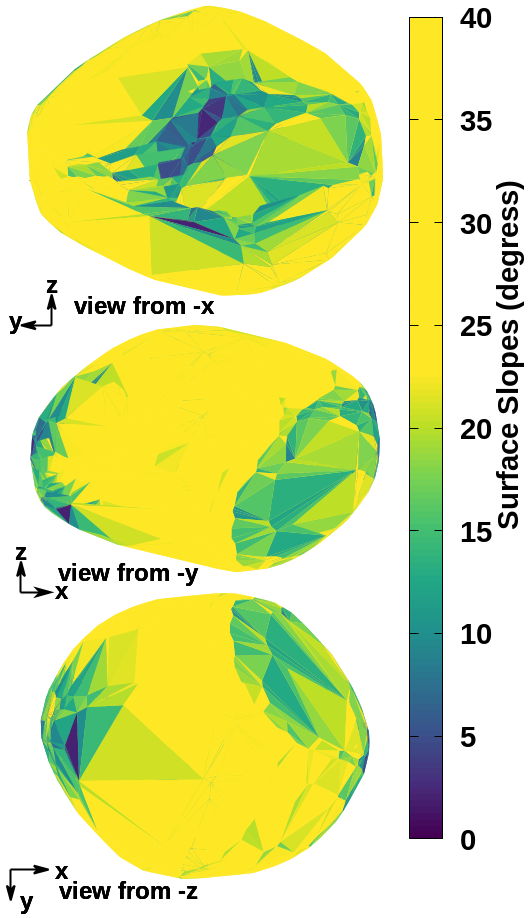}
    \caption{Global behavior of the slopes as a function of particle size of $9.0\times10^{-2}\,\mu m$, at approximately 3.17 au. The color scale shows slopes from 0$^{\circ}$ to 40$^{\circ}$}.
    \label{mbac}
\end{figure}

Figures \ref{mbaa}, \ref{mbab}, and \ref{mbac},  we fix the apocenter of the asteroid Justitia's orbit, and we vary the particle sizes from $3\times10^{-1}\,\mu m$ to $9.0\times10^{-2}\,\mu m$, at the centroid of each face of the convex model. 
We find that particles $<\sim1.5\times10^{-1}\,\mu m$ on the surface of asteroid Justitia are significantly influenced by the SRP at the apocenter of the asteroid's orbit, while particles $> 3.0\times10^{-1}\,\mu m$ are not significantly influenced by the SRP. For comparison purposes, the solar radiation pressure does not significantly affect particles across the surface of comet 67P with sizes $\sim> 10^{-3}$\,cm at apocenter and $\sim>10^{-1}$\, cm at pericenter \citep{Braga2025}.

\begin{figure}
    \centering
    \includegraphics[width=1.0\linewidth]{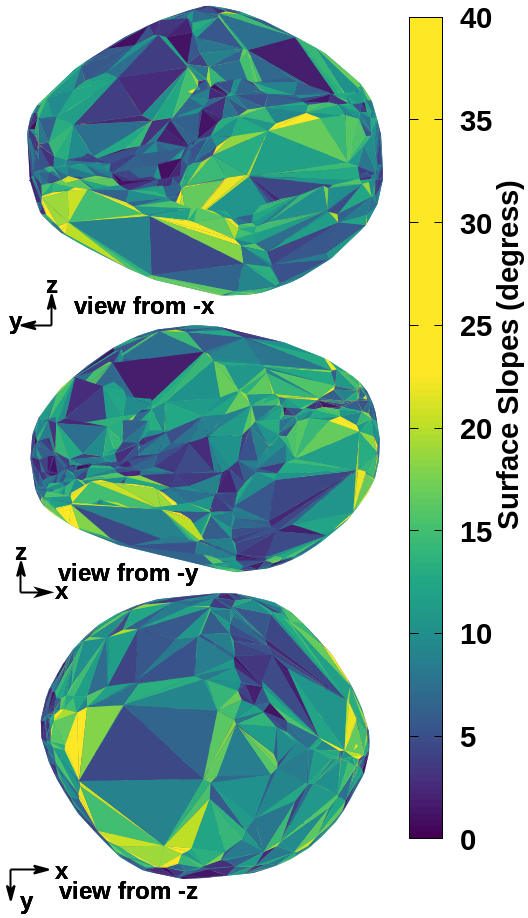}
    \caption{Global behavior of the slopes as a function of particle size of $5.0\times10^{-3}\,\mu m$, at approximately 40 au. The color scale shows slopes from 0$^{\circ}$ to 40$^{\circ}$}.
    \label{kbpa}
\end{figure}

\begin{figure}
    \centering
    \includegraphics[width=1.0\linewidth]{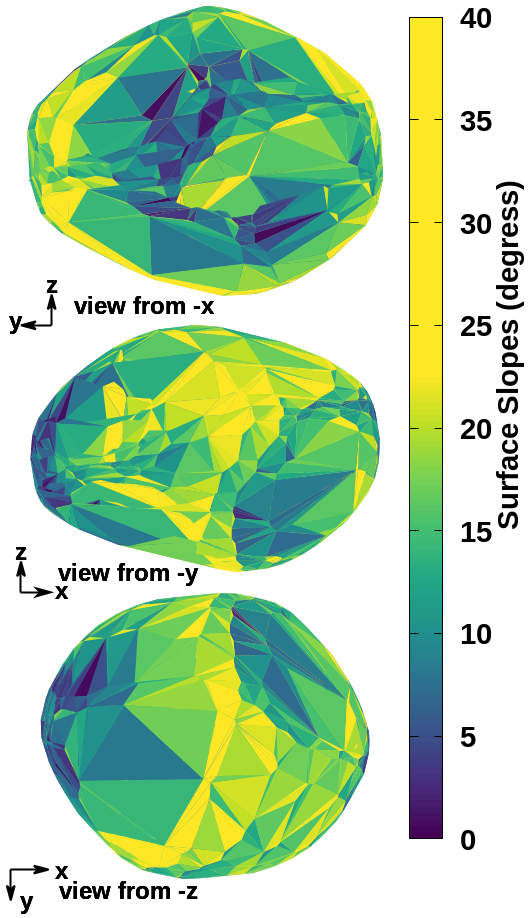}
    \caption{Global behavior of the slopes as a function of particle size of $1.0\times10^{-3}\,\mu m$, at approximately 40 au. The color scale shows slopes from 0$^{\circ}$ to 40$^{\circ}$}.
    \label{kbpb}
\end{figure}

\begin{figure}
    \centering
    \includegraphics[width=1.0\linewidth]{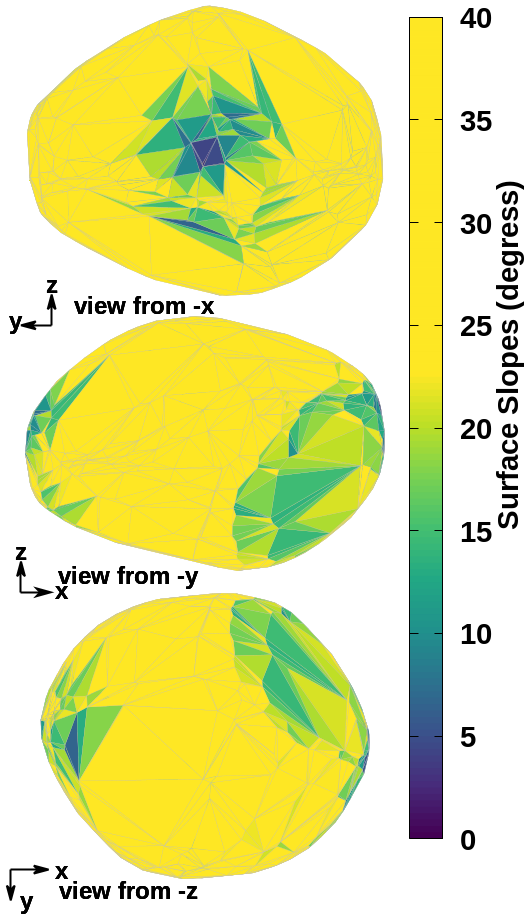}
    \caption{Global behavior of the slopes as a function of particle size of $5.0\times10^{-4}\,\mu m$, at approximately 40 au. The color scale shows slopes from 0$^{\circ}$ to 40$^{\circ}$}.
    \label{kbpc}
\end{figure}

Figures \ref{kbpa}, \ref{kbpb}, and \ref{kbpc}, we set Justitia at 40 au and vary the particle sizes from $5.0\times10^{-3}\,\mu m$ to $5.0\times10^{-4}\,\mu m$.
We identified that particles $<\sim1.0\times10^{-3}\,\mu m$ are significantly influenced by SRP, while particles $> 5.0\times10^{-3}\,\mu m$
are not significantly influenced by SRP.
However, for this case, it is important to emphasize that other electromagnetic forces influence particles smaller than $10^{-2}\, \mu m$
around irregular bodies, causing the gravitational field to become a disturbance \citep{mingard1982, horanyi, amarante2022}. In this work, we do not consider these forces.

\subsection{The disturbing effects of a Third-Body}
\label{TBE}
We analyze the slopes on the surface of Justitia, considering the gravitational perturbations of Mars and Jupiter (Mars (M$_a$), Jupiter (J)) on a particle located
at the centroid of each face of the 3-D shape model, using Eq.\eqref{eqofmotionpertubed}:

\begingroup
\begin{equation}
\mathbf{a_{\text{pert}}}(\mathbf{r}) = G M_{\text{Ma/J}}
\left(
\frac{\mathbf{r}_{\mathbf{\text{Ma/J}}} - \mathbf{r}}{\lVert \mathbf{r}_{\text{Ma/J}} - \mathbf{r}\rVert^3}
-
\frac{\mathbf{r}_{\text{Ma/J}}}{\lVert \mathbf{r}_{\text{Ma/J}}\rVert^3}
\right)
\label{eqofmotionpertubed}
\end{equation}
\endgroup
\noindent where $M_\text{Ma} = 6.4\times10^{23}$\,kg is the total mass of Mars, $M_\text{J} = 1.9\times10^{27}$\,kg is the total mass of Jupiter, $\mathbf{r_{M_a}}$ is the position vector of Mars concerning Justitia, $\mathbf{r_J}$ is the position vector of Jupiter concerning Justitia, $\|\mathbf{r_{M_a}-r}\|$
is the distance from the barycenter vector of each face to Mars, and $\|\mathbf{r_J-r}\|$ is the distance from the barycenter vector of each face to Jupiter.

The new slopes are now computed from Eq. \eqref{eq:novaslope}:

\begingroup
\begin{equation}
\theta = \dfrac{180^\circ}{\pi}
\bigg[
1 - \arccos\!\left(
\frac{
-\mathbf{a}_{\mathbf{total}}(\mathbf{r}) \cdot \mathbf{\hat{n}}
}{
\lVert -\mathbf{a}_{\mathbf{total}}(\mathbf{r}) \rVert
}
\right)
\bigg],
\label{eq:novaslope}
\end{equation}
\endgroup

\noindent where $\mathbf{{a}_{{total}}=a_{pert}{(r)-\nabla V(\mathbf{r})}}$, and $\mathbf{\hat{n}}$ is the normal vector calculated at the centroid of each face.

\begin{figure*}
    \centering
	\plotone{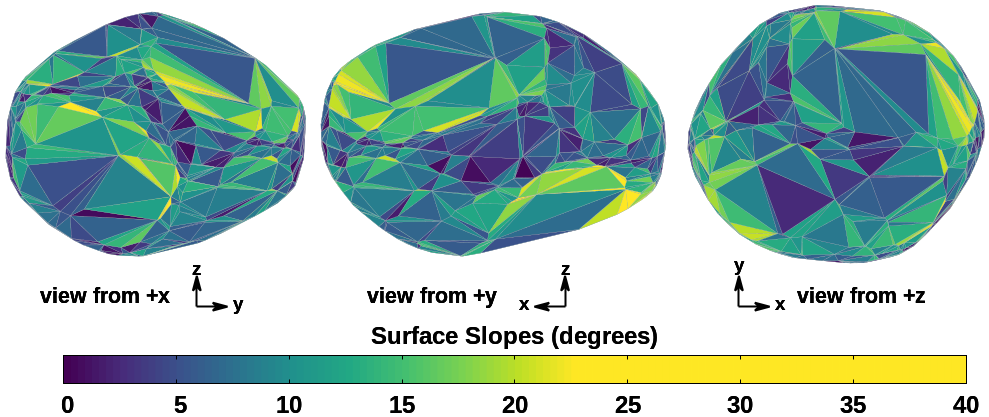}
    \caption{Surface slopes mapped on the surface of Justitia considering the Mars gravitational perturbation at 1.08 au distance from asteroid Justitia, respectively.}
    \label{mars}
\end{figure*}

\begin{figure*}
    \centering
	\plotone{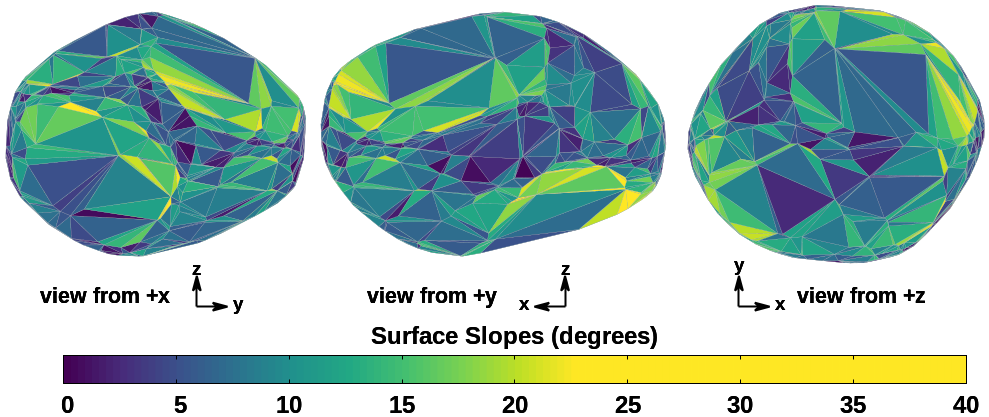}
    \caption{Surface slopes mapped on the surface of Justitia considering the Jupiter gravitational perturbation at 2.61 au distance from asteroid Justitia, respectively.}
    \label{jup}
\end{figure*}

Figures \ref{mars}, \ref{jup} show the behavior of the slopes due to perturbations from Mars and Jupiter, respectively. As we can see, the two bodies do not significantly affect the surface slopes of Justitia.

\section{Surface Tilt}
\label{sec:2.5}


Let $\mathbf{r}$ be an arbitrary point on the surface of Justitia. The orientation of the surface at this point, in the reference frame fixed to the asteroid, is determined by the direction of the normal unit vector $\mathbf{\hat{n}}$. We adopt as the reference direction the vector that originates at the center of mass of the asteroid and points to the arbitrary location $\mathbf{r}$ on its surface. Then, the angle between the normal vector $\hat{n}$ to the surface and the barycenter vector $\mathbf{r}$ is specified as tilt \citep{scheeresetal2016}. 

\begin{figure*}
    \centering
	\plotone{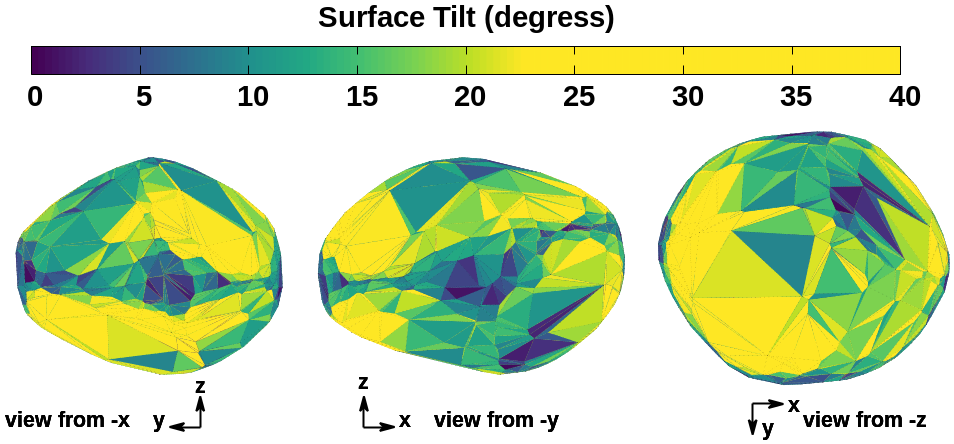}
    \caption{Surface tilt is mapped across the surface of Justitia, considering its 3-D polyhedral shape model with 574 vertices and 1,144 faces. The color code gives the surface tilt in degrees.}
    \label{fig:9}
\end{figure*}

Figure \ref{fig:9} shows the tilt distribution across the surface of asteroid Justitia. We observe that the variation of the tilt does not exceed 50$^{\circ}$. The maximum values for tilt are found around 50$^{\mathbf{\circ}}$, while the minimum values are are found at approximately 0$\mathbf{^\circ}$. 

Furthermore, this entirely geometric quantity can guide the landing of a spacecraft if it wishes to reach the object's surface at a 90$^{\circ}$ angle. Since the tilt is an entirely geometric quantity, it will be the same regardless of the density value.
\section{Escape Speed}
\label{sec:2.6}
The escape speed over the surface of asteroid Justitia can be calculated by combining its angular speed, surface orientation, and gravitational potential. Equation \eqref{eq:4} gives the escape speed v$_e$ \citep{scheeresetal2016}.
\begin{equation}
v_e = -\hat{\mathbf{n}} \cdot (\boldsymbol{\omega} \times \mathbf{r}) + \sqrt{\left[\hat{\mathbf{n}} \cdot (\boldsymbol{\omega} \times \mathbf{r})\right]^2 
- 2U_{\text{min}} - \lVert\boldsymbol{\omega} \times \mathbf{r}\rVert^2},
\label{eq:4}
\end{equation}
\noindent where $\mathbf{r}$ is the radius vector from the asteroid's center of mass to the local surface, 
\( U_{\text{min}} = \min \left[ U, -\frac{GM}{\mid{\mathbf{r}}\mid} \right] \), $G = 6.67430\times10^{-20}$\,km$^{3}$\,kg$^{-1}$\,s$^{-2}$ is the gravitational constant, $M = 1.02\times10^{17}$\,kg is the mass of the Justitia for a density of 1.0 g\,cm$^{-3}$, $\boldsymbol{\omega}$ is the angular velocity vector and $\hat{\mathbf{n}}$ is the normal unit vector.

\begin{figure*}
	\plotone{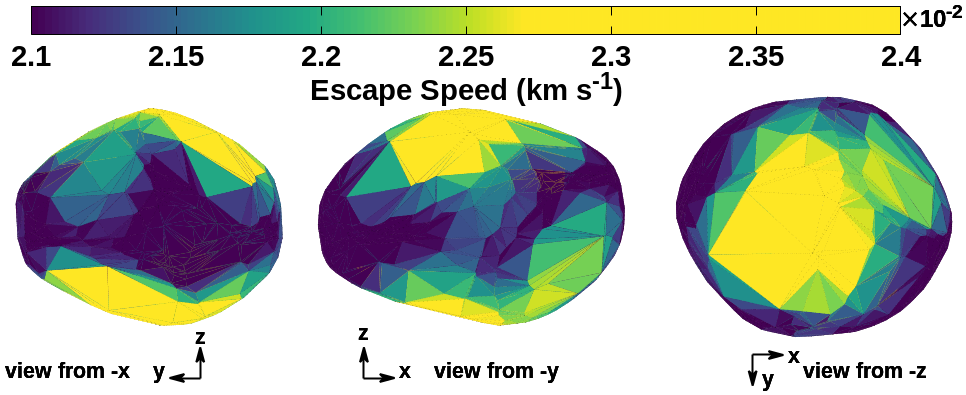}
    \caption{Local normal escape speed v$_e$ calculated over the surface of Justitia, in km\,s$^{-1}$, from three perspectives: $-$x, $-$y, and $-$z. The color box denotes the values of escape speed.}
    \label{fig:10}
\end{figure*}

If the particle's speed launched exceeds the escape speed of asteroid Justitia, the particle will escape. We assume that the particle is launched in the direction normal to the local triangular facet of the asteroid.
Fig. \ref{fig:10} shows the results of escape speed mapped across the surface of the asteroid Justitia.

Comparing Fig.\,\ref{fig:10} and Fig. \ref{fig:2}, the regions of maximum escape speed correspond to the regions with minimum geopotential. This global behavior of escape speed also appears in other slowly rotating minor bodies, such as (4179) Toutatis \citep{scheeres1998} and KBO Arrokoth \citep{amaranteandwinter2020}.

\section{Equilibrium Points}
\label{2.7}
In this section, we numerically computed the external and inner equilibrium points of the asteroid Justitia using the following Eq. \eqref{eq:6}.
\begin{equation}
\nabla V(\mathbf{r}) = 0.
\label{eq:6}
\end{equation}
\begin{figure*}
	\plotone{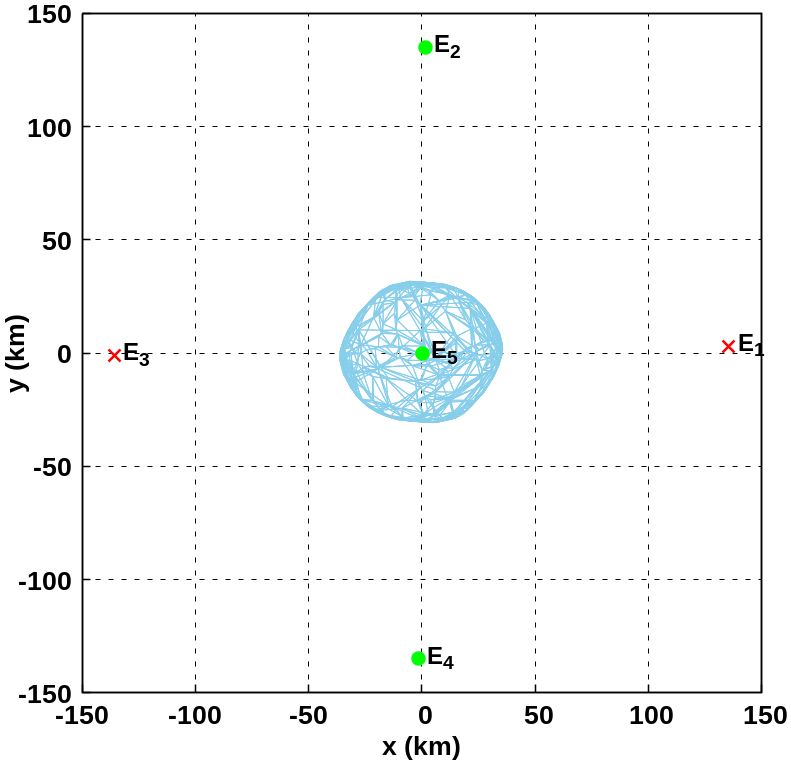}
    \caption{Location and stability of the five equilibrium points about asteroid Justitia are projected onto the xOy plane. The crosses points denote saddle–center–center points, which are hyperbolically unstable due to their saddle-like topology. In contrast, the circles points represent center–center–center points, indicating linear stability.}
    \label{fig:12}
\end{figure*} 
\noindent We found five equilibrium points for the rotational period of 33.12962\,h and a density of 1 g\,cm$^{-3}$ for Justitia (Fig. \ref{fig:12}). Two of them are unstable (E$_1$, E$_3$) and three are linearly stable (E$_2$, E$_4$, E$_5$), where E$_5$ is located internally to the asteroid. In Table \ref{tab:1}, we show the locations of the equilibrium points around Justitia for different density values. We observe that as the density increases, the gravitational attraction becomes stronger, while the centrifugal force remains unchanged. As a result, the equilibrium points shift outward, seeking positions where the centrifugal force balances the increased gravitational pull. It is interesting to note that the internal equilibrium point E$_5$ shifts inward as the density increases.

Equilibrium points are also classified by stability. To analyze these stabilities, we applied a linearization method using the eigenvalues of the characteristic equation, as described in cases by \citet{Jiang2015}.
We defined these classifications as follows: equilibrium points E$_2$, E$_4$, and E$_5$ have three pairs of imaginary eigenvalues, meaning they are classified as Case 1 (linearly stable). On the other hand, the equilibrium points E$_1$ and E$_3$ are classified as Case 2, as they present two pairs of imaginary eigenvalues (unstable).

Table \ref{tab:2} (see Appendix \ref{appendixB}) gives the eigenvalues for each equilibrium point about Justita, considering different density values. Additionally, we have assigned these topological classifications, which remain unchanged as the density increases. Based on these eigenvalues, E$_1$ and E$_3$ have a saddle-center-center configuration, characterizing hyperbolic instability, while E$_2$, E$_4$, and E$_5$ follow a center-center-center structure, being linearly stable. Thus, the asteroid Justitia can be characterized based on the distribution of its external equilibrium points as an asteroid of type I \citep{scheeres1994}.
\begin{table*}
\centering
\caption{Location of equilibrium points around Justitia. Equilibria are computed numerically through \textsc{Minor-Equilibria} package \citep{amaranteandwinter2020}, considering the polyhedron method and an accuracy of 10$^{-5}$ for the Newton-Raphson method, assuming different densities for Justitia and a rotational period of 33.12962\,h.}
\label{tab:1}
\resizebox{1.0\textwidth}{!}{%
\begin{tabular}{l c c c c}
\hline\hline
Equilibrium Point & Density (g\,cm$^{-3}$) & \multicolumn{1}{c}{$x$ (km)} & \multicolumn{1}{c}{$y$ (km)} & \multicolumn{1}{c}{$z$ (km)} \\
\hline
\multirow{6}{*}{E\textsubscript{1}} 
& 1.0 & 135.56449 & 3.06619 & 0.01870 \\
& 1.2 & 143.98391 & 2.96037 & 0.01621 \\
& 1.4 & 151.51594 & 2.87370 & 0.01435 \\
& 1.6 & 158.36302 & 2.80081 & 0.01292 \\
& 1.8 & 164.66247 & 2.73825 & 0.01177 \\
& 2.0 & 170.51225 & 2.68369 & 0.01083 \\
\hline
\multirow{6}{*}{E\textsubscript{2}} 
& 1.0 & 1.49254 & 135.03974 & -0.03475 \\
& 1.2 & 1.38354 & 143.48737 & -0.03060 \\
& 1.4 & 1.29792 & 151.04212 & -0.02749 \\
& 1.6 & 1.22820 & 157.90809 & -0.02506 \\
& 1.8 & 1.16991 & 164.22363 & -0.02311 \\
& 2.0 & 1.12013 & 170.08734 & -0.02150 \\
\hline
\multirow{6}{*}{E\textsubscript{3}} 
& 1.0 & -135.62925 & -1.05365 & -0.00842 \\
& 1.2 & -144.04152 & -0.94723 & -0.00646 \\
& 1.4 & -151.56810 & -0.86034 & -0.00511 \\
& 1.6 & -158.41086 & -0.78740 & -0.00414 \\
& 1.8 & -164.70680 & -0.72490 & -0.00341 \\
& 2.0 & -170.55364 & -0.67045 & -0.00285 \\
\hline
\multirow{6}{*}{E\textsubscript{4}} 
& 1.0 & -1.68310 & -135.01066 & -0.03519 \\
& 1.2 & -1.58656 & -143.46152 & -0.03098 \\
& 1.4 & -1.51020 & -151.01873 & -0.02783 \\
& 1.6 & -1.44766 & -157.88665 & -0.02537 \\
& 1.8 & -1.39512 & -164.20377 & -0.02339 \\
& 2.0 & -1.35007 & -170.06881 & -0.02175 \\
\hline
\multirow{6}{*}{E\textsubscript{5}} 
& 1.0 & 0.13756 & -0.14938 & 0.05430 \\
& 1.2 & 0.13728 & -0.14911 & 0.05431 \\
& 1.4 & 0.13707 & -0.14892 & 0.05431 \\
& 1.6 & 0.13692 & -0.14878 & 0.05432 \\
& 1.8 & 0.13680 & -0.14867 & 0.05432 \\
& 2.0 & 0.13671 & -0.14858 & 0.05432 \\
\hline
\end{tabular}%
}
\end{table*}

\subsection{Evolution of the Equilibrium Points}
In this section, we propose that Justitia may have had a faster spin in the past. From this, we analyze the evolution of the equilibrium points around Justitia for different rotational periods. 

\begin{figure*}
	\plotone{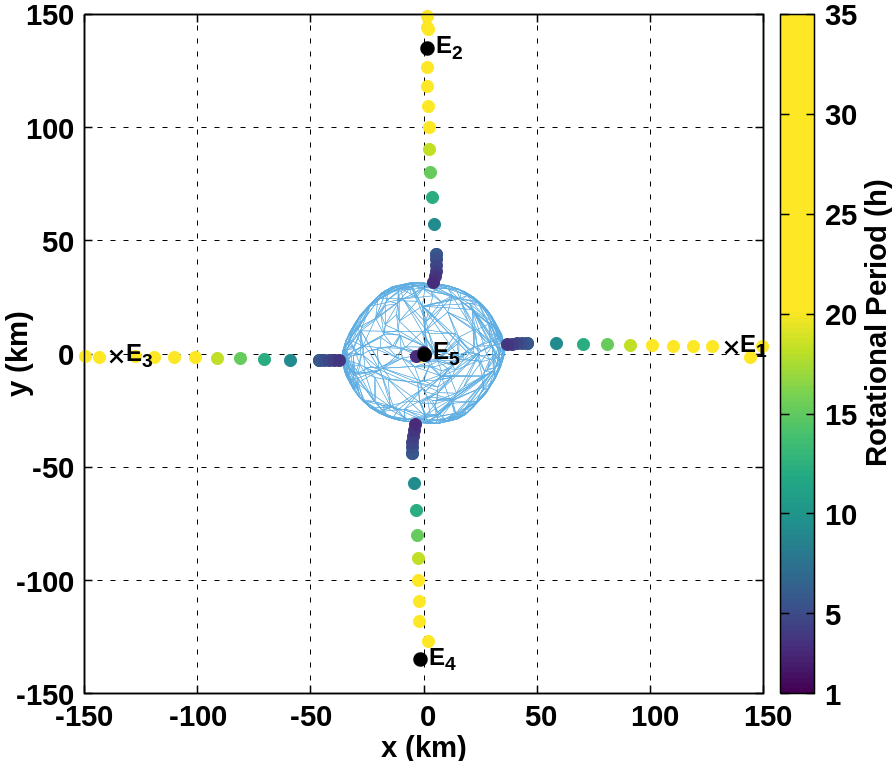}
    \caption{Distribution of equilibrium points around a slowly rotating body, showing how they approach the body as the rotational period increases. The positions of the five original equilibrium points around Justitia are in black. Each colored point represents the equilibrium position for a given rotational period, as indicated by the colored bar on the right (in hours). In the center of the image, a representation of the surface of Justitia.}
    \label{JKB}
\end{figure*}

Fig.\,\ref{JKB} illustrates the locations of the different equilibrium points calculated around Justitia. As we can see, as we decrease the rotational period of 33 hours, the equilibrium points move closer to the surface of the body, a common characteristic of slowly rotating bodies, as explained in the previous section on density variation. In other words, as we decrease the rotational period, the centrifugal force increases. For the balance between gravity and this force to be limited, the equilibrium point needs to be closer to the body's surface, where the gravitational force is stronger.

\section{Zero-Velocity Curves}
\label{3.0}
Zero-velocity curves divide space into regions of allowed and not allowed motion. In general, they are any surfaces that ensure that a particle is trapped or limited by the surface of the ellipsoid \citep{scheeres1994}. The dynamical equations governing particle motion in the vicinity of Justitia, assuming uniform rotation about its $z$-axis, can be expressed as \citet{jiang2014}:
\begin{equation}
    \ddot{x} - 2\omega\dot{y} + \frac{\partial V}{\partial x} = 0,\,\ddot{y} + 2\omega\dot{x} + \frac{\partial V}{\partial y} = 0,\,\ddot{z} + \frac{\partial V}{\partial z} = 0. 
    \label{eq:7}
\end{equation}
\noindent It is important to emphasize that these equations are related to the return speed, following the inequality:
\begin{align}
J - V(r) \geq 0.
\label{eq:8}
\end{align}
\noindent Thus, based on the value of the Jacobi constant $J$, the space around the asteroid can be divided into regions where the particle's movement is allowed or prohibited, reducing areas where its trajectory will be confined. Once this inequality is violated, there will be regions where movement is prohibited. Consequently, when $V(\mathbf{r}) = J$, the limits of these regions are called the zero-velocity curves \citep{murray1999}. 

\begin{figure*}
    \includegraphics[width=0.5\textwidth]{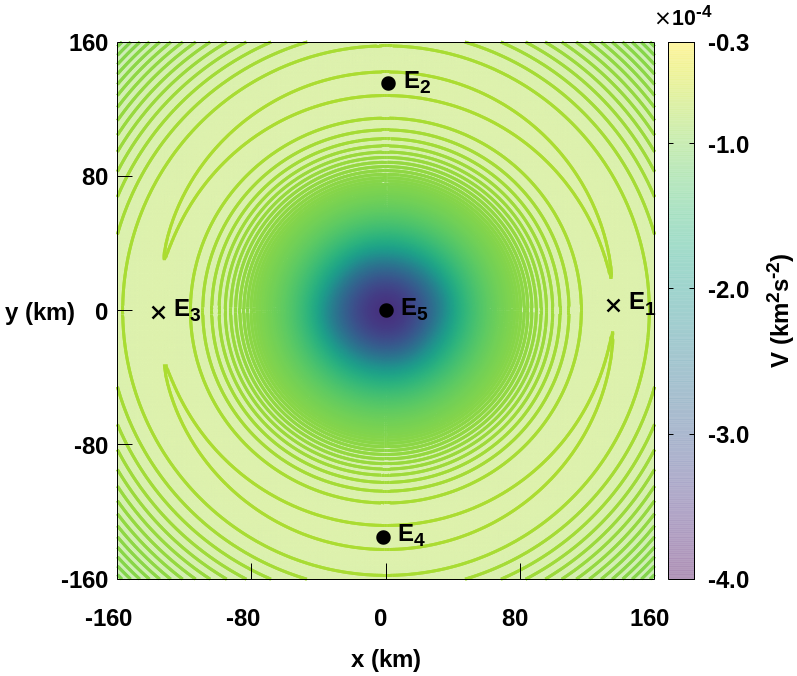}
    \includegraphics[width=0.5\textwidth]{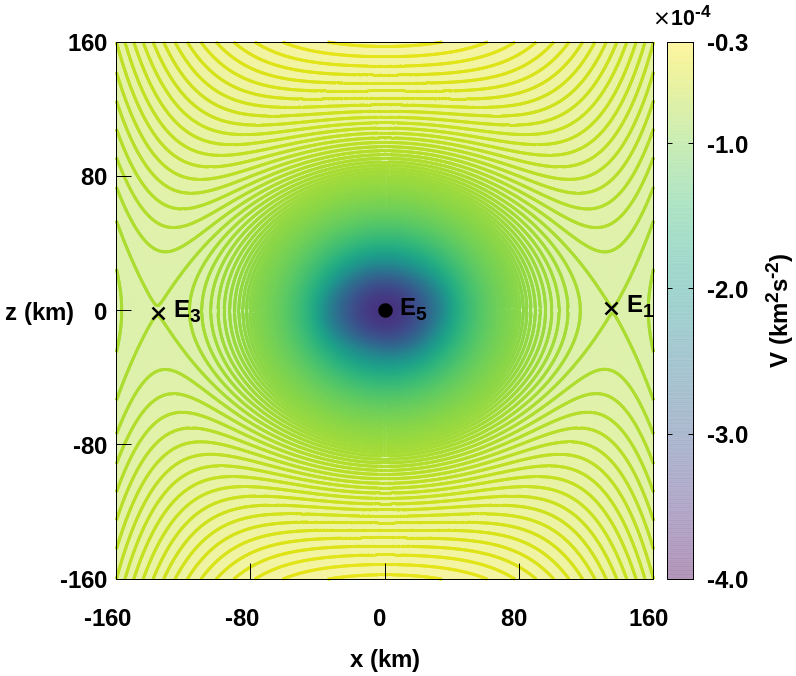}
    \includegraphics[width=0.5\textwidth]{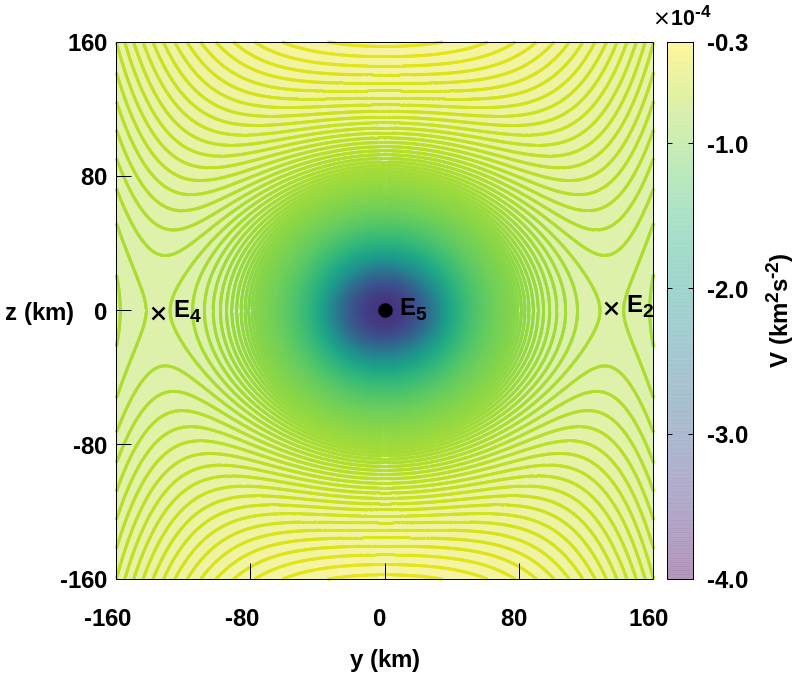}
    \caption{Plot of Justitia's $V(\mathbf{r})$ in the $xOy$, $xOz$, and $yOz$ planes, respectively. The color bar illustrates the zero-velocity curves of Justitia's Jacobi constant $J$, in km$^{2}$\,s$^{-2}$. The dots represent the equilibrium point locations for a rotational period of \textbf{33.12962}\,h and a density of 1\,g\,cm$^{-3}$.}
    \label{fig:15}
\end{figure*}

Figure \ref{fig:15} presents the geopotential $V(\mathbf{r}) = J$ distribution of Justitia across three projection planes: xOy, xOz, and yOz planes, with contour lines representing constant values of the Jacobi integral. The topological structure of the equilibrium points can be determined by examining zero-velocity curves in these projection planes. The topological structure of the xOy plane confirms the existence of five equilibrium points about the asteroid Justitia. The equilibrium points E$_1$ and E$_3$ in projection planes xOy and xOz exhibit saddle-type geometries (unstable). In contrast, equilibrium points E$_2$ and E$_4$ display central-type characteristics (linearly stable) in the xOy plane. However, they present saddle-type geometries in the yOz plane. Additionally, equilibrium point E$_5$ is a central-type point in the three projection planes.

\section{Return Speed}
\label{3.1}

The curve that encompasses the asteroid Justitia and intersects at the equilibrium point with minimum Jacobi constant value $J'$ defines the return speed $v_r$ as Eq. \eqref{eq:5} \citep{scheeres2012}:
\begin{equation}
v_r = \sqrt{2\left(J' - V(\mathbf{r})\right)},
\label{eq:5}
\end{equation}
\noindent where $V(\mathbf{r})$ represents the geopotential at location $\mathbf{r}$.]

The necessary condition for a particle to escape from Justitia's surface is called its return speed. However, this condition is necessary but not sufficient; in other words, if the particle has a speed greater than the return speed, it will not immediately return to the surface, but may enter a temporary orbit around the asteroid and eventually escape.

\begin{figure*}
	\plotone{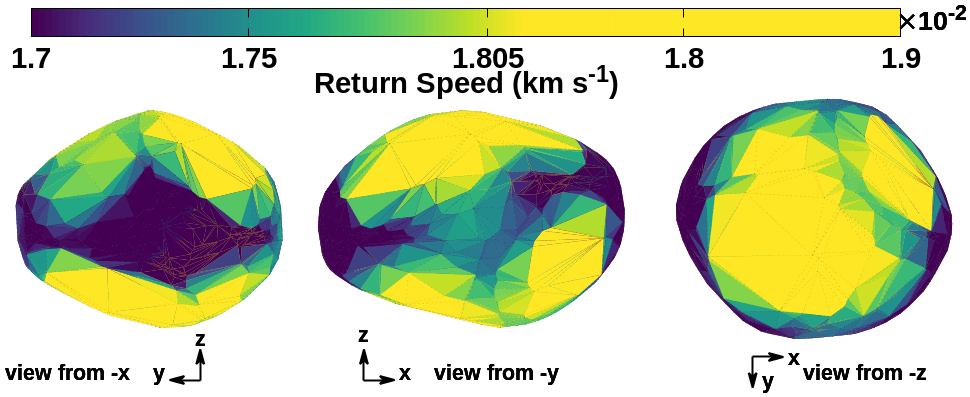}
    \caption{Return speed computed across the surface of asteroid Justitia. The color box denotes the values of Return speed, in km\,s$^{-1}$.}
    \label{fig:16}
\end{figure*}

The Fig \ref{fig:16} shows the mapping of the return speed on the surface of Justitia, considering the bulk density of 1.0 g\,cm$^{-3}$. The value of $J'$ used is 7.60$\times$10$^{-5}$ km$^{2}$\,s$^{-2}$ and is associated with the equilibrium point E$_3$. Note that the minimum values of the return speed (views from -x, -y) are located in the regions of minimum escape speed (Fig. \ref{fig:10}). In contrast, the maximum values (View from -z) are distributed over the regions associated with maximum values of escape speed across the surface of asteroid Justitia. 

However, these characteristics describe physically distinct phenomena. While escape speed represents the minimum speed a particle must have to completely overcome the body's gravitational 
field and move away from it, the return speed corresponds to the highest initial speed a particle can have when launched from the surface of Justitia and remain gravitationally bound to the body before returning to its surface.  

\section{Orbits Around Asteroid (269) Justitia}\label{sec14}
The analysis of the orbital dynamics around the asteroid Justitia was conducted based on a simplified model of its shape, in which an equivalent ellipsoid approximates its gravitational potential.

We use the inertia momentum constrained by the 3-D polyhedral surface and computed numerically to infer the ellipsoidal dimensions ($a$, $b$, and $c$ semi-axes) of the asteroid Justitia, given an ``equivalent ellipsoidal'' shape \citep{Dobrovolskis1998}. Also, the parameters $\mathcal{K}$, $M$, and $R$, discussed in the following paragraphs, were obtained from the original 3-D polyhedral shape model. To this end, the potential was expanded as a series of spherical harmonics up to fourth order, providing an approximate representation of the gravitational perturbations associated with its irregular shape. The adoption of the simplified model reduces the computational cost involved in the analysis. 

Writing the effective potential $V$ in Cartesian coordinates, as given by Eq.~\eqref{potefetivo}, where $r = \sqrt{x^2 + y^2 + z^2}$ represents the distance from the point to the center of the asteroid, and $C_{nm}$, with $n$ and $m$ being the degree and order of the expansion, respectively, represent the spherical harmonic coefficients derived from the polyhedral model described in Section~\ref{shapemodel} and listed in Table~\ref{Coeficientes}. The dimensionless parameter $\mathcal{K}$ corresponds to the ratio between gravitational acceleration and centrifugal acceleration.
\begin{align}
    \mathcal{K}=\frac{GM}{\omega^2 R^3} \in (0,\infty),
\end{align} \label{paramk}
\noindent where $\mathcal{K} < 1$ means fast rotation of the asteroid, and $\mathcal{K} > 1$ means slow rotation \citep{Riaguas2001}. $G$ the gravitational constant, $M$ the total mass of the asteroid, $R$ the equatorial radius, and $\omega$ the angular speed are all presented in Tab. \ref{Parametros}.
\resizebox{0.6\textwidth}{!}{%
\hspace{-3cm}
\begin{minipage}{\textwidth}
\begin{align}
V =\; & \frac{1}{2}(x^2 + y^2) \nonumber\\
& + \frac{\mathcal{K}}{r} \left[ 
1 - \left( \frac{1}{r} \right)^2 \left(
\frac{C_{20}(x^2 + y^2 - 2z^2)}{2r^2} 
+ \frac{3C_{21}xz}{r^2} 
- \frac{3C_{22}(x^2 - y^2)}{r^2}
\right) \right. \nonumber\\
& + \left( \frac{1}{r} \right)^3 \left(
\frac{C_{30}z(-3x^2 - 3y^2 + 2z^2)}{2r^3} 
+ \frac{3C_{31}x(x^2 + y^2 - 4z^2)}{2r^3}
\right. \nonumber\\
& \left.
+ \frac{15C_{32}(x^2 - y^2)}{r^3} 
- \frac{15C_{33}x(x^2 - 3y^2)}{r^3}
\right) \nonumber\\
& + \left( \frac{1}{r} \right)^4 \left(
\frac{C_{40}\left(3(x^2 + y^2)^2 - 24(x^2 + y^2)z^2 + 8z^4 \right)}{8r^4} 
\right. \nonumber\\
& \left. 
+ \frac{5C_{41}xz(3x^2 + 3y^2 - 4z^2)}{2r^4} 
- \frac{15C_{42}(x^2 - y^2)(x^2 + y^2 - 6z^2)}{2r^4} \right. \nonumber\\
& \left. \left.
- \frac{105C_{43}xz (x^2 - 3y^2)}{r^4} 
+ \frac{105C_{44}(x^4 - 6x^2y^2 + y^4)}{r^4}
\right) \right].
\label{potefetivo}
\end{align}
\end{minipage}%
}

\begin{table}
    \centering
    \caption{Normalized spherical harmonics coefficients of (269) Justitia's gravity field up to degree and order four, derived from the polyhedral shape model \citep{WERNER19971071}. The coefficients are computed using a reference radius of $R = \mathbf{29.0}$ km and density of 1.0 g\,cm$^{-3}$. The reference frame is centered at the center of mass and aligned with the principal axes of inertia.}
    \label{Coeficientes}
\begin{tabular}{c c c}
    \hline\hline
    Order (n) & Degree (m) & $C_{n,m}$\\
    \hline
    0 & 0 & $1.0$  \\
    1 & 0 & $0$  \\
    1 & 1 & $0$  \\
    2 & 0 & $-5.44305\times10^{-2}$  \\
    2 & 1 & $0$  \\
    2 & 2 & $2.410451\times10^{-2}$  \\
    3 & 0 & $2.33221\times10^{-3}$  \\
    3 & 1 & $4.32010\times10^{-2}$   \\
    3 & 2 & $3.04401\times10^{-3}$   \\
    3 & 3 & $-1.26468\times10^{-3}$  \\
    4 & 0 & $1.39609\times10^{-2}$  \\
    4 & 1 & $-5.76107\times10^{-3}$  \\
    4 & 2 & $-5.38729\times10^{-3}$  \\
    4 & 3 & $1.85068\times10^{-3}$  \\
    4 & 4 & $8.98077\times10^{-3}$  \\
\hline
\end{tabular}
\end{table}

\begin{table}
    \caption{Parameters of (269) Justitia, being $M$ the total mass of the asteroid, $R$ the equatorial radius, and $\mathcal{K}$ the dimensionless parameter.}
    \label{Parametros}
    \centering
    \begin{tabular}{lccccr}
        \hline
        $M$ (kg) & $R$ (km) & $\omega$ (rad/s) & $\mathcal{K}$\\
        \hline
        $1.021604\times10^{17}$ & $29.0$ & $5.26818\times10^{-5}$ & $100.73314$  \\
    \hline
    \end{tabular}
\end{table}
Considering the rotational frame system, where the asteroid is centered on its center of mass and aligned with its main axis of inertia, being the equation that describes the movement dimensionless of a point around the asteroid, after the transformations $t\rightarrow \omega\tau$ and $\textbf{r} \rightarrow R_\textbf{r}$, given by Eq. \eqref{EqMotion}. Note that the time and length scales chosen in the transformation are equivalent to $1/\omega$ as the unit of time and $R$ as the unit of length.
\begin{equation}
    \begin{aligned}
        \ddot{x}-2 \dot{y} & = \frac{\partial V}{\partial x}, \\
        \ddot{y}+2 \dot{x} & = \frac{\partial V}{\partial y}, \\
        \ddot{z} & =\frac{\partial V}{\partial z}.
    \end{aligned}
    \label{EqMotion}
\end{equation}
\\
In this study, we aim to explore planar symmetric periodic orbits (SPO) around asteroid Justitia using the Grid Search Method \citep{markellos1974grid, russell2006global, tsirogiannis2009improved, barrio2009systematic}. From a mathematical standpoint, a system exhibits periodic solutions if, for the differential equation as in Eq.~\eqref{EqMotion}, there exists a particular solution satisfying $(x(t_0), \dot{x}(t_0)) = (x(t0+T), \dot{x}(t0+T))$, where $T > 0$ represents the period. This means that the initial state $(x(t_0), \dot{x}(t_0))$ repeats after time $T$, such that $(x(t) = x(t+T), \dot{x}(t) = \dot{x}(t+T))$. Specifically, when $\dot{x}(t_0) = \dot{x}(t_0+T) = 0$, the solution represents a symmetric periodic orbit, with the velocity being perpendicular to the axis of motion. For the equation \eqref{EqMotion}, the system is symmetric under the transformation $(x, y, \dot{x}, \dot{y}; t) \rightarrow (x, -y, -\dot{x}, \dot{y}; -t)$. Thus, an orbit with initial conditions $(x_0, y_0, \dot{x}_0, \dot{y}_0) = (x_0, 0, 0, \dot{y}_0)$ will cross the $x$-axis at $t = 0$ and again at $t = T/2$, under identical conditions, meaning $y(x_0, 0, 0, \dot{y}_0; T/2) = x(x_0, 0, 0, \dot{y}_0; T/2) = 0$. Consequently, the orbit is closed and symmetric with respect to the $x$-axis. 
\\
\\
\noindent \textit{Grid Search Method} – After identifying a symmetric periodic orbit, various methods can be applied to determine the corresponding family \citep{deprit1967natural, lara1995numerical}. The Grid Search Method provides a straightforward and systematic approach for computing the whole families of periodic orbits in non-integrable systems within a specific region of the relevant initial condition space. This method can be enhanced by additional data-processing techniques for classifying periodic orbits \citep{markellos1974grid, russell2006global, tsirogiannis2009improved, barrio2009systematic}.

Starting with the Jacobi constant \(J = 2V(x, y) - (\dot{x}^2 + \dot{y}^2)\), a grid was created based on the integral of motion (\(J\)) and spatial variables. In this context, the pair \((x, J)\) defines a complete set of initial conditions, where \(y_0 = \dot{x}_0 = 0\) and \(\dot{y}(x, J)\). For each set of initial conditions \((x, J)\), the equation of motion \eqref{EqMotion} is numerically integrated to compute the Poincaré map (crossing the \(x\)-axis with \(\dot{y} > 0\)). Upon obtaining the Poincaré map, the sign of \(\dot{x}\) is checked, and using continuity, if \(\dot{x}(t_0 + n\Delta t) < 0\) and \(\dot{x}(t_0 + (n+1)\Delta t) > 0\), a root-finding procedure is applied in conjunction with the numerical integrator \citep{brent1971algorithm}. Once convergence is reached, the initial conditions for the symmetric periodic orbit are considered satisfied. The final check involves integrating the initial conditions up to the full period \(T\), and verifying whether \( \left\| (x, 0, 0, \dot{y}) - (x(pT), \dot{x}(pT), y(pT), \dot{y}(pT)) \right\| < P_{\text{tol}} \), where \(p\) is a multiple of the complete period \(T\) and \(P_{\text{tol}}\) is the tolerance. Initial conditions that do not meet this final check are discarded.

\vspace{1em}
\noindent \textit{Topological Classification} - The topological classification facilitates the identification of the main characteristics of the families without requiring direct visualization of the orbits \citep{abad2023evolution}. It reveals the features standard to all orbits within a family, as well as the variations that occur at the endpoints of each curve, based on the specific points encircled by the orbit—namely, the asteroid (denoted by $B$) and the equilibrium points, whose locations correspond to those depicted in Fig. \ref{fig:15}(a), with $E_2$ and $E_4$ representing the triangular points, collectively denoted by the single letter $E_{2,4}$.

\begin{figure*}
  \centering
  \plotone{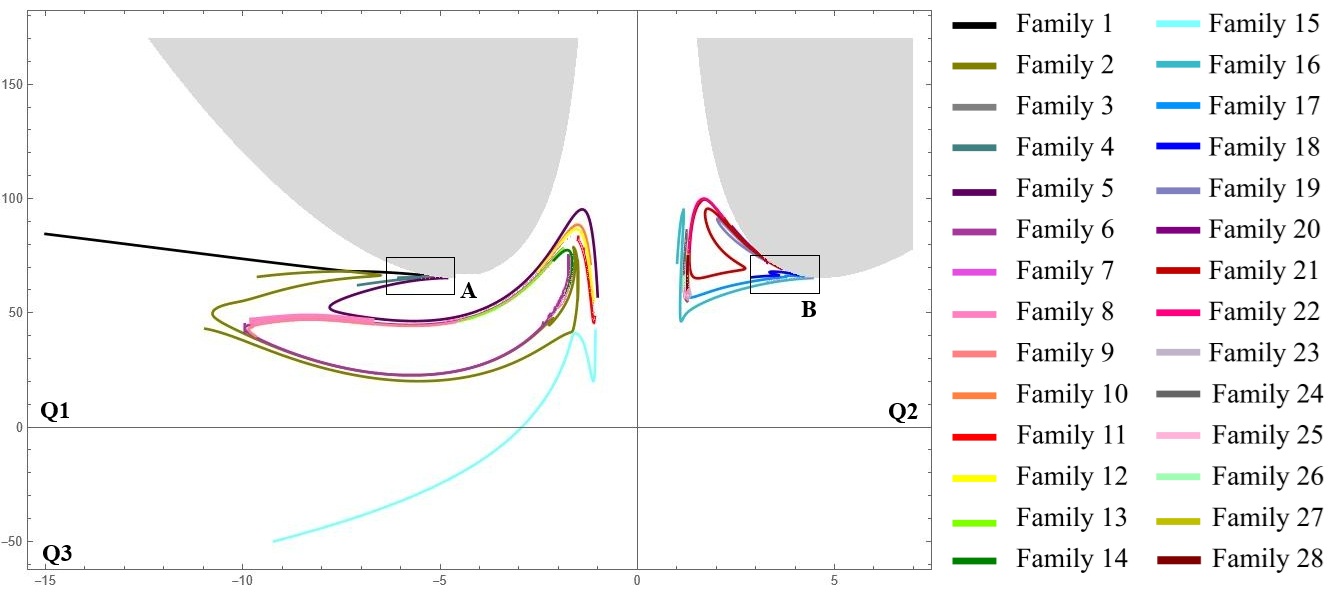}
 \caption{Characteristic curves of the asteroid (269) Justitia, considering the density equal to 1.0 g cm$^{-3}$. Each curve, represented by a distinct color and named according to the legend, corresponds to a family of symmetric periodic orbits (SPO). Each point along a given SPO family represents the initial condition of an orbit. The $x$-axis is the initial position at $x_0$ of the orbit at time $t=0$; the $y$-axis, the Jacobi constant $(J)$. The map is divided into quadrants $Q1$, $Q2$ and $Q3$, with regions $A$ (Fig. \ref{zoom:a}) and $B$ (Fig. \ref{zoom:b}) indicating close view areas.}
  \label{MapaCI}
\end{figure*}
Additionally, this classification system considers the orbital motion's sense: ``Direct ($\mathcal{D(\quad)}$)'', refers to cases where $x_0 > x_{T/2}$, with $x_{T/2}$ defined as the $x$-coordinate at half the period $T$ where the trajectory crosses the $x$-axis perpendicularly, and ``Retrograde ($\mathcal{R(\quad)}$)'', applies to cases where $x_0 < x_{T/2}$. Consequently, the notation $[\quad]$ indicates that all orbits within the family exhibit similar features; $[\rightarrow\;...]$ denotes that the orbits at one end retain the general family characterization while differing at the opposite end; and $[...\;\rightarrow\;...]$ signifies that both ends diverge from the typical characterization. It is worth noting that, in Tab. \ref{tab:Topologica}, the common term of each representation is emphasized in boldface.
\begin{table*}
\centering
\caption{Topological characterization of families of SPO for the asteroid Justitia}
\label{tab:Topologica}
\begin{tabularx}{\textwidth}{lll}
\cmidrule(r){1-3}
\textbf{Region} & \textbf{Curves} & \textbf{Topological characterization} \\
\cmidrule(r){1-3}
& Families 1, 3, 4  & $\boldsymbol{\mathcal{R}(E_3\ B\ E_1\ E_{2,4})}\ [\quad]$ \\
$Q1$ & Family 2    & $\boldsymbol{\mathcal{R}(B)}\ [\{E_1,E_3,E_{2,4}\}\rightarrow\ \{E_1,E_3\}\rightarrow \{E_1,E_{2,4}\}\rightarrow \ \{E_1\} \rightarrow \quad \quad \rightarrow  \{E_3\} \rightarrow  \{E_3,E_{2,4}\}]$ \\
     & Family 5  & $\boldsymbol{\mathcal{R}(\quad)}\ [E_3\ \rightarrow \quad]$ \\
     & Family 6  & $\boldsymbol{\mathcal{R}(B)}\ [\quad  \rightarrow \{E_3\}  \rightarrow \{E_1,E_{2,4}\}]$ \\
     & Families 7, 8  & $\boldsymbol{\mathcal{R}(E_3\ B\ E_{2,4})}\ [\quad]$ \\
     & Family 9  & -  \\
     & Family 10  & $\boldsymbol{\mathcal{D}(\quad)}\ [\quad]$ \\
     & Family 11  & $\boldsymbol{\mathcal{R}(B)}\ [\quad \rightarrow \{E_{2,4}\}]$ \\
     & Family 12, 14  & $\boldsymbol{\mathcal{R}(\quad)}\ [\quad]$ \\
     & Family 13  & $\boldsymbol{\mathcal{R}(B)}\ [\quad]$ \\
\cmidrule(r){1-3}
$Q1$ and $Q3$ & * Family 15 & $\boldsymbol{\mathcal{R}(B)}[\{E_1, E_3, E_{2,4}\} \rightarrow \quad\quad \rightarrow \{E_1\}  \rightarrow \{E_1, E_{2,4}\}]$  \\
\cmidrule(r){1-3}
    & Family 16  & $\boldsymbol{\mathcal{R}(\quad)}\ [\quad \rightarrow \{E_3\}]$\\
$Q2$ & Families 17-19, 21-23, 26  & $\boldsymbol{\mathcal{D}(B)}\ [\quad]$ \\
    & Families 20,24-25,27-28  & $\boldsymbol{\mathcal{D}(\quad)}\ [\quad]$ \\
     \cmidrule(r){1-3}
\end{tabularx}
\end{table*}




\vspace{1em}
\textit{Stability of the orbit} - The linear stability of an orbit refers to the analysis of the system's response to small perturbations around a periodic solution, made through the linearization of the equations of motion (Eq.~\eqref{EqMotion}), that is, it indicates whether minor variations in the position or velocity of the asteroid are amplified, remain constant, or decay over time.

The stability of a periodic orbit is determined by analyzing the eigenvalues of the monodromy matrix associated with the orbit, which correspond to the modes that govern the linear behavior near the periodic trajectory.

The monodromy matrix of an orbit, $M = \Phi(t_0+T,t_0)$, is obtained by integrating the state transition matrix $\dot \Phi = A(t)\Phi(t,t_0)$ for a complete period $T$ \citep{parker2014low}, since $A(t)$ is the Jacobian matrix (Eq.~\eqref{MatA}), $V_{xx}, V_{xy}, V_{yx}, V_{yy}$ the second derivative of the effective potential (Eq.~\eqref{potefetivo}) and the state transition matrix in the initial time equal to the identity matrix, $\Phi(t_0,t_0)=I_{4\times4}$. It provides information about all regions the particle traverses along its orbital path.
\begin{align}
	A(t)=	\left[\begin{array}{cccc}
		0 & 0 & 1 & 0 \\
		0 & 0 & 0 & 1 \\
		V_{xx} & V_{xy} &  0 & 2 \\
		V_{xy} & V_{yy} & -2 & 0
	\end{array}\right]. \label{MatA}
\end{align}
In the planar case, the monodromy matrix has four eigenvalues and their corresponding eigenvectors. Due to the symplectic structure of the system, the eigenvalues appear in reciprocal pairs, and, as a consequence of the periodicity of the solution, two of them are equal to one, so $\lambda_1, \lambda_2=\frac{1}{\lambda_1}, \lambda_3=1, \lambda_4=1$. Therefore, the characteristic polynomial is
\begin{align}
	P(\lambda)=\lambda^4-(k-2)\lambda^3+(2+2k)\lambda^2-(2+k)\lambda+1, \label{PolCMM}
\end{align}
since $k=\mid \lambda_1+\frac{1}{\lambda_1}\mid$ the stability index. This index is a single metric to assess the stability of the system \citep{parker2014low}:

- $k > 2$ the orbit is unstable;

- $k = 2$ the orbit is neutrally stable;

- $k < 2$, the orbit is linearly stable.

\subsection{Symmetric Periodic Solutions}
\begin{figure}
    \centering
    \begin{subfigure}{0.45\textwidth}
        \includegraphics[width=\linewidth]{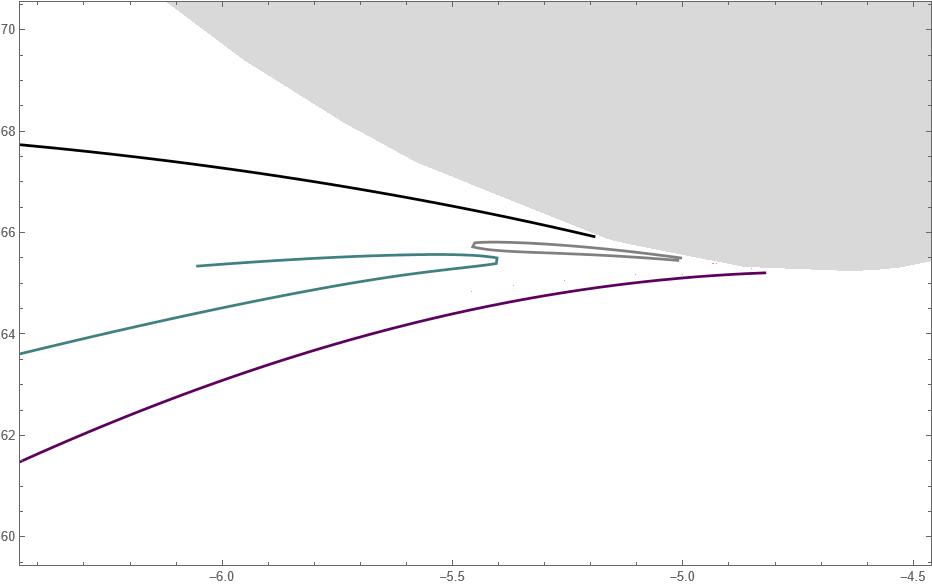}
        \caption{} 
        \label{zoom:a}
    \end{subfigure}
    \hfill
    \begin{subfigure}{0.45\textwidth}
        \includegraphics[width=\linewidth]{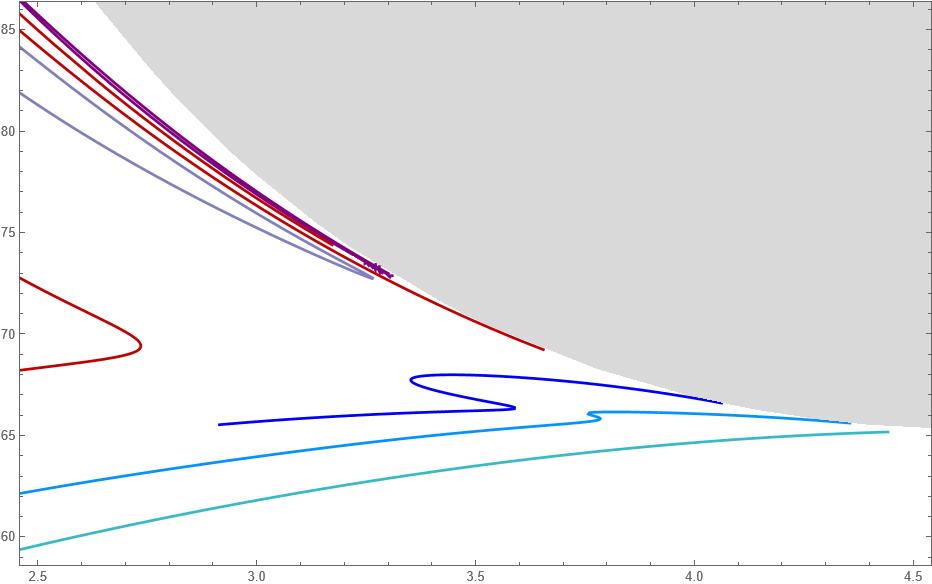}
        \caption{} 
        \label{zoom:b}
    \end{subfigure}
    \caption{Closer view of the regions highlighted by (a) rectangle A, and (b) rectangle B, in Fig. \ref{MapaCI}.}
    \label{zoom}
\end{figure}

In autonomous Hamiltonian systems, periodic orbits appear in families, represented by characteristic curves (smooth one-parameter continuous curves), as in Fig. \ref{MapaCI}. Each point in the families, represented on the axes $(x_0,J)$, is the initial condition of an SPO. Note that 28 families were found, represented by different colors, and distributed between quadrants $Q1$, $Q2$, and $Q3$. The asteroid is arranged on the $x$-axis, from $-1$ to $1$, since the unit of distance is its equatorial radius. The gray region, limited by $J=2V$ for $y=0$, is physically forbidden due to the negative speed of the particle orbiting the asteroid. The rectangles $A$ and $B$, highlighted on the map, are shown at higher magnification in Figs. \ref{zoom:a} and \ref{zoom:b}, respectively.

\begin{figure}
    \centering
    \begin{subfigure}{0.25\textwidth}
        \includegraphics[width=\linewidth]{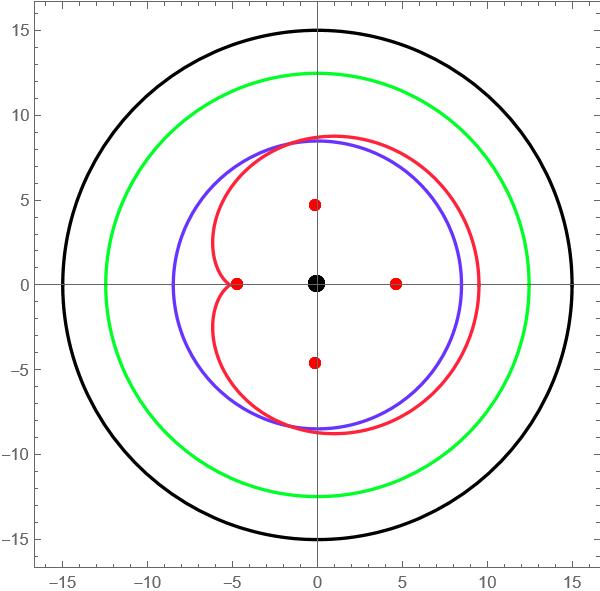}
        \caption{} 
        \label{evoa}
    \end{subfigure}
     \hfill
    \begin{subfigure}{0.2\textwidth}
        \includegraphics[width=\linewidth]{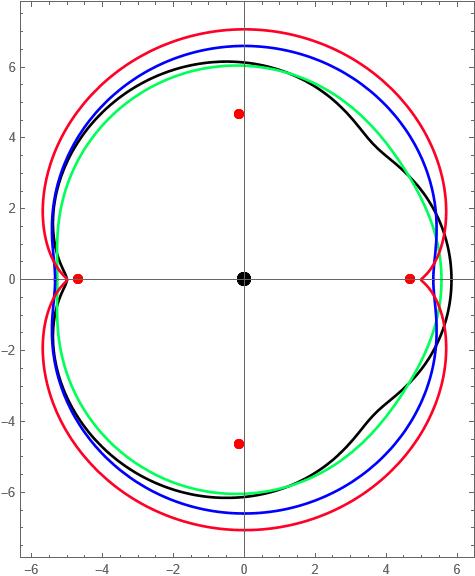}
        \caption{} 
        \label{evob}
    \end{subfigure}
    \hfill
     \begin{subfigure}{0.23\textwidth}
        \includegraphics[width=\linewidth]{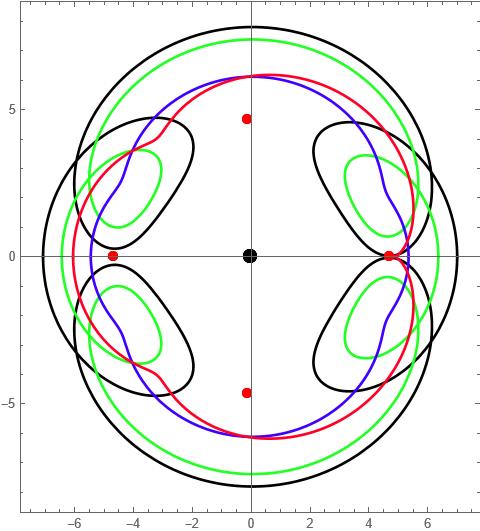}
        \caption{} 
        \label{evoc}
    \end{subfigure}
    \caption{Evolution of the orbits of families (a) 1, (b) 3, and (c) 4. Black and red orbits are the ends of the families, and green and blue are intermediate points. The black dot represents the asteroid Justitia (not to scale), and the red dots are the equilibrium points. The abscissa and ordinate axes are the components of the particle's position around the asteroid, in x and out y, respectively.}
    \end{figure}

Note, in Tab. \ref{tab:Topologica}, that Families 1, 3, and 4 have the same topological characterization; all of them surround the collinear and triangular equilibrium points and the asteroid. The difference between them lies in the loops that appear when families change. A loop occurs when the periodic orbit presents one or more self-intersections in the configuration space, that is, there exists at least one pair of instants $t_1 \neq t_2 \in [0,T]$ such that, 
$$x(t_1)=x(t_2),$$
since the same point is crossed more than once, but in different directions, characterizing the loop. To exemplify this behavior, see the evolution of the orbits of Families 1, 3, and 4 in Figs. \ref{evoa}, \ref{evob}, \ref{evoc}. This work focuses on the search and presentation of orbit families; therefore, a study of the reasons for their occurrence and characterization will not be presented. 

\begin{figure}
    \centering
    \begin{subfigure}{0.23\textwidth}
        \includegraphics[width=\linewidth]{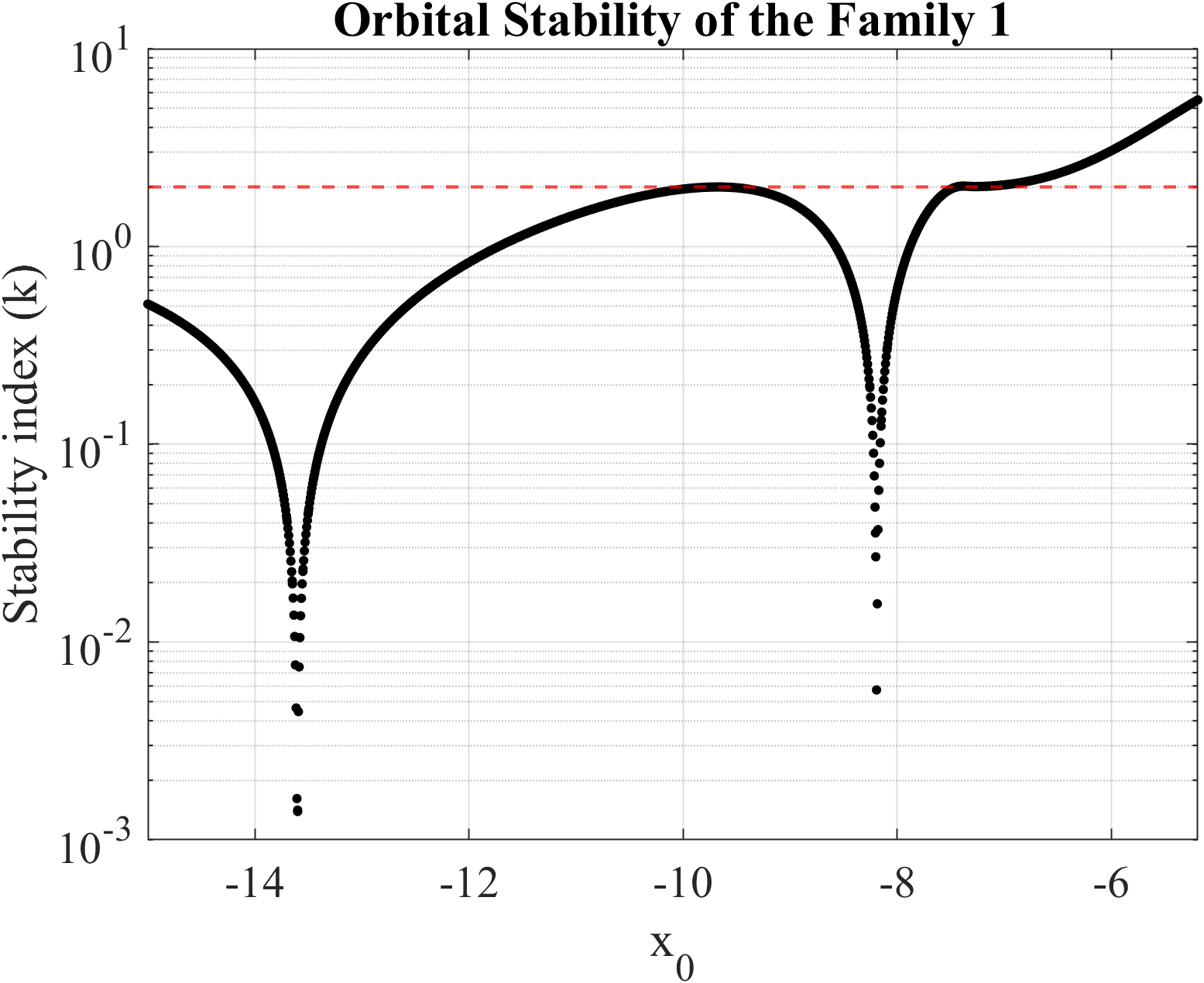}
        \caption{} 
        \label{Stabla1}
    \end{subfigure}  
  \hfill
    \begin{subfigure}{0.23\textwidth}
        \includegraphics[width=\linewidth]{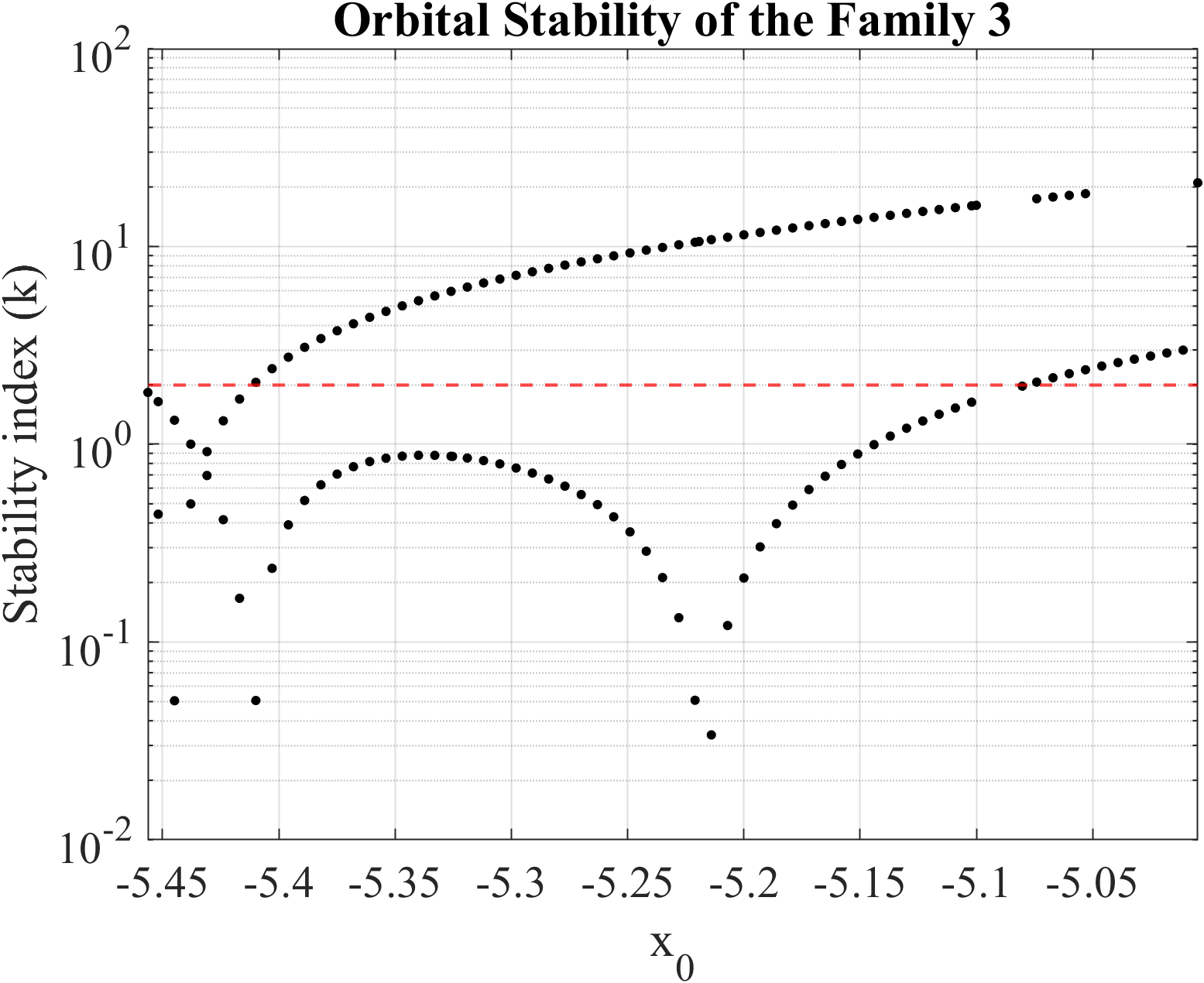}
        \caption{} 
        \label{Stabla3}
    \end{subfigure}
    \hfill
     \begin{subfigure}{0.23\textwidth}
        \includegraphics[width=\linewidth]{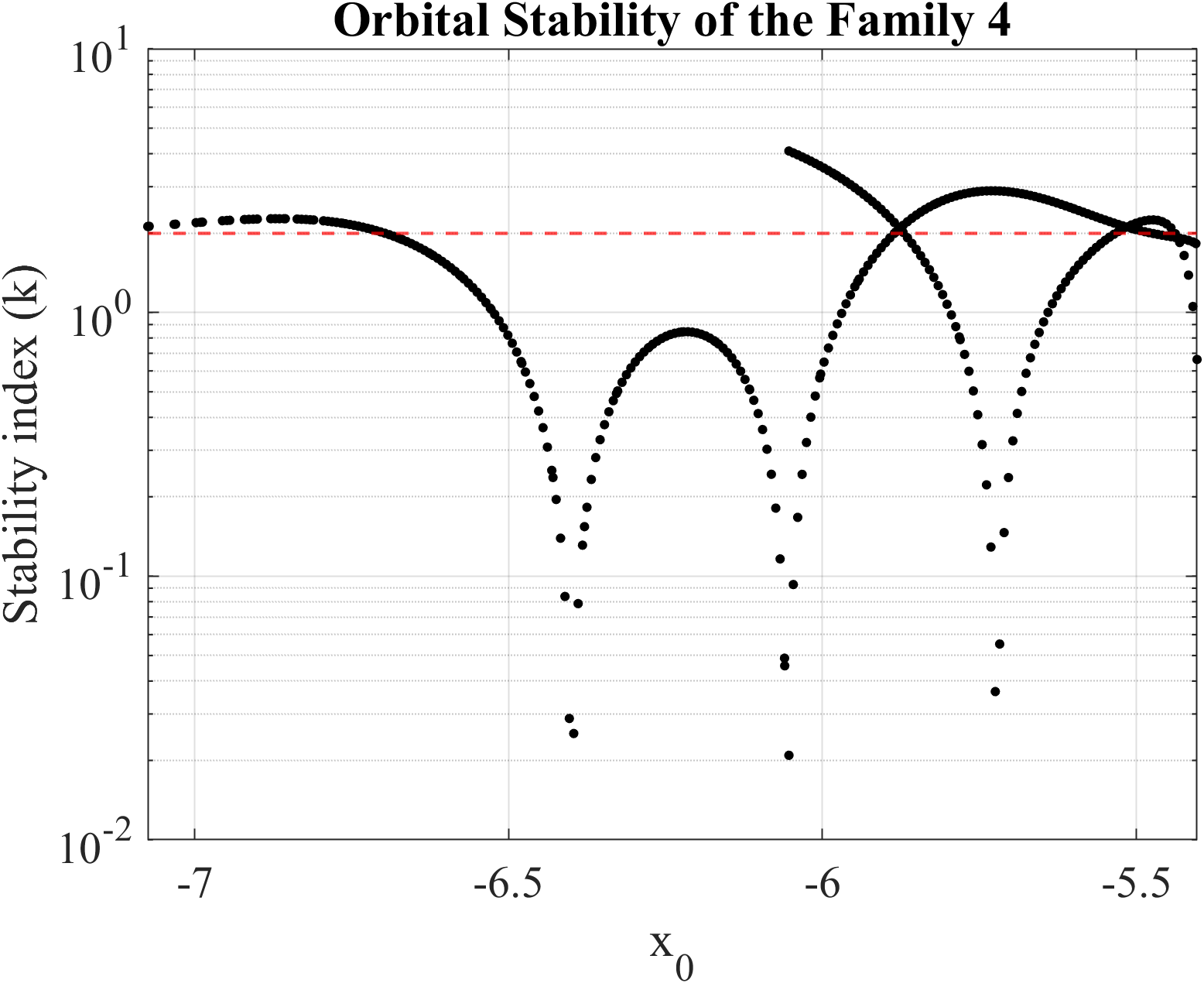}
        \caption{} 
        \label{Stabla4}
    \end{subfigure}
    \caption{Stability index referring to families: (a) 1, (b) 3, and (c) 4. The dashed red line at $k=2$.}
    \end{figure}

Regarding the linear stability of the orbits, Figs. \ref{Stabla1}, \ref{Stabla3}, \ref{Stabla4}, for the cases in Figs. \ref{evoa}, \ref{evob}, \ref{evoc}, presents the stability index $k$ as a function of the initial condition $x_0$, with the red dashed line on the graph at the limit $k=2$. When the stability index $k=2$, the periodic orbit is in a critical condition, which marks the limit between linear stability $(k<2)$ and instability $(k>2)$. In this situation, it does not characterize instability but rather a transition point where small variations in the system parameters can qualitatively change the behavior of the solutions. Note that the families in Figs. \ref{Stabla1}, \ref{Stabla3}, \ref{Stabla4} have a significant part of stable orbits, with some unstable parts, with $k$ around $10$. 


\begin{figure}
    \centering
    \begin{subfigure}{0.25\textwidth}
        \includegraphics[width=\linewidth]{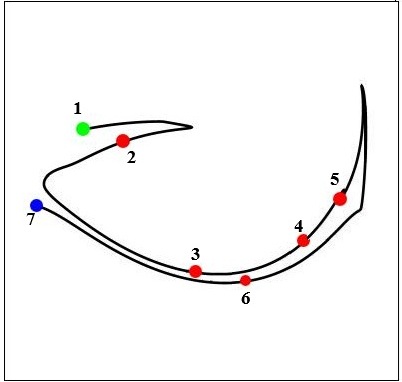}
        \caption{Family 2} 
        \label{evo0}
    \end{subfigure}
     \hfill
    \begin{subfigure}{0.2\textwidth}
        \includegraphics[width=\linewidth]{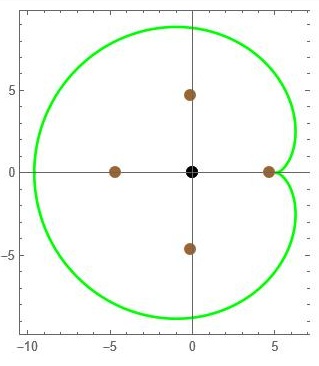}
        \caption{Orbit 1} 
        \label{evo1}
    \end{subfigure}
    \hfill
     \begin{subfigure}{0.23\textwidth}
        \includegraphics[width=\linewidth]{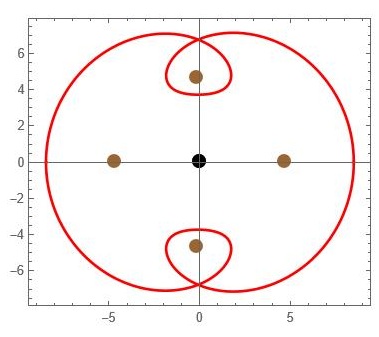}
        \caption{Orbit 2} 
        \label{evo2}
    \end{subfigure}
        \hfill
     \begin{subfigure}{0.23\textwidth}
        \includegraphics[width=\linewidth]{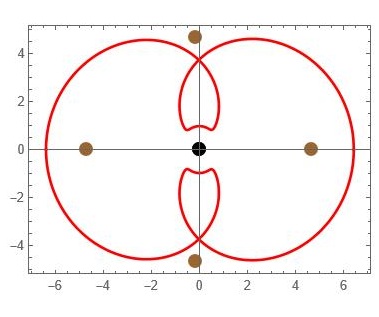}
        \caption{Orbit 3} 
        \label{evo3}
    \end{subfigure}
        \hfill
     \begin{subfigure}{0.23\textwidth}
        \includegraphics[width=\linewidth]{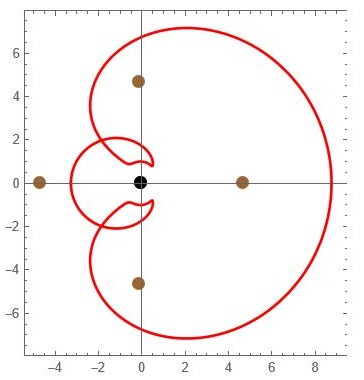}
        \caption{Orbit 4} 
        \label{evo4}
    \end{subfigure}
        \hfill
     \begin{subfigure}{0.23\textwidth}
        \includegraphics[width=\linewidth]{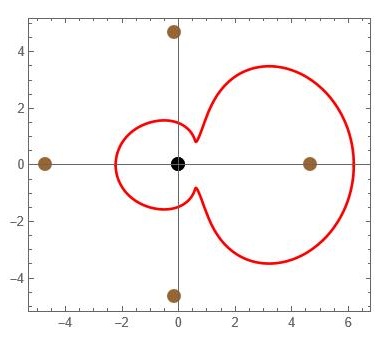}
        \caption{Orbit 5} 
        \label{evo5}
    \end{subfigure}
        \hfill
     \begin{subfigure}{0.23\textwidth}
        \includegraphics[width=\linewidth]{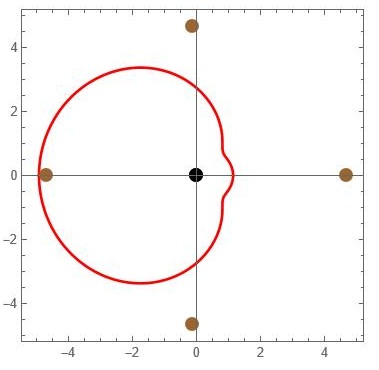}
        \caption{Orbit 6} 
        \label{evo6}
    \end{subfigure}
        \hfill
     \begin{subfigure}{0.23\textwidth}
        \includegraphics[width=\linewidth]{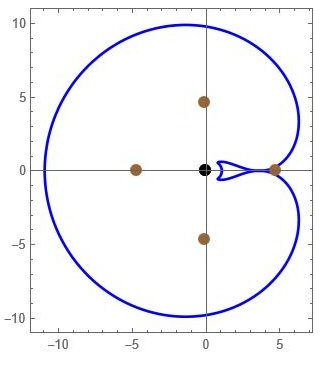}
        \caption{Orbit 7} 
        \label{evo7}
    \end{subfigure}
    \caption{Evolution of the orbits of Family 2. Orbit 1 (green) and Orbit 2 (blue) are at the ends of the family, and red are intermediate points. The black dot represents the asteroid Justitia (not to scale), and the brown dots are the equilibrium points. The abscissa and ordinate axes are the components of the particle's position around the asteroid, in x and out y, respectively.}\label{fig:evogeral}
    \end{figure}

Figure \ref{fig:evogeral} shows the evolution of Family 2, which has an initial condition at point 1 (green) in Fig.~\ref{evo0}, and the orbit evolves according to Fig.~\ref{evo1}. Points 2 to 7 in Fig.~\ref{evo0} (red points) are the initial conditions of the orbits in Fig.~\ref{evo2} to Fig.~\ref{evo6}. Notice that Orbit 2 closes a loop at the triangular equilibrium points, and subsequently, in Orbit 3, the loop no longer encircles these points. Continuing the evolution, in Orbit 4, the left side of the orbit ($x<0$) decreases significantly in relation to the right side ($x>0$) until the orbit encircles points $E_1$ and the triangular points, as well as the body, leaving out equilibrium point $E_3$. Continuing, in Orbit 5, the loop disappears, the amplitude of the right side of the orbit decreases until it no longer involves the triangular equilibrium points, remaining only around $E_1$. Then the amplitude of the left side increases, while the amplitude of the right side of the orbit decreases, encircling only the body and point $E_3$. Finally, the family is completed by Orbit 7 (in blue), represented in Fig.~\ref{evo7}. Note that this family has seven variations of topological orbital features, in addition to the common feature (see Tab.~\ref{tab:Topologica}). The topological classification of this case, as well as Families 6 and 15, which also vary similarly, differs slightly from \cite{abad2023evolution} characterization method, which is limited to two variations in addition to the common part of the orbit; therefore, an adaptation of this method was made for these cases.

\begin{figure}
\centering
   \plotone{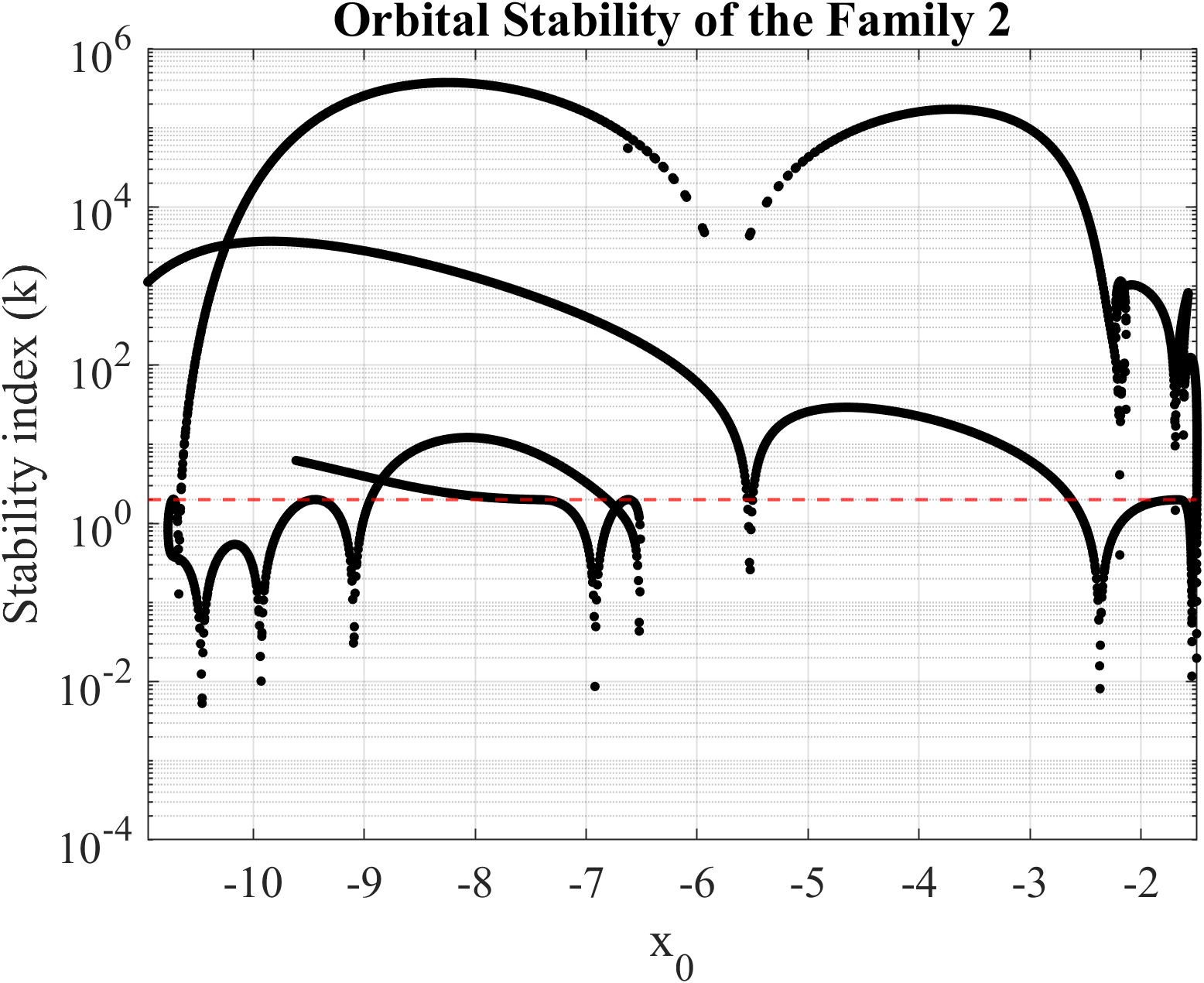}
    \caption{Stability index referring to Family 2. The dashed red line at $k=2$.}
    \label{Stab2}
\end{figure}

Regarding the stability of Family 2, see Figure \ref{Stab2}, which shows that there are stable and unstable cases, with the majority being unstable. The orbits resulting from initial conditions 1 to 7, shown in Fig.\ref{evo0}, are all unstable; the initial conditions of the stable orbits are small segments between points 1 and 2 and another segment between points 5 and 6.

\begin{figure}
    \centering
    \begin{subfigure}{0.4\textwidth}
        \includegraphics[width=\linewidth]{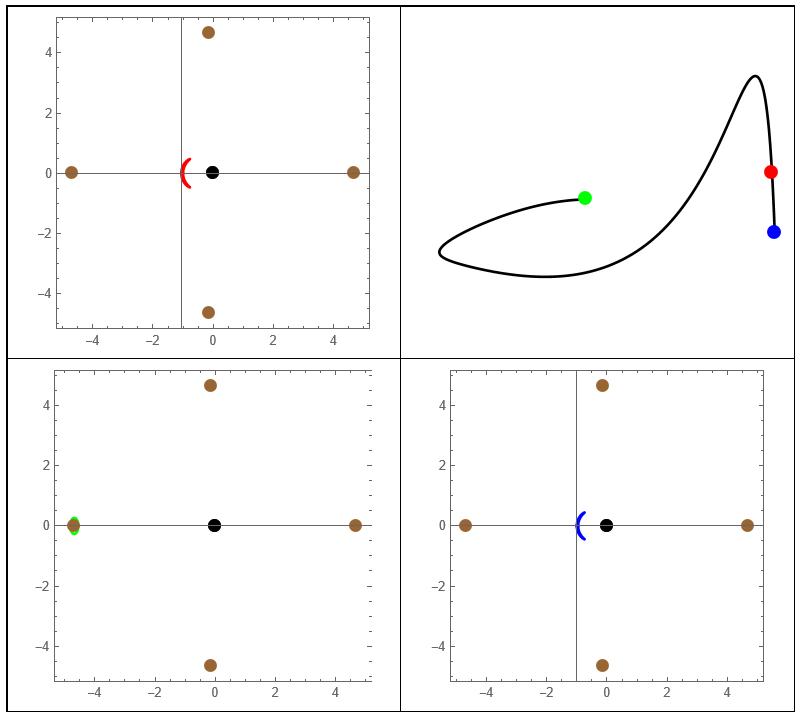}
        \caption{} 
        \label{fig:O5detalhada}
    \end{subfigure}
     \hfill
    \begin{subfigure}{0.4\textwidth}
        \includegraphics[width=\linewidth]{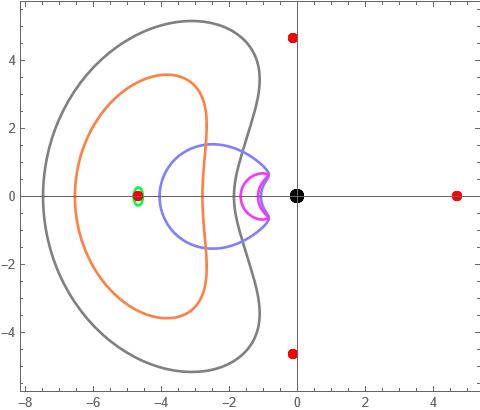}
        \caption{} 
        \label{fig:Orb5}
    \end{subfigure}
    \caption{Evolution of the orbits of Family 5. (a) Top right side: The complete Family 5 curve (black curve). The red point marks the limit at which the collision cases begin. The green and blue points represent the orbits at the ends of the family. Top left: orbit of the family whose initial conditions are the green point. Bottom left: the orbit of the family whose initial conditions are the blue point. The brown points are the equilibrium points; (b) It shows five different orbits of Family 5, which are between the orbit of the initial end of the family (black) and the pink one (but free collision); the red points are the equilibrium points. The black dot represents the asteroid Justitia (not to scale). The abscissa and ordinate axes are the components of the particle's position around the asteroid, in x and out y, respectively.}\label{fig:O5}
    \end{figure}

The behavior of the Family 5 is shown in Fig.~\ref{fig:O5}. In Fig.~\ref{fig:O5detalhada}, at the top right of the figure, we have the complete Family 5 (black curve). The red point represents the limit where collision cases begin; that is, the part of the curve between the green and red points corresponds to the families considered in the topological classification. The part between the red and blue points represents all the initial conditions of the orbits that, at some point, diverged from the initial one and collided with the asteroid. At the top left, we have the evolution of the collision limit orbit.

\begin{figure}
\centering
   \plotone{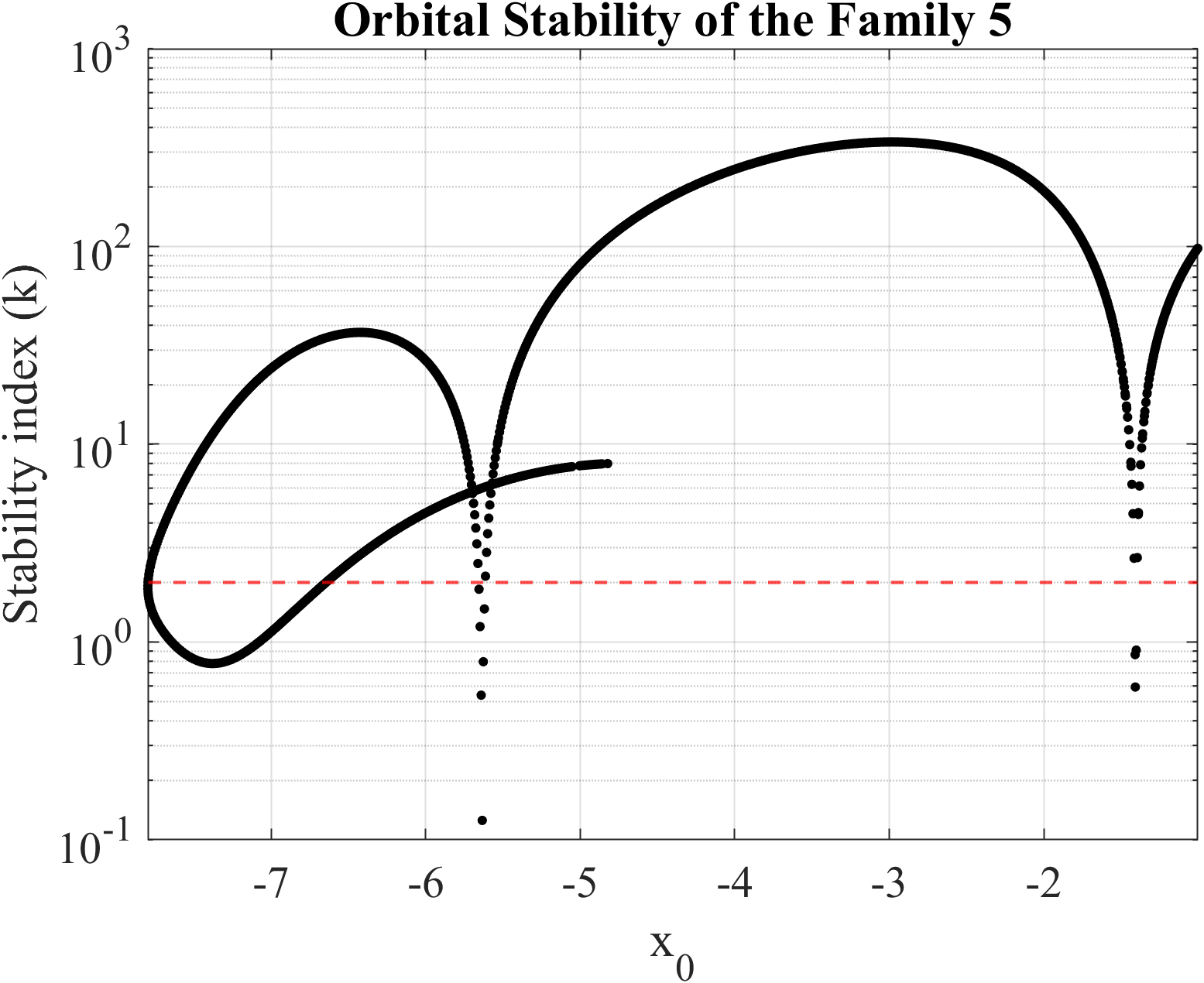}
    \caption{Stability index referring to Family 5. The dashed red line at $k=2$.}
    \label{Stab5}
\end{figure}

In contrast, at the bottom, we show the evolution of the family's initial orbit (on the left) and its final orbit (on the right), corresponding to the green and blue points of the complete curve; therefore, only the blue orbit is a collision orbit. The black and pink orbits, and all those between them (some shown in Fig.~\ref{fig:Orb5}), are collision-free orbits, some of which are Lyapunov type. Lyapunov orbits, in this case, emerge from the collinear Lagrange point $E_3$, are retrograde, and mostly unstable (as in Fig.~\ref{Stab5}). Family 16 exhibits behavior similar to Family 5, with Lyapunov-type orbits emerging from the equilibrium point $E_1$ and without collision orbits; however, there are some cases in which it passes within 100 meters of the body's surface. Regarding stability, there is a greater number of stable orbits compared to Fig.~\ref{Stab5}, in some cases of the Lyapunov type.

\begin{figure}
\centering
   \plotone{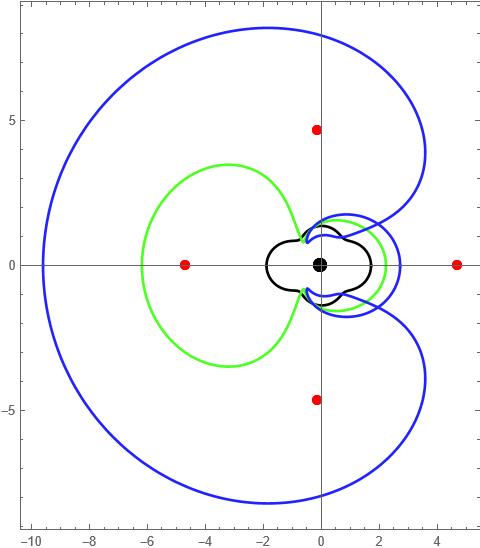}
    \caption{Evolution of the orbits of Family 6. Black and red orbits are the ends of the families, and green and blue are intermediate points. The black dot represents the asteroid Justitia (not to scale), and the red dots are the equilibrium points. The abscissa and ordinate axes are the components of the particle's position around the asteroid, in x and y, respectively.}
    \label{O6}
\end{figure}

\begin{figure}
\centering
   \plotone{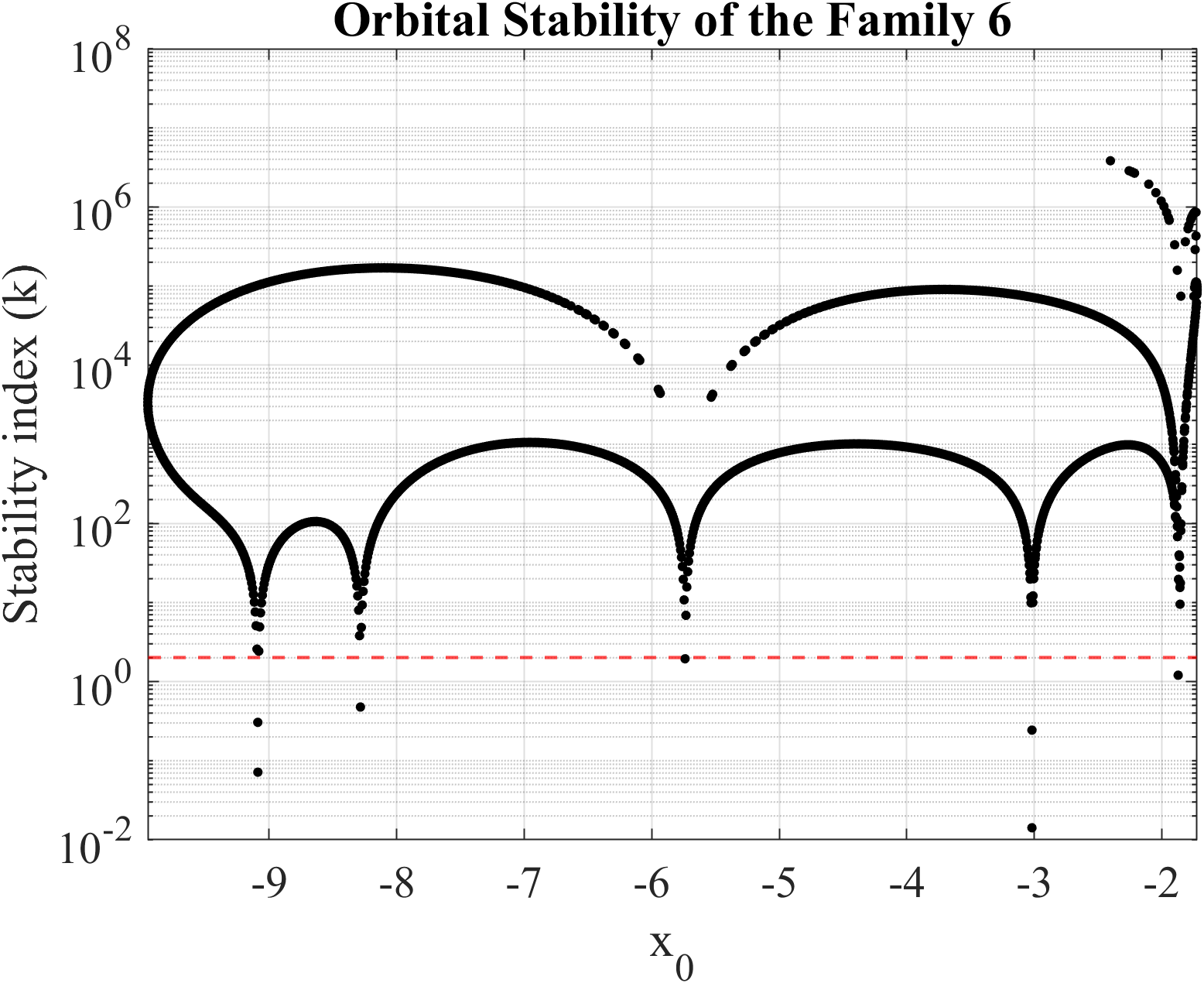}
    \caption{Stability index referring to Family 6. The dashed red line at $k=2$.}
    \label{Stab6}
\end{figure}

Family 6 is formed by retrograde and highly unstable orbits (see Fig.~\ref{Stab6}). According to the topological characterization (Tab.~\ref{tab:Topologica} and Fig.~\ref{O6}), the first orbit of the family is the black one, which only circles the asteroid (black dot). Representing the second characterization of the family, we have the green orbit that involves the body and the point $E_3$, and the blue orbit that circles the body, $E_3$, and the triangular points $E_{2,4}$. All evolution from the blue orbit resulted in a collision with the asteroid.

\begin{figure}
    \centering
    \begin{subfigure}{0.23\textwidth}
        \includegraphics[width=\linewidth]{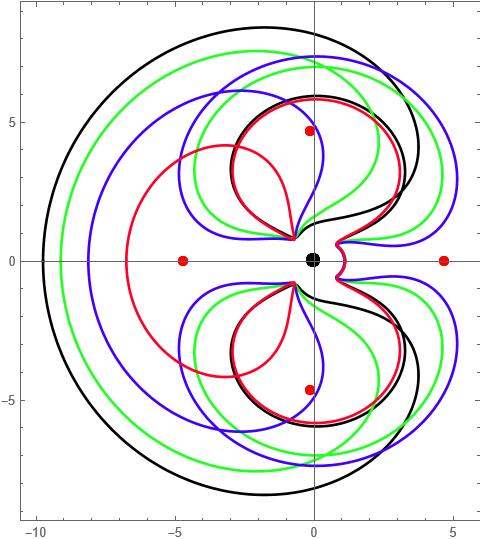}
        \caption{} 
        \label{fig:O7}
    \end{subfigure}
     \hfill
    \begin{subfigure}{0.23\textwidth}
        \includegraphics[width=\linewidth]{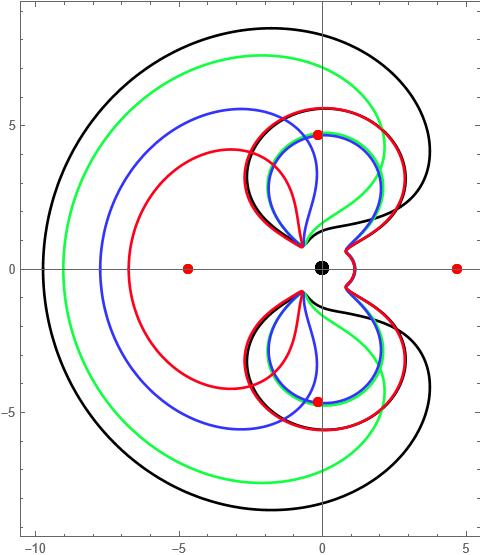}
        \caption{} 
        \label{fig:O8}
    \end{subfigure}
    \caption{Evolution of the orbits of families (a) 7, and (b) 8. Black and red orbits are the ends of the families, and green and blue are intermediate points. The black dot represents the asteroid Justitia (not to scale), and the red dots are the equilibrium points. The abscissa and ordinate axes are the components of the particle's position around the asteroid, in x and y, respectively.}\label{fig:O7e8}
    \end{figure}

\begin{figure}
    \centering
    \begin{subfigure}{0.23\textwidth}
        \includegraphics[width=\linewidth]{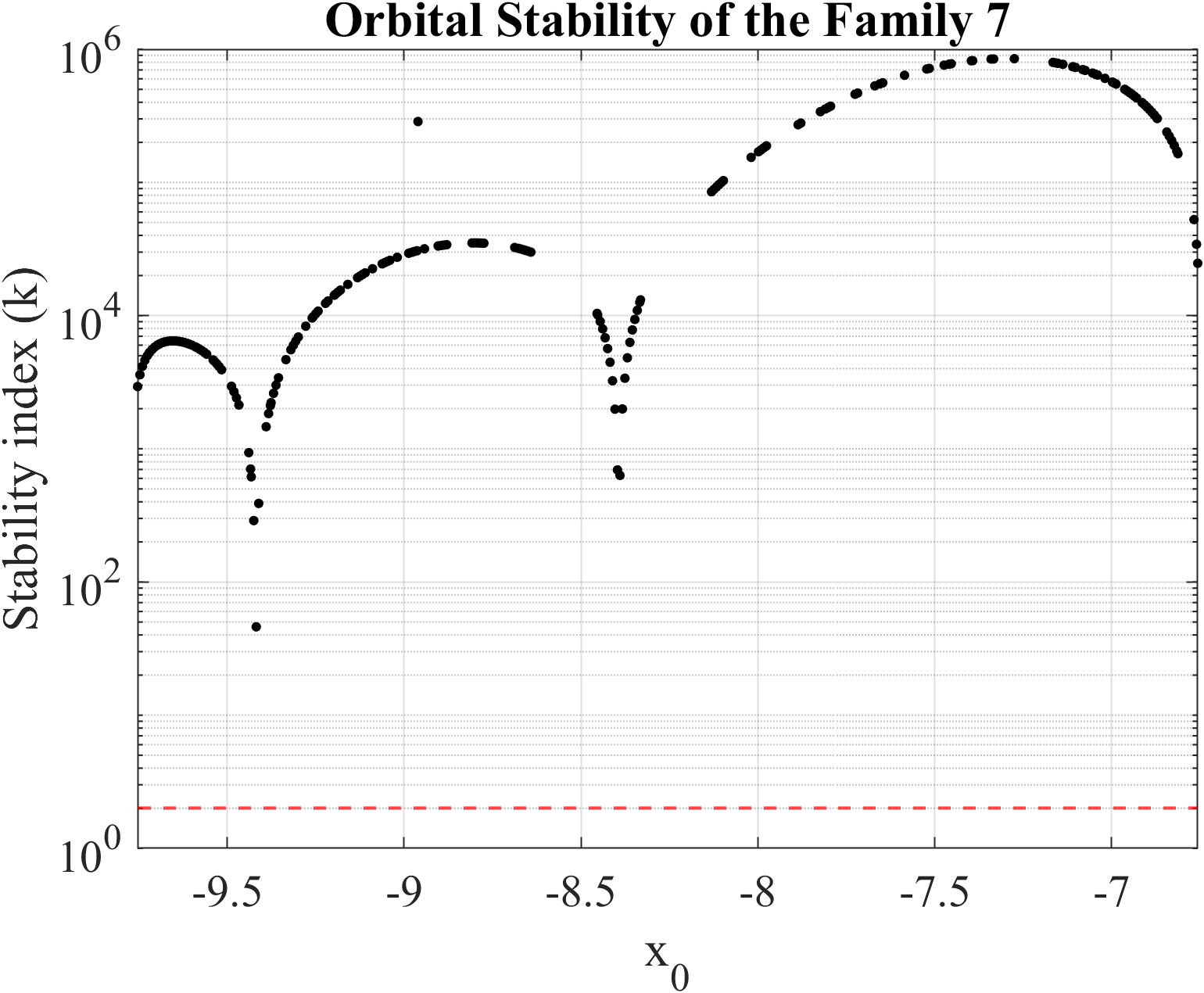}
        \caption{} 
        \label{Stab7}
    \end{subfigure}
     \hfill
    \begin{subfigure}{0.23\textwidth}
        \includegraphics[width=\linewidth]{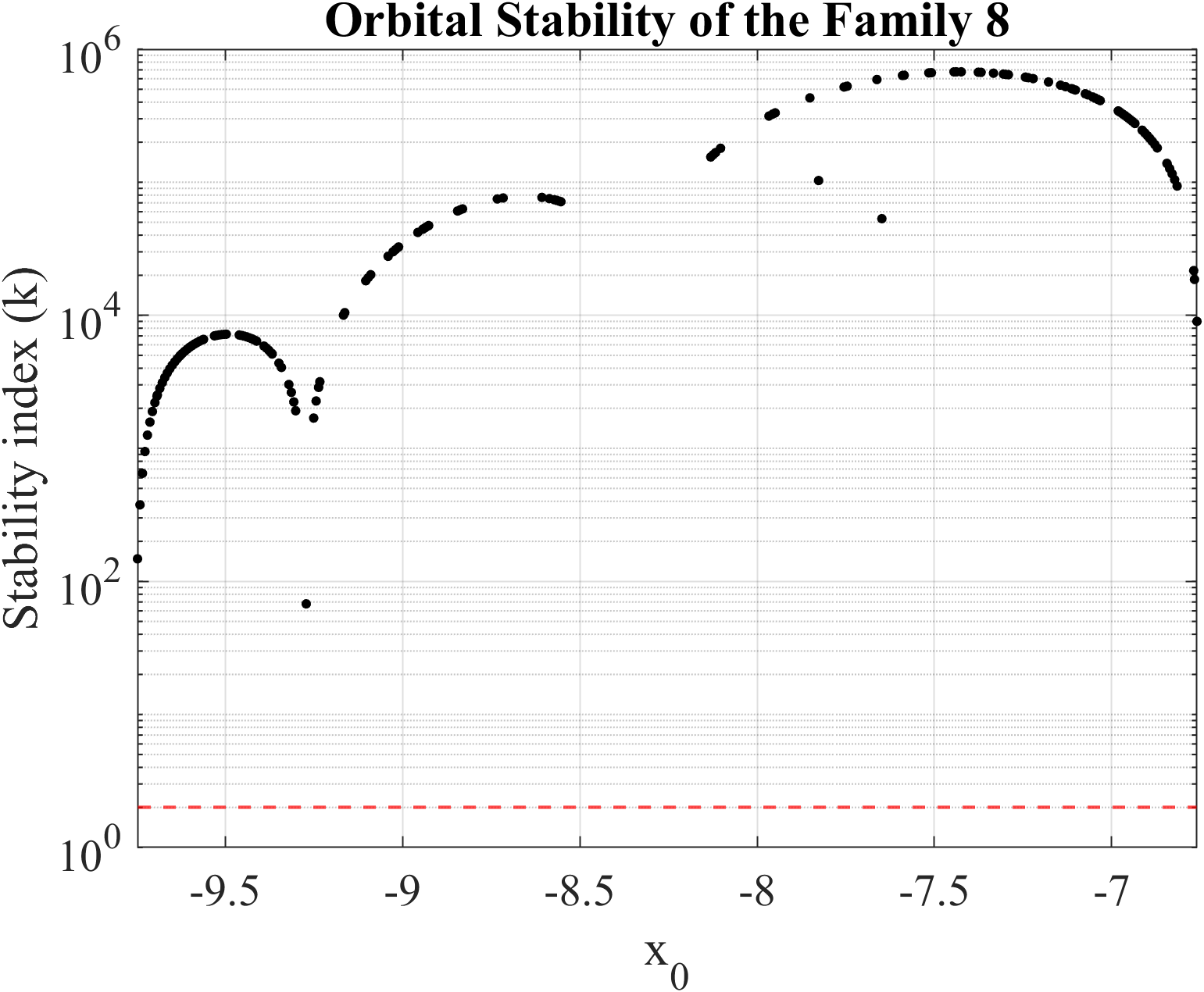}
        \caption{} 
        \label{Stab8}
    \end{subfigure}
    \caption{Stability index referring to Families (a) 7, and (b) 8. The dashed red line at $k=2$.}\label{StabO7e8}
    \end{figure}

Families 7 and 8 (Fig. ~\ref{fig:O7e8}) constantly orbit the body, the equilibrium point $E_3$, and the triangular points. Throughout the family, this characterization remains unchanged. These are families with retrograde and highly unstable orbits. For stability, see Fig.~\ref{StabO7e8}.

In Table \ref{tab:Topologica}, there is no name for the topological characterization of Family 9, because all orbits in this family, at some point other than the initial one, collide with the asteroid. In the map of curve characteristics (Fig. \ref{MapaCI}), the Jacobi constant is around 40 or more. In these cases, the particle will have low available kinetic energy, tending to remain confined to a small region near the asteroid or, as occurred, to collide with the body. The stability indices in these cases are much greater than 2.0, which indicates highly unstable cases.

\begin{figure}
    \centering
    \begin{subfigure}{0.19\textwidth}
        \includegraphics[width=\linewidth]{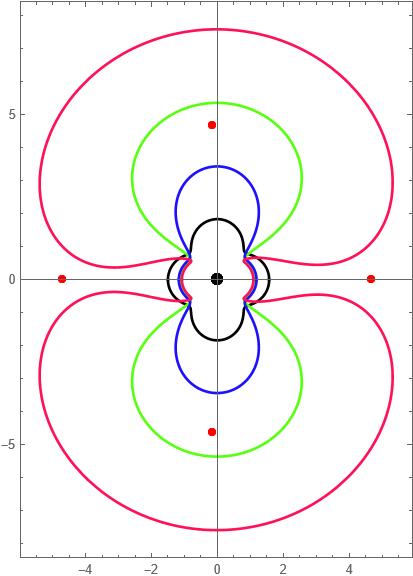}
        \caption{} 
        \label{fig:O11}
    \end{subfigure}
     \hfill
    \begin{subfigure}{0.25\textwidth}
        \includegraphics[width=\linewidth]{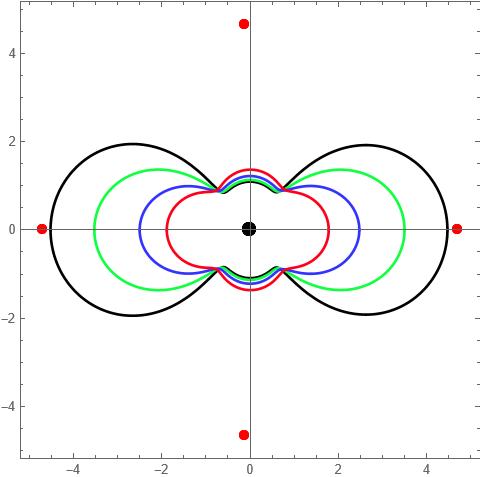}
        \caption{} 
        \label{fig:O13}
    \end{subfigure}
    \caption{Evolution of the orbits of families (a) 11, and (b) 13. Black and red orbits are the ends of the families, and green and blue are intermediate points. The black dot represents the asteroid Justitia (not to scale), and the red dots are the equilibrium points. The abscissa and ordinate axes are the components of the particle's position around the asteroid, in x and y, respectively.}\label{fig:O11e13}
    \end{figure}

\begin{figure}
    \centering
    \begin{subfigure}{0.23\textwidth}
        \includegraphics[width=\linewidth]{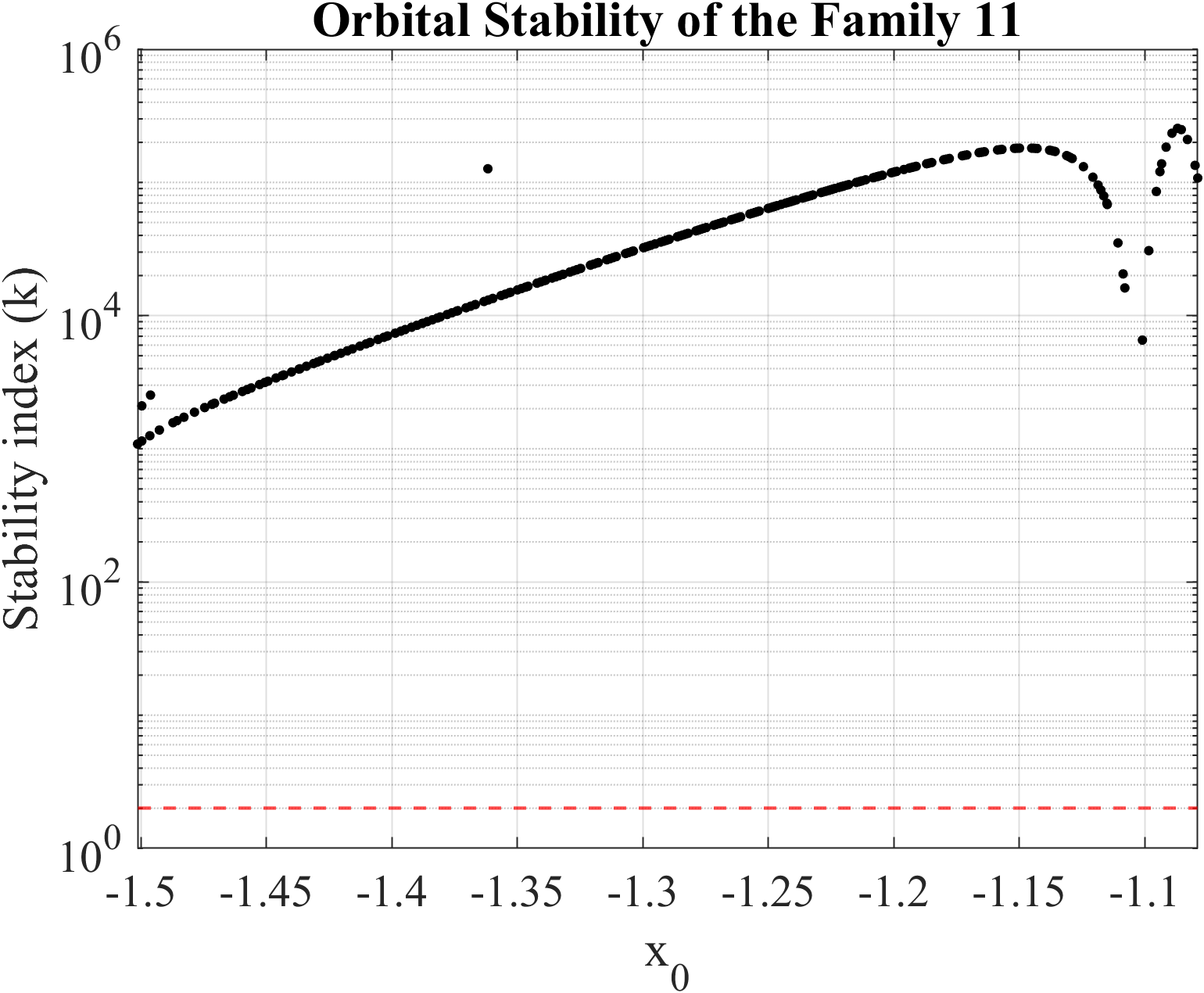}
        \caption{} 
        \label{Stab11}
    \end{subfigure}
     \hfill
    \begin{subfigure}{0.23\textwidth}
        \includegraphics[width=\linewidth]{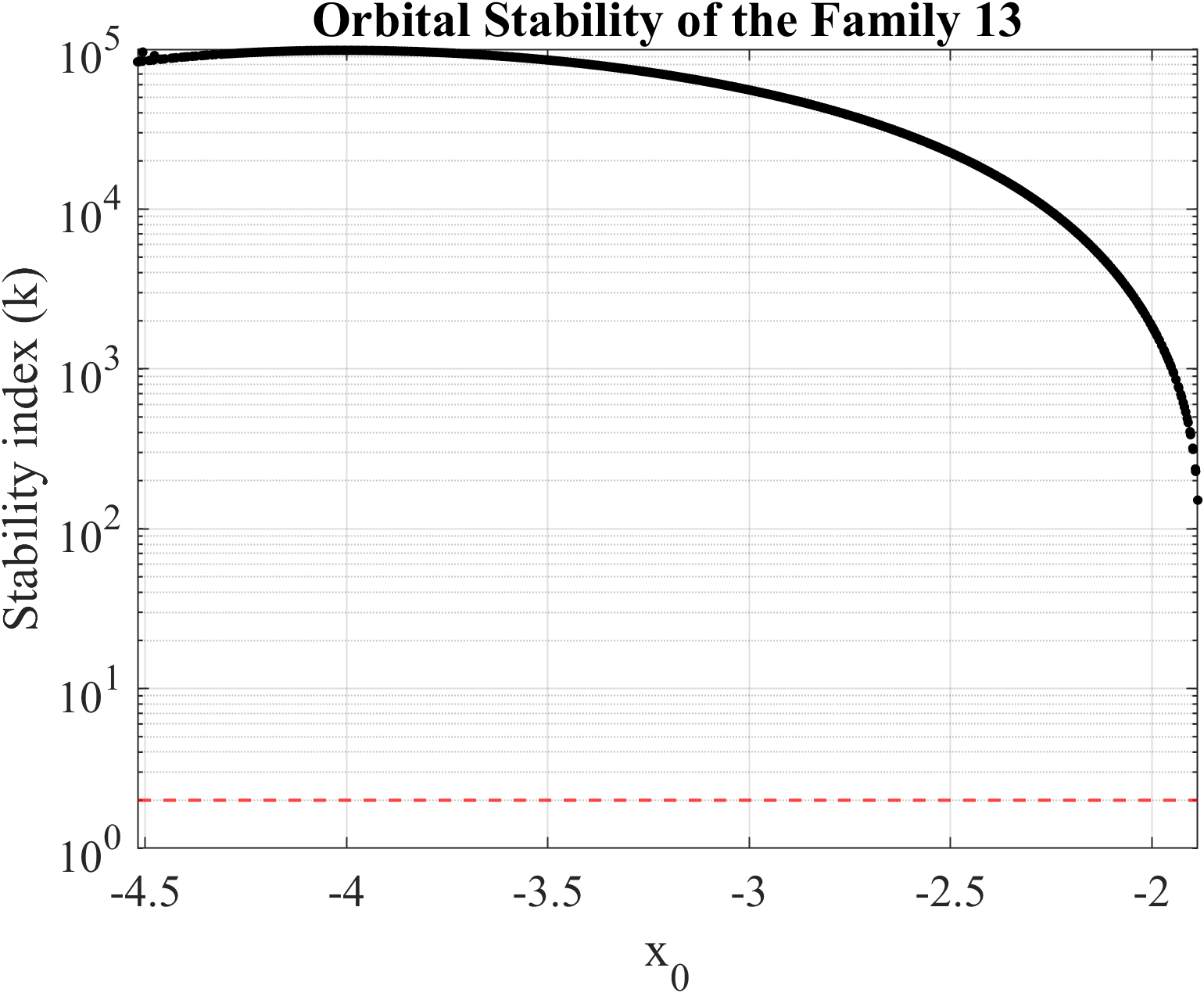}
        \caption{} 
        \label{Stab13}
    \end{subfigure}
    \caption{Stability index referring to Families (a) 11, and (b) 13. The dashed red line at $k=2$.}\label{Stab11e13}
    \end{figure}

Family 11 (Fig. \ref{fig:O11}) has a topological characterization $\boldsymbol{\mathcal{R}(B)}\ [\quad \rightarrow \{E_{2,4}\}]$, showing that as it evolves, the orbits tend to circle the triangular equilibrium points $E_{2,4}$, in addition to the body. Family 13 (Fig. \ref{fig:O13}), on the other hand, orbits only the asteroid in all cases. Both families have retrograde and highly unstable orbits (see Fig. \ref{Stab11e13}).

\begin{figure}
    \centering
    \begin{subfigure}{0.23\textwidth}
        \includegraphics[width=\linewidth]{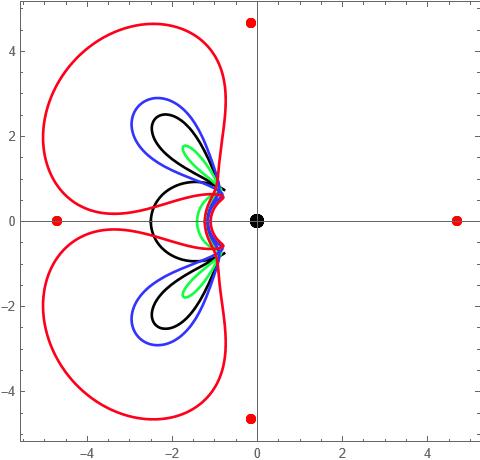}
        \caption{} 
        \label{fig:O12}
    \end{subfigure}
     \hfill
    \begin{subfigure}{0.23\textwidth}
        \includegraphics[width=\linewidth]{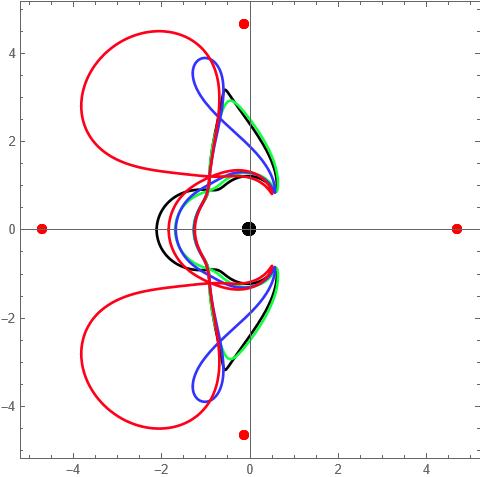}
        \caption{} 
        \label{fig:O14}
    \end{subfigure}
    \caption{Evolution of the orbits of families (a) 12, and (b) 14. Black and red orbits are the ends of the families, and green and blue are intermediate points. The black dot represents the asteroid Justitia (not to scale), and the red dots are the equilibrium points. The abscissa and ordinate axes are the components of the particle's position around the asteroid, in x and y, respectively.}\label{fig:O12e14}
\end{figure}

\begin{figure}
    \centering
    \begin{subfigure}{0.23\textwidth}
        \includegraphics[width=\linewidth]{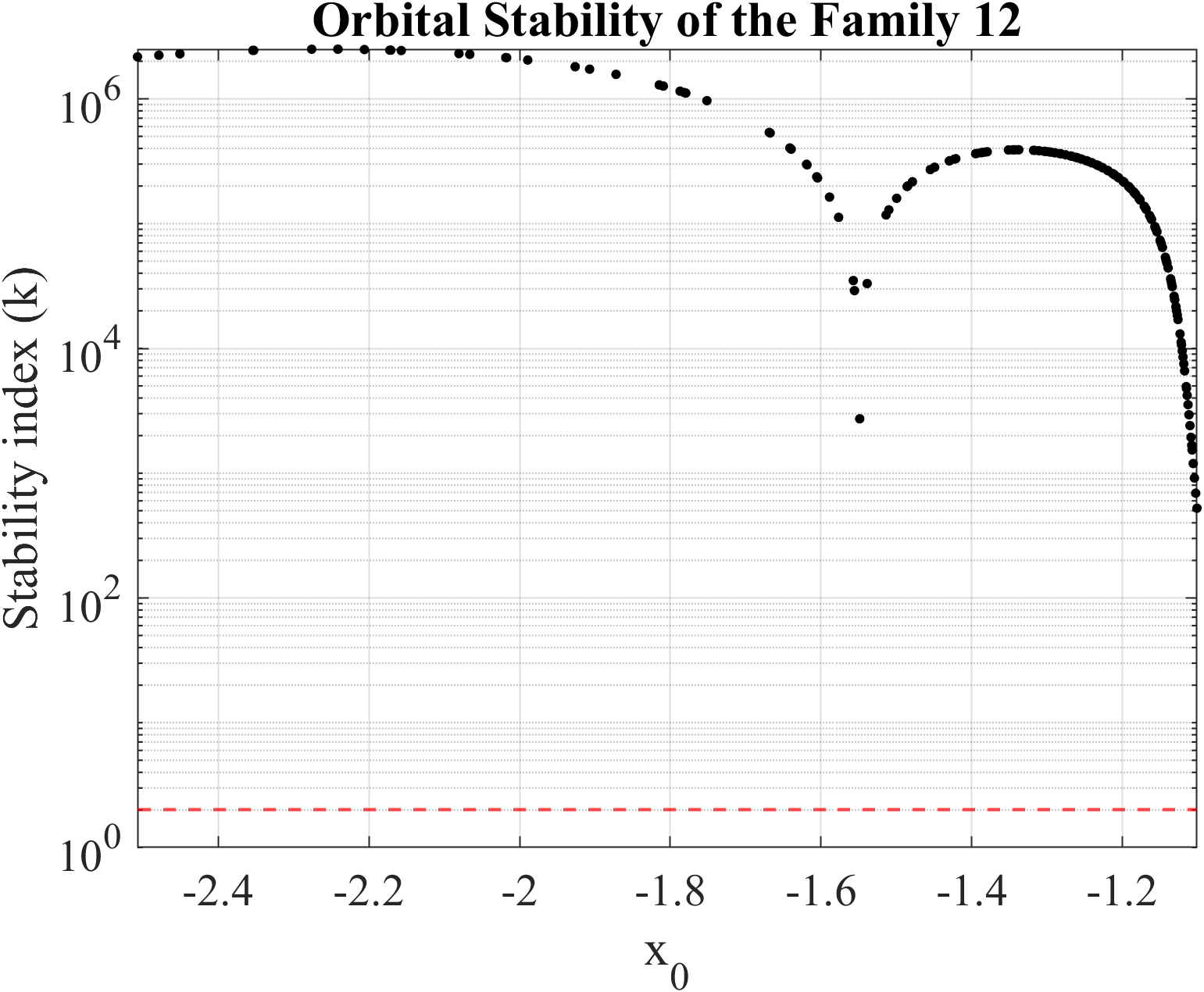}
        \caption{} 
        \label{Stab12}
    \end{subfigure}
     \hfill
    \begin{subfigure}{0.23\textwidth}
        \includegraphics[width=\linewidth]{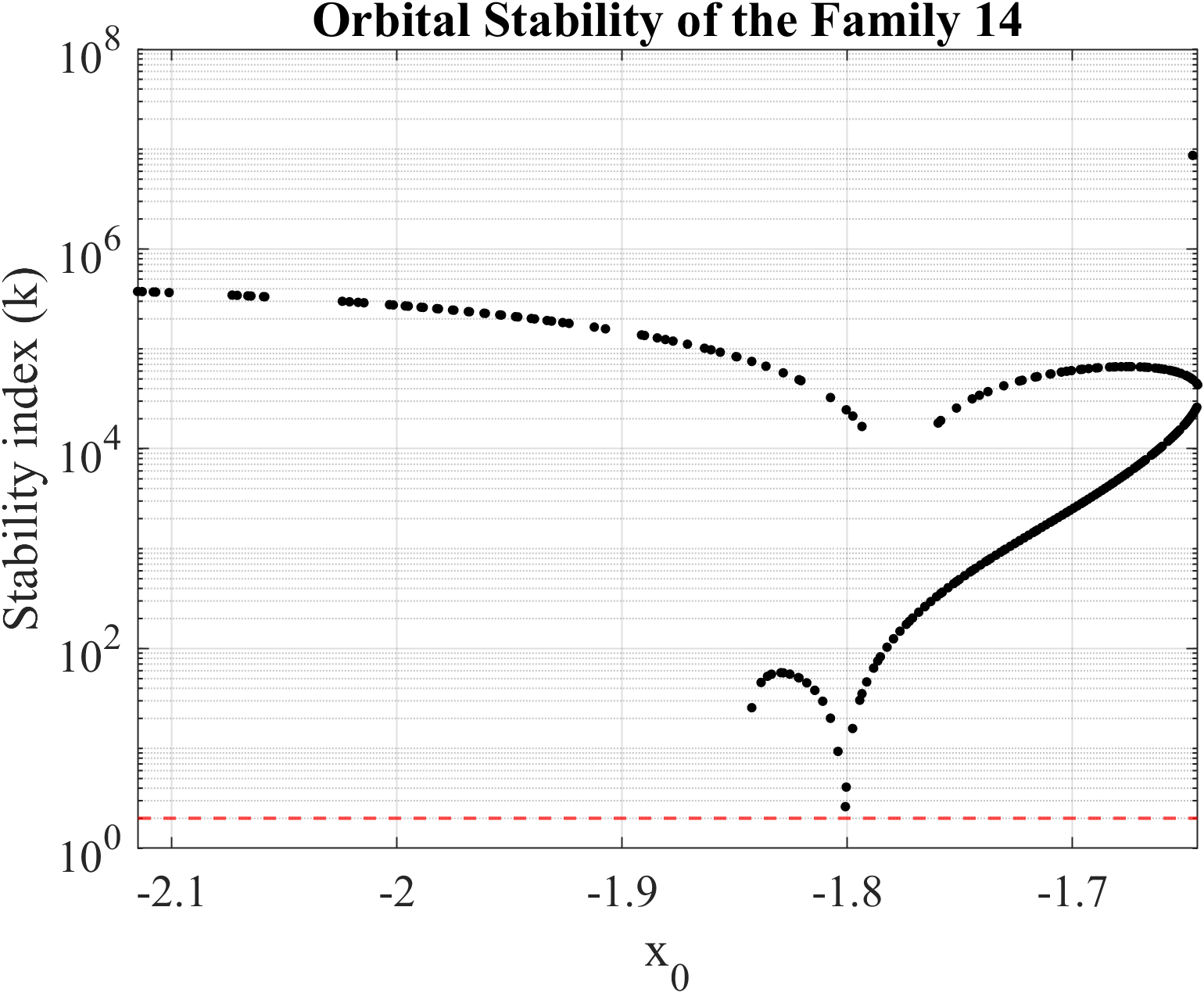}
        \caption{} 
        \label{Stab14}
    \end{subfigure}
    \caption{Stability index referring to Families (a) 12, and (b) 14. The dashed red line at $k=2$.}\label{Stab12e14}
    \end{figure}

Families 12 and 14 have the same topological characterization $\boldsymbol{\mathcal{R}(\quad)}\ [\quad]$ (see Fig. \ref{fig:O12e14}), both do not encircle the body nor the equilibrium points, are retrograde and unstable (see Fig. \ref{Stab12e14}).

\begin{figure}
\centering
   \plotone{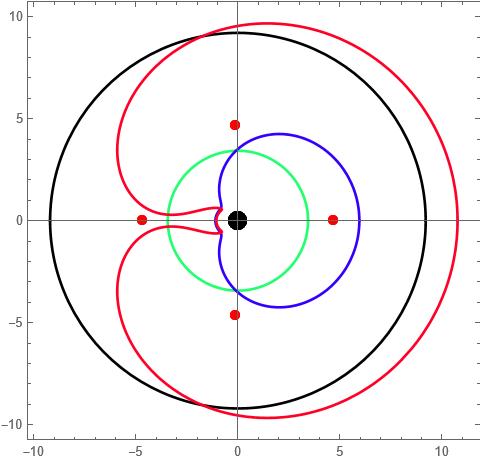}
    \caption{Evolution of the orbits of Family 15. Black and red orbits are the ends of the families, and green and blue are intermediate points. The black dot represents the asteroid Justitia (not to scale), and the red dots are the equilibrium points. The abscissa and ordinate axes are the components of the particle's position around the asteroid, in x and y, respectively.}
    \label{fig:O15}
\end{figure}

\begin{figure}
\centering
   \plotone{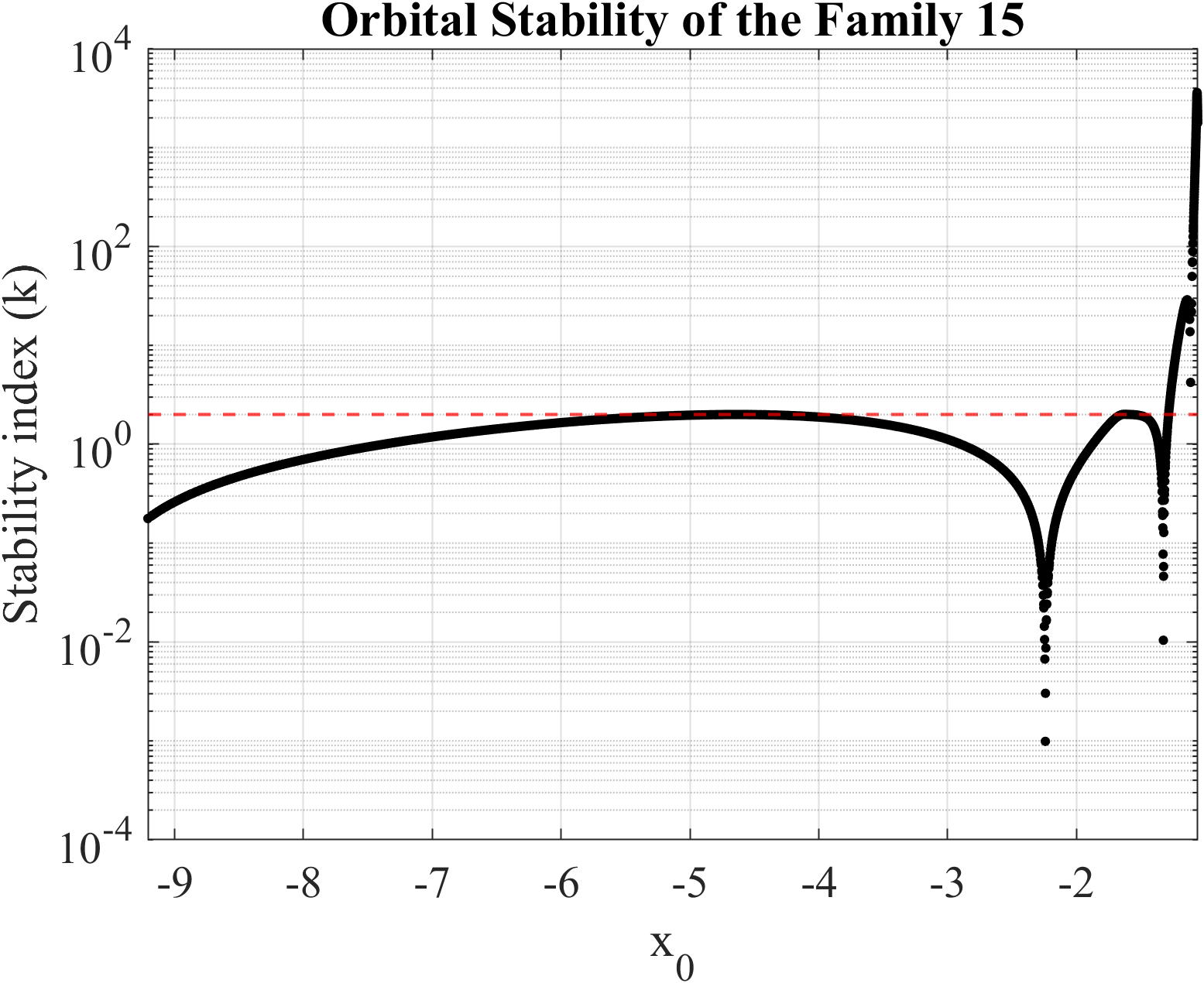}
    \caption{Stability index referring to Family 15. The dashed red line at $k=2$.}
    \label{Stab15}
\end{figure}

Up to Family 14, the characteristic curves or SPO families belonged to quadrant $Q1$, the initial conditions map (Fig. \ref{MapaCI}). Regarding Family 15 (Fig. \ref{fig:O15}), which is divided between quadrants $Q1$ and $Q3$, the topological classification (Tab. \ref{tab:Topologica}) has five parts in addition to the common part, in which all orbits circle the asteroid. In the first part, the black orbit in Fig. \ref{fig:O15}, involves, in addition to the body, the equilibrium points $E_3$ and $E_{2,4}$; then, the green curve orbits only the asteroid, next the body and point $E_1$ (blue orbit); and finally, the red orbit circles the body, $E_1$, and the triangular points $E_{2,4}$. In this orbit, there are points where the distance from the body's surface is less than 100 meters. The family is retrograde and with most of its orbits stable (see Fig. \ref{Stab15}).

\begin{figure}
    \centering
    \begin{subfigure}{0.23\textwidth}
        \includegraphics[width=\linewidth]{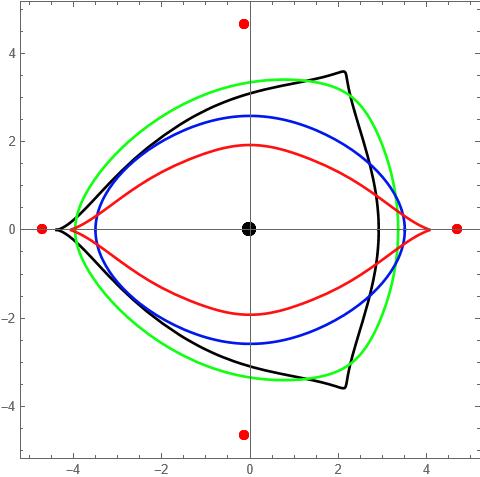}
        \caption{} 
        \label{O18a}
    \end{subfigure}
     \hfill
    \begin{subfigure}{0.23\textwidth}
        \includegraphics[width=\linewidth]{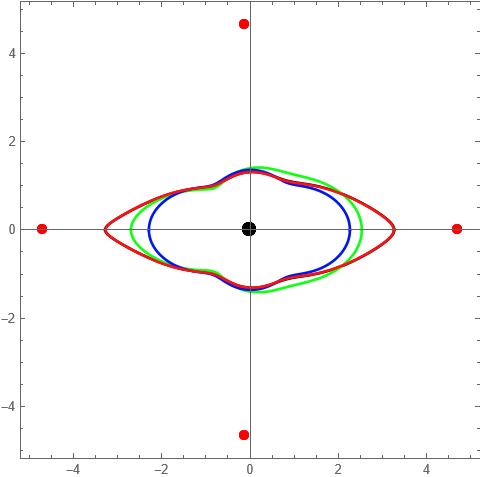}
        \caption{} 
        \label{O19b}
    \end{subfigure}
    \hfill
     \begin{subfigure}{0.23\textwidth}
        \includegraphics[width=\linewidth]{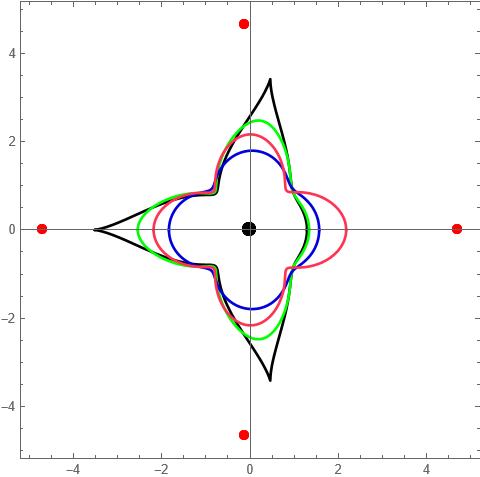}
        \caption{} 
        \label{O22c}
    \end{subfigure}
    \hfill
      \begin{subfigure}{0.23\textwidth}
        \includegraphics[width=\linewidth]{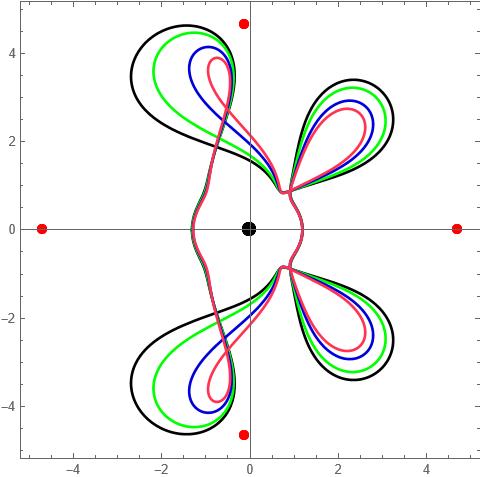}
        \caption{} 
        \label{O26d}
    \end{subfigure}
    \caption{Evolution of the orbits of families (a) 18, (b) 19, (c) 22, and (d) 26. Black and red orbits are the ends of the families, and green and blue are intermediate points. The black dot represents the asteroid Justitia (not to scale), and the red dots are the equilibrium points. The abscissa and ordinate axes are the components of the particle's position around the asteroid, in x and y, respectively.}
    \label{Fig:Orbit17a26}
    \end{figure}

\begin{figure}
    \centering
    \begin{subfigure}{0.23\textwidth}
        \includegraphics[width=\linewidth]{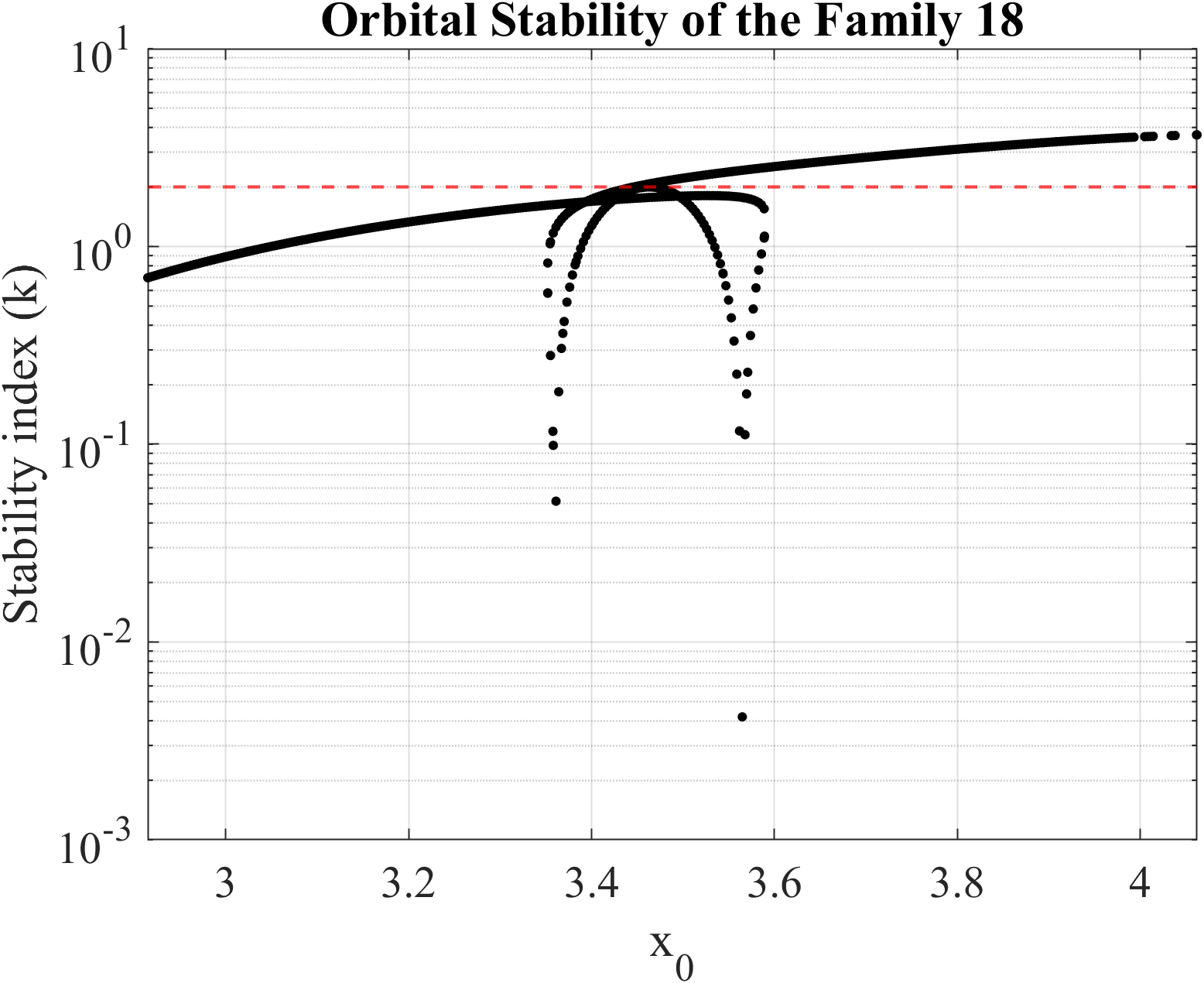}
        \caption{} 
        \label{Stab18}
    \end{subfigure}
     \hfill
    \begin{subfigure}{0.23\textwidth}
        \includegraphics[width=\linewidth]{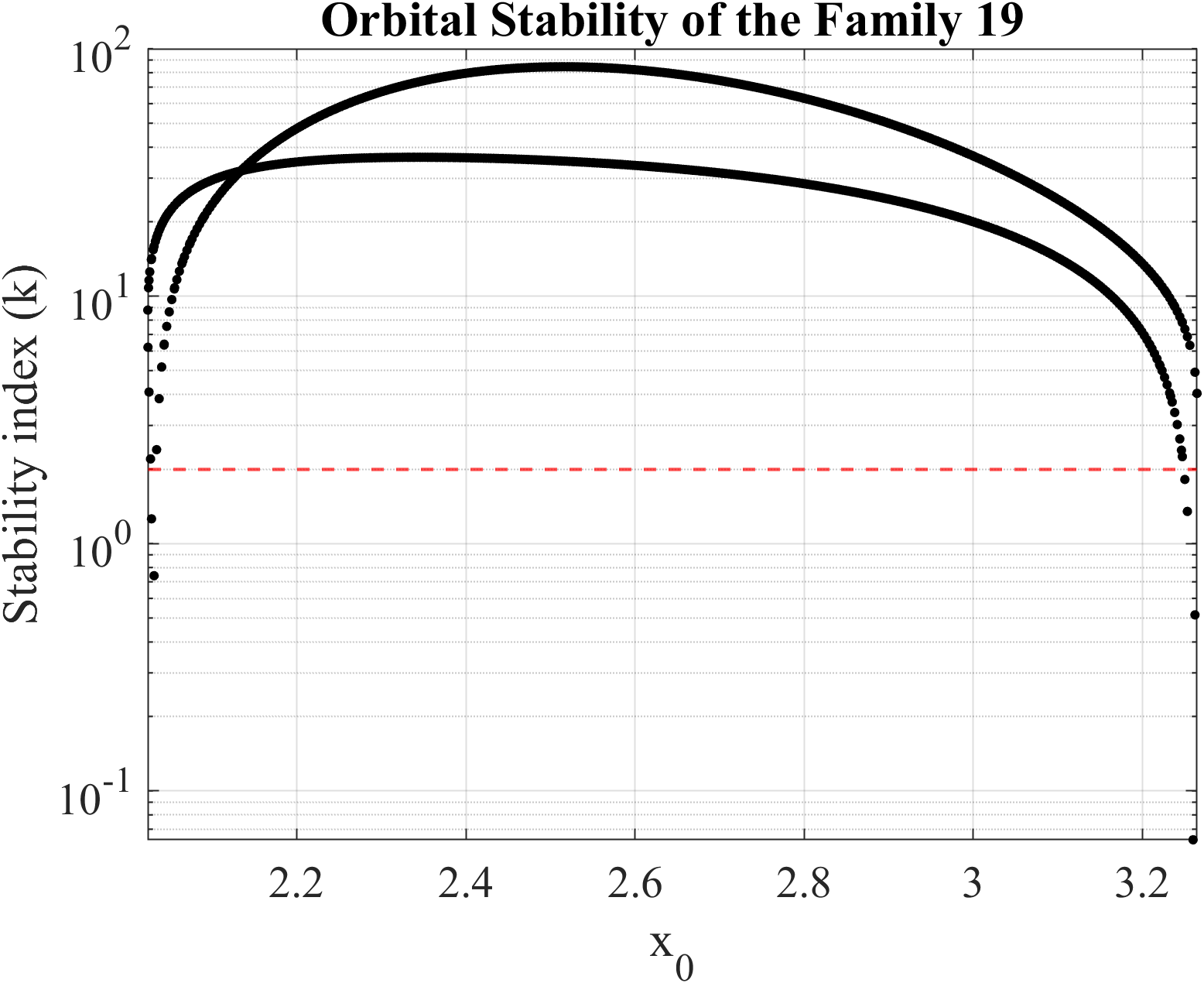}
        \caption{} 
        \label{Stab19}
    \end{subfigure}
    \hfill
     \begin{subfigure}{0.23\textwidth}
        \includegraphics[width=\linewidth]{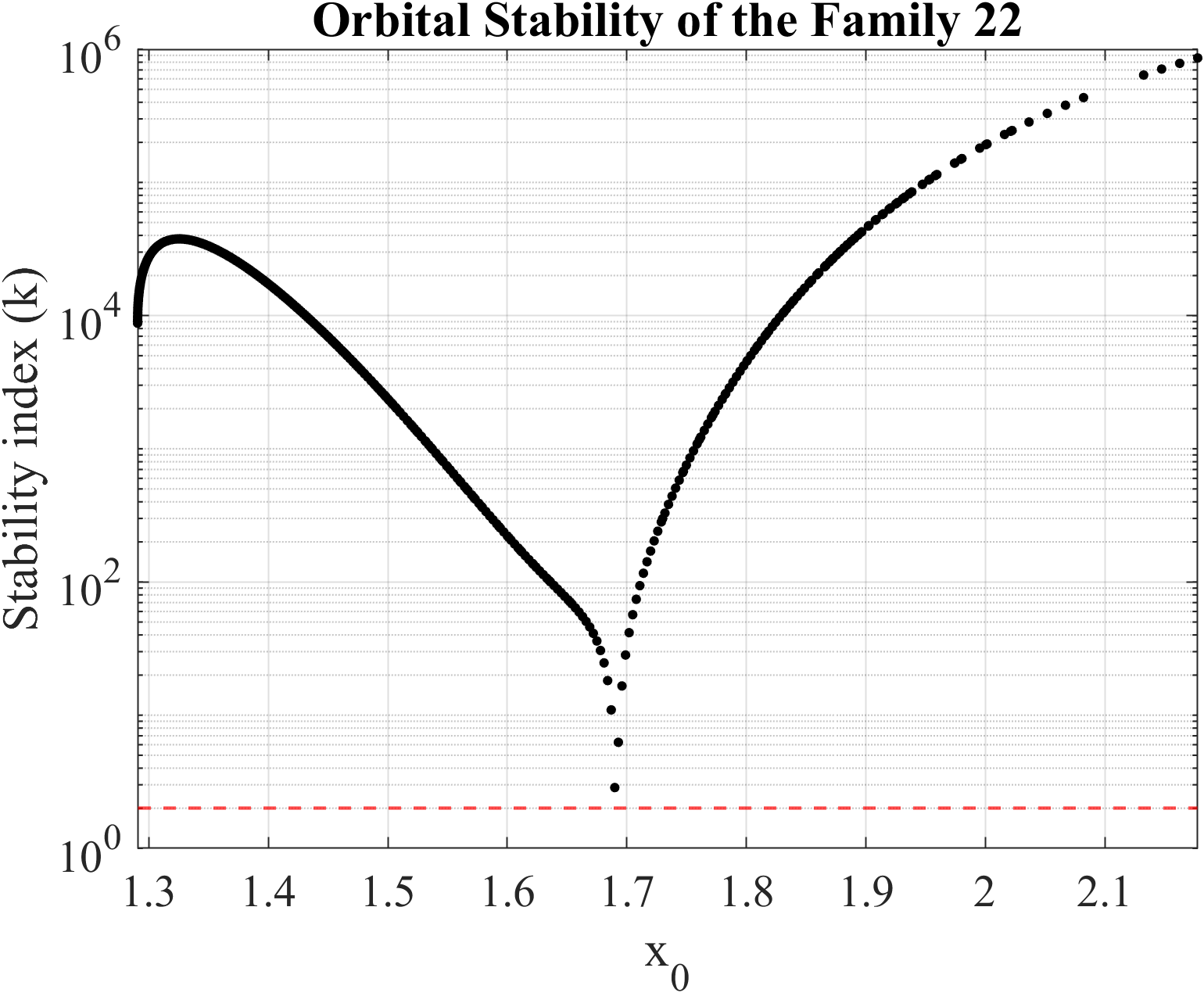}
        \caption{} 
        \label{Stab22}
    \end{subfigure}
    \hfill
      \begin{subfigure}{0.23\textwidth}
        \includegraphics[width=\linewidth]{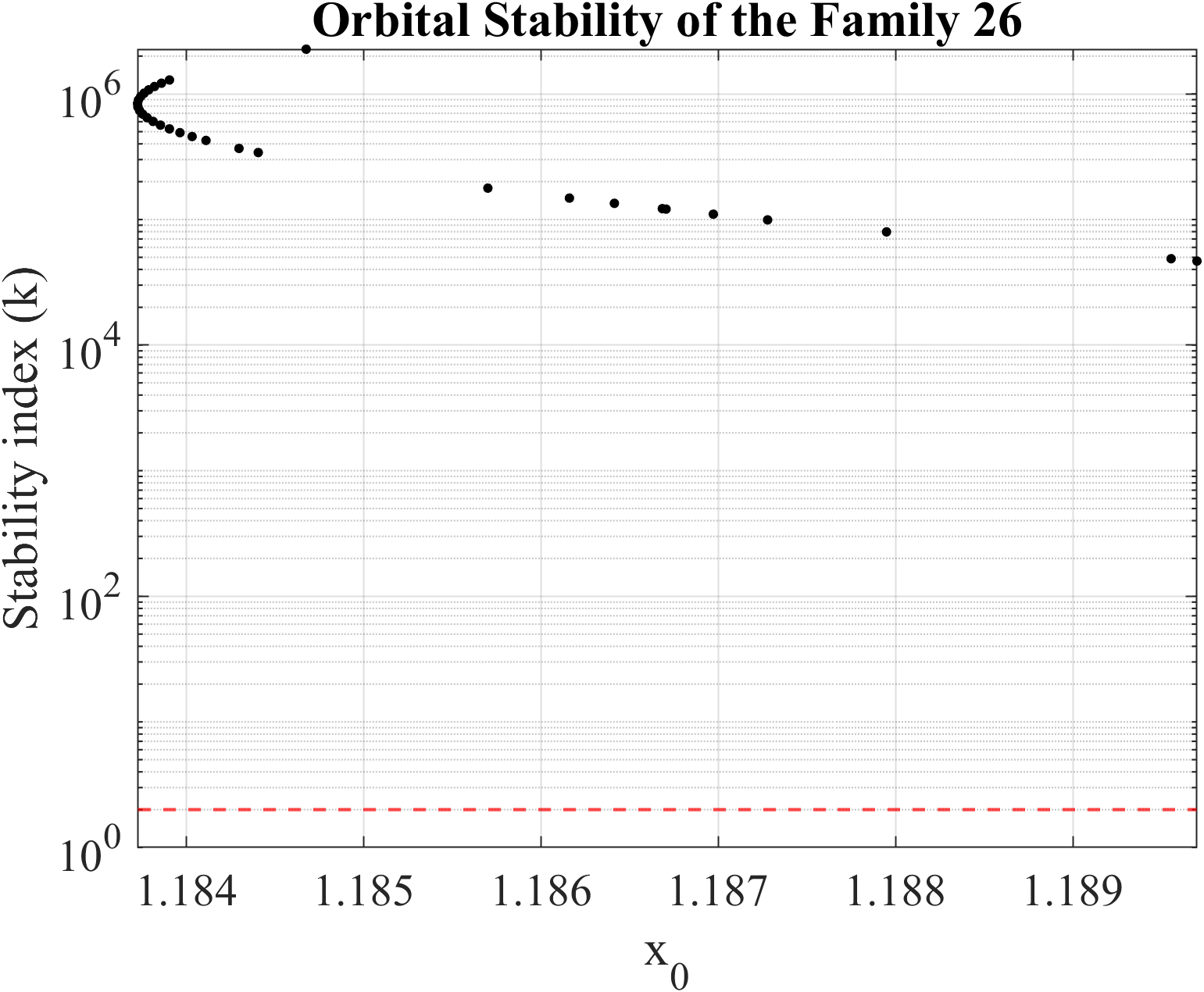}
        \caption{} 
        \label{Stab26}
    \end{subfigure}
    \caption{Stability index referring to families: (a) 18, (b) 19, (c) 22, and (d) 26. The dashed red line at $k=2$.}\label{Stab18e19e22e26}
    \end{figure}

Families 17, 18, 19, 21, 22, 23, and 26 form a set of orbits that circle the asteroid Justitia only. The orbits are prograde, and some of them also tend to loop. The Fig. \ref{Fig:Orbit17a26} shows the evolution of some of these families, being Fig. \ref{O18a} for Family 18, \ref{O19b} for Family 19, \ref{O22c} for Family 22, and \ref{O26d} for Family 26. Regarding the linear stability of the orbits, note in Fig. \ref{Stab18e19e22e26} that there are stable orbits in Family 18. They are not shown here, but most orbits in Families 17 and 21, which are also in this topological characterization group, are stable.

\begin{figure}
    \centering
    \begin{subfigure}{0.23\textwidth}
        \includegraphics[width=\linewidth]{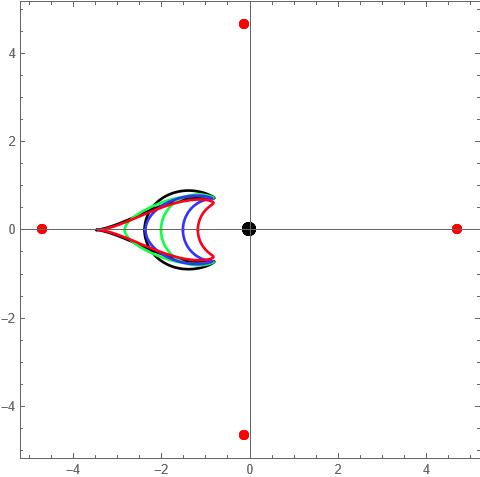}
        \caption{} 
        \label{O10}
    \end{subfigure}
     \hfill
    \begin{subfigure}{0.23\textwidth}
        \includegraphics[width=\linewidth]{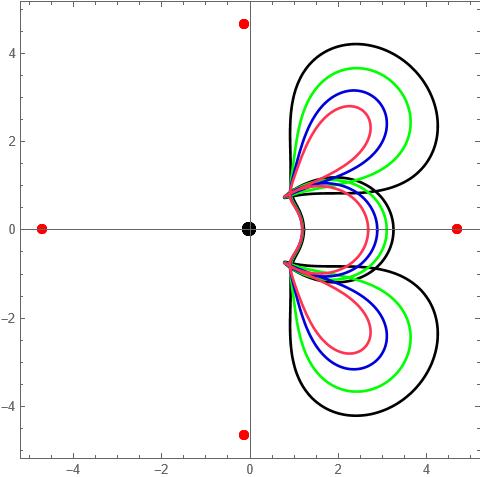}
        \caption{} 
        \label{O25}
    \end{subfigure}
    \hfill
     \begin{subfigure}{0.23\textwidth}
        \includegraphics[width=\linewidth]{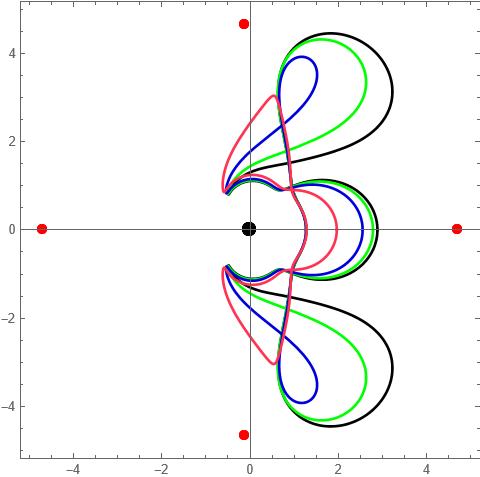}
        \caption{} 
        \label{O28}
    \end{subfigure}
    \caption{Evolution of the orbits of families (a) 10, (b) 25, and (c) 28. Black and red orbits are the ends of the families, and green and blue are intermediate points. The black dot represents the asteroid Justitia (not to scale), and the red dots are the equilibrium points. The abscissa and ordinate axes are the components of the particle's position around the asteroid, in x and y, respectively.}
    \label{fig:O10e25e28}
    \end{figure}

\begin{figure}
    \centering
    \begin{subfigure}{0.23\textwidth}
        \includegraphics[width=\linewidth]{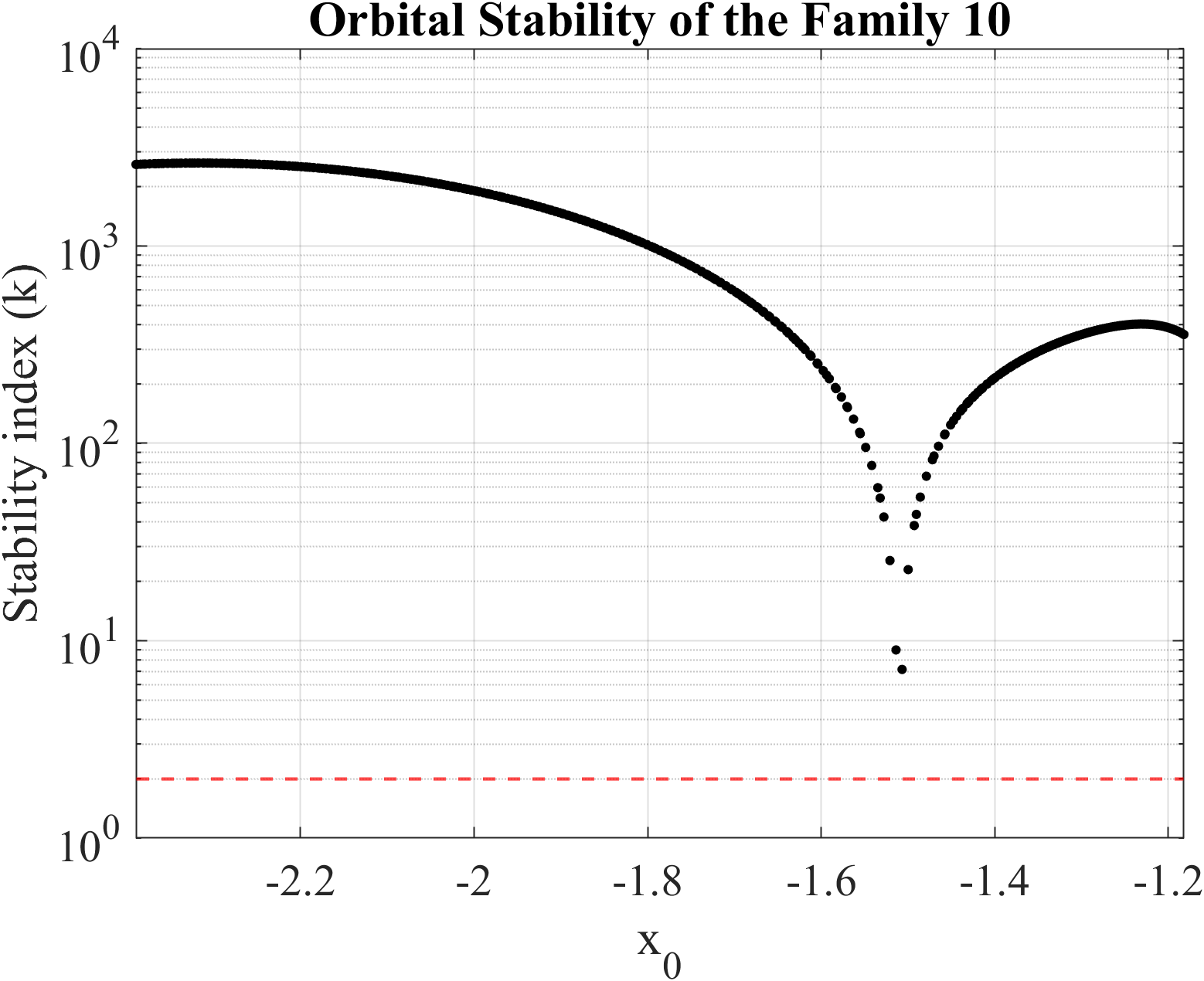}
        \caption{} 
        \label{Stab10}
    \end{subfigure}
     \hfill
    \begin{subfigure}{0.23\textwidth}
        \includegraphics[width=\linewidth]{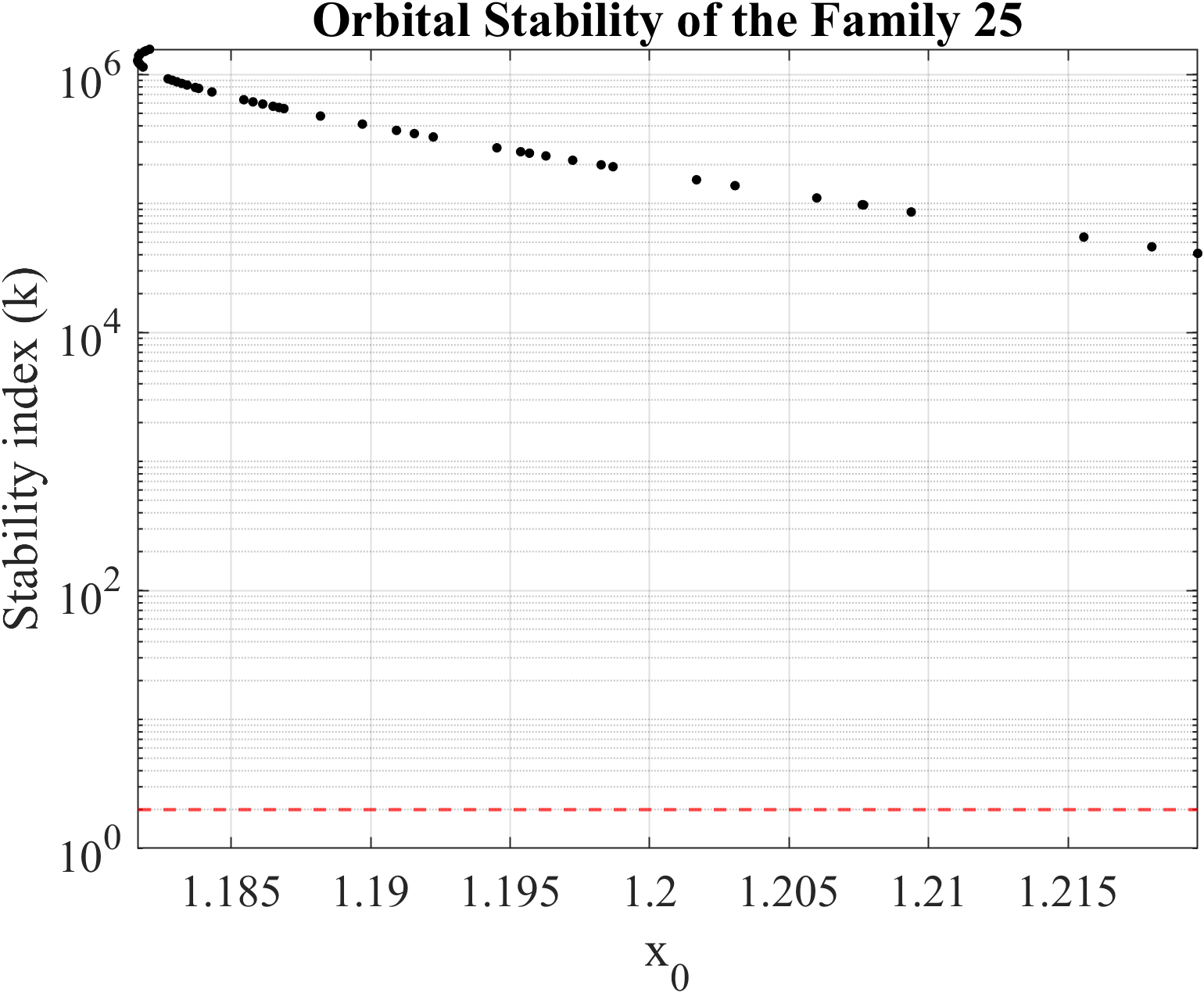}
        \caption{} 
        \label{Stab25}
    \end{subfigure}
    \hfill
     \begin{subfigure}{0.23\textwidth}
        \includegraphics[width=\linewidth]{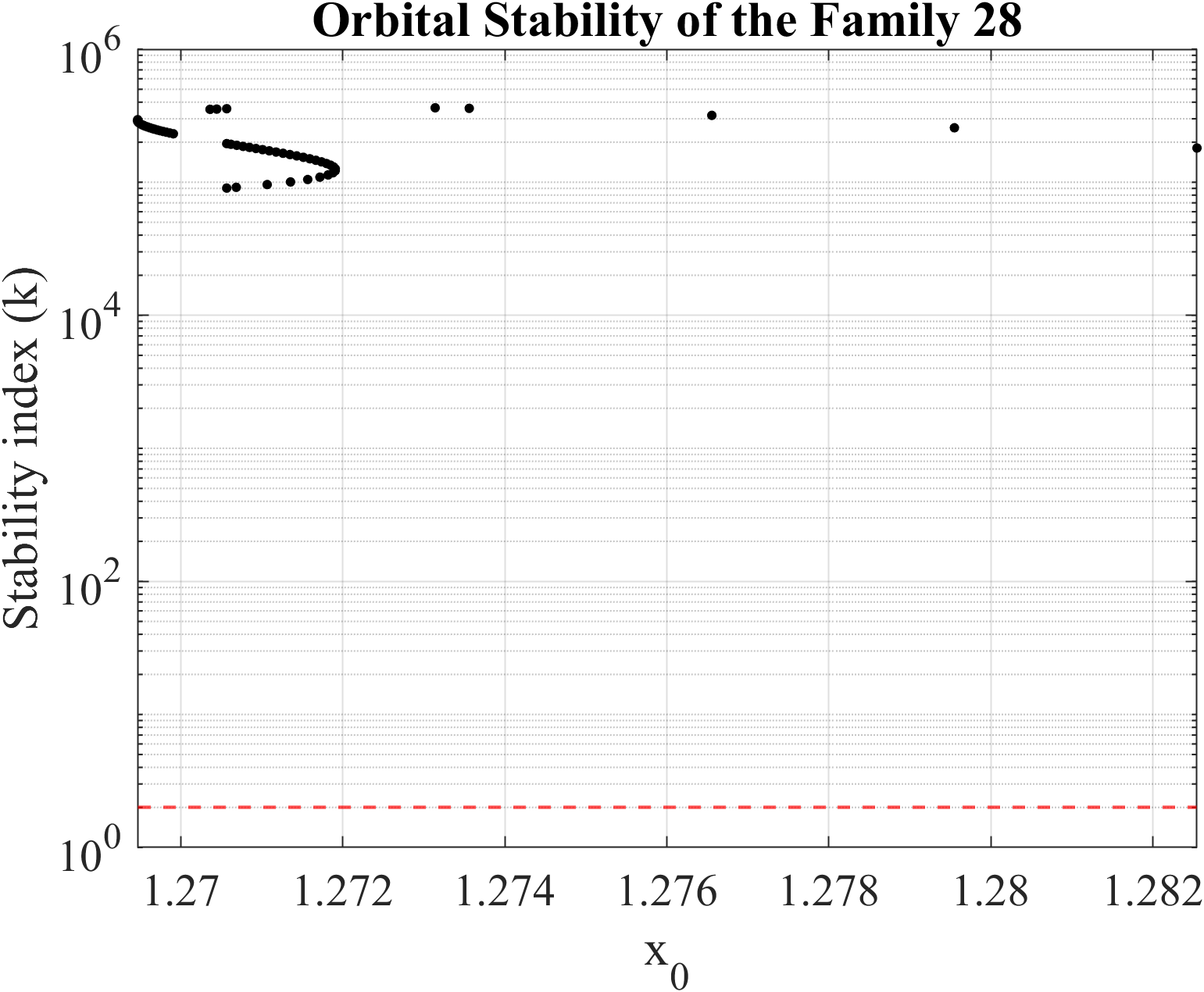}
        \caption{} 
        \label{Stab28}
    \end{subfigure}
    \caption{Stability index referring to families: (a) 10, (b) 25, and (c) 28. The dashed red line at $k=2$.}\label{Stab10e25e28}
    \end{figure}

Finally, we have Families 10, 20, 24, 25, 27, and 28 with prograde, unstable orbits that do not encircle the body and equilibrium points. Orbits 10 and 20 are similar, but mirror each other across the $y$-axis. Orbits 24, 25, 27, and 28 are also similar to each other, with some variations; they are composed of loops. To exemplify them, some will be presented in the Fig.~\ref{fig:O10e25e28}. In this group, all families are composed of highly unstable orbits, i.e., $k>>2.0$, as shown in Fig.~\ref{Stab10e25e28} for Families 10, 25, and 28.

\begin{figure*}
  \centering
  \plotone{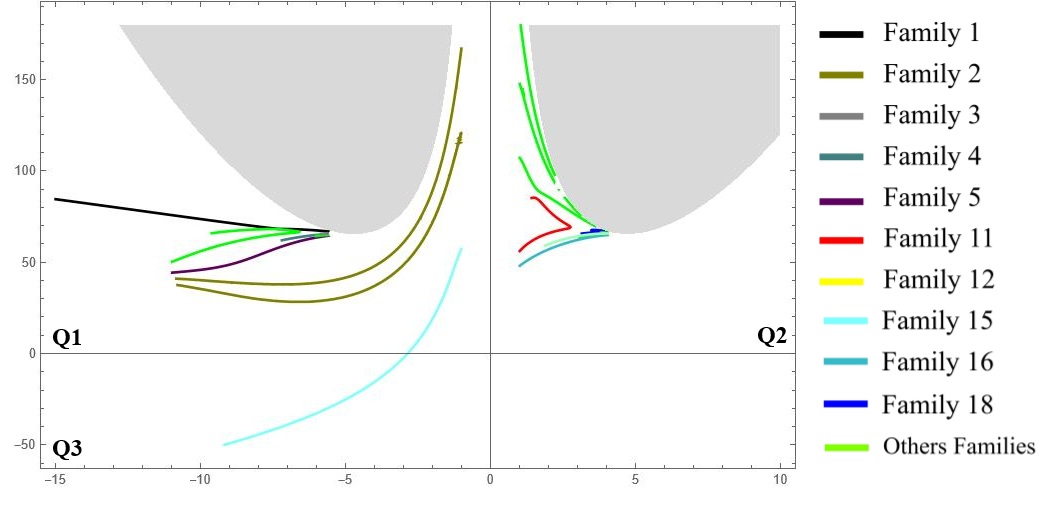}
 \caption{Characteristic curves of the asteroid (269) Justitia, considering the density equal to $2.0$ g cm$^{-3}$. Each curve, represented by a distinct color and named according to the legend, corresponds to a family of symmetric periodic orbits (SPO). Each point along a given SPO family represents the initial condition of an orbit. The $x$-axis is the initial position at $x_0$ of the orbit at time $t=0$; the $y$-axis, the Jacobi constant $(J)$. The map is divided into quadrants $Q1$, $Q2$, and $Q3$.}
  \label{MapaCI_d2}
\end{figure*}

The results presented so far for orbits around Justitia refer to the case with a density of $1.0$ g cm$^{-3}$. When the density is considered to be $2.0$ g cm$^{-3}$, changes are observed in the families of symmetric periodic orbits. Figure~\ref{MapaCI_d2} shows the map of initial conditions of the SPO for a density of $2.0$, showing a smaller number of families (24 in total) and some with different sizes compared to Fig.~\ref{MapaCI}. In this figure, we highlight, using the same colors and numerical identifications, the families that exhibit behavior similar (though not identical) to that of the case with density $1.0$. Of the 24 families shown in Fig.~\ref{MapaCI_d2}, only 10 maintain comparable characteristics (these are Families 1, 2, 3, 4, 5, 11, 12, 15, 16, 18), and some of them have different sizes and orbits; the remaining families, represented in green, correspond to new solutions, distinct from those obtained for density $1.0$. Note also that some families have disappeared (for a density of $1.0$, there were 28 families, and for a density of $2.0$, there are 24). 

\section{3-D Thermophysical model}\label{thermophysical}

As we mentioned before, the EMA mission will investigate the origin and evolution of water- and carbon-rich main-belt asteroids by using three infrared sensors: a thermal infrared spectrometer (EMBIRS), a mid-wave infrared imaging spectrometer (MIST-A), and a thermal infrared imaging system (IR-cam), along with others operating in the visible wavelength range \citep{2025epsc.conf.1435A}. Justitia is a primitive asteroid with an extremely red spectral slope, consistent with either space weathering effects or an outer Solar System origin \citep{2025ApJ...992..125H}. In this sense, EMA will use infrared observations to determine the physical and compositional properties of the surface materials on Justitia, providing insights into its origins and geologic history.

We can estimate the surface and subsurface temperature distribution by applying the TEMPEST software  \citep{1989Icar...78..337S,2024EPSC...17.1121L, lyster2025tempest}. This code calculates surface temperatures by solving the one-dimensional heat conduction equation for each facet of a shape model, using a surface energy balance that includes solar flux, thermal emission, vertical heat conduction, shadowing, radiative exchange, and (optionally) self-heating. It simulates diurnal temperature variations of a Solar System body based on a given shape model and iteratively computes insolation and temperature arrays for each facet until convergence.
These results are useful for constraining the thermal properties of Justitia, such as thermal inertia, and provide insights into surface characteristics, including grain size and regolith structure.

Here, we present results of a 3-D thermophysical model incorporating the best-available shape model for asteroid Justitia, based on rotational lightcurve and stellar occultation measurements \citep{marciniak2025}. For this analysis, we adopted an albedo of 0.06, a heliocentric distance of $r_\odot = 2.6$~au (semi-major axis), a rotational period of $P_{\mathrm{rot}} = 33.13$~h, and ecliptic pole orientation coordinates of $73^\circ$ and $-81^\circ$ \citep{marciniak2025}. We also used a bolometric emissivity of 0.9 and a bulk density of 1~g/cm$^3$, according to \citep{buie2025}. We then investigated a range of thermal inertia ($\Gamma$) values - 40, 100, and 200~J~m$^{-2}$~K$^{-1}$~s$^{-1/2}$ - spanning a wide range for primitive asteroids \citep{2010Icar..205..505M, 2015Icar..256..101H}. Thermal inertia is usually determined with considerable uncertainty. In \cite{marciniak2025}, it is reported as $41^{+110}_{-40}$ SI units for Pole direction 1 (used in this simulation). According to the authors, this large uncertainty may be related to the relatively wide variations in Justitia’s heliocentric distance, since thermal inertia depends on heliocentric distance through its influence on surface temperatures. The parameter values used in the thermal analyses are summarized in Table 8.

\begin{figure}[h]
\centering
    \includegraphics[width=1.0\linewidth]{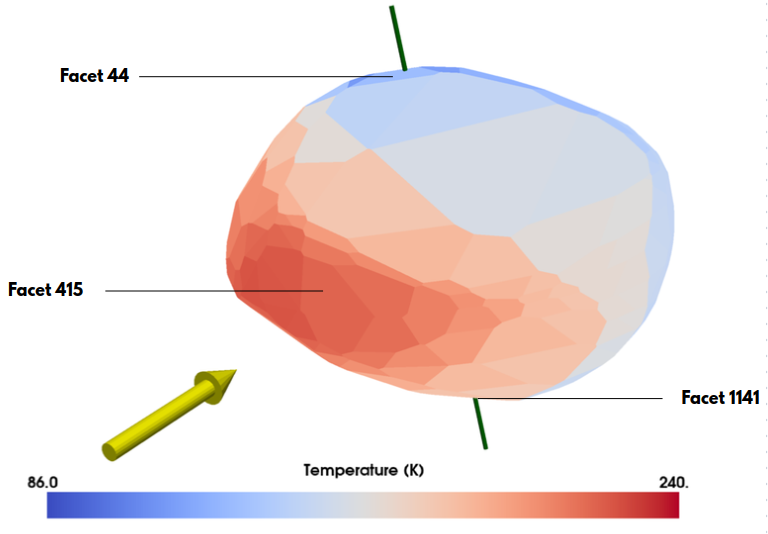}
    \caption{Surface temperature map predicted by TEMPEST for a thermal inertia of 40~J~m$^{-2}$~K$^{-1}$~s$^{-1/2}$, at a heliocentric distance corresponding to the semi-major axis of 2.6~au. Facets 44, 415, and 1141 on the surface represent the north pole, equator, and south pole, respectively.}
    \label{temperatures1}
\end{figure}

\begin{figure}[h]
\centering
    \includegraphics[width=1.0\linewidth]{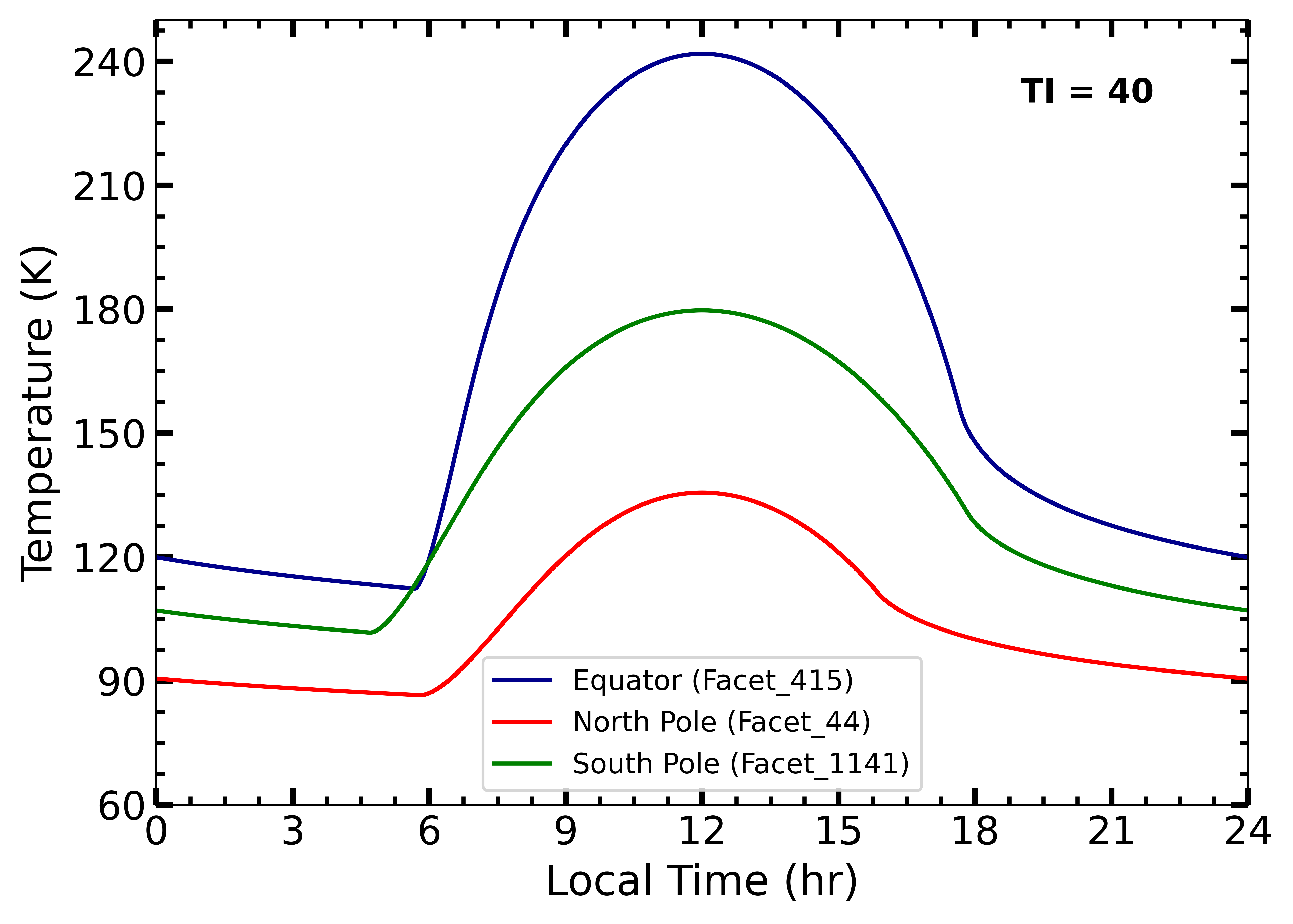}

    \vspace{0.2em} 
    \includegraphics[width=1.0\linewidth]{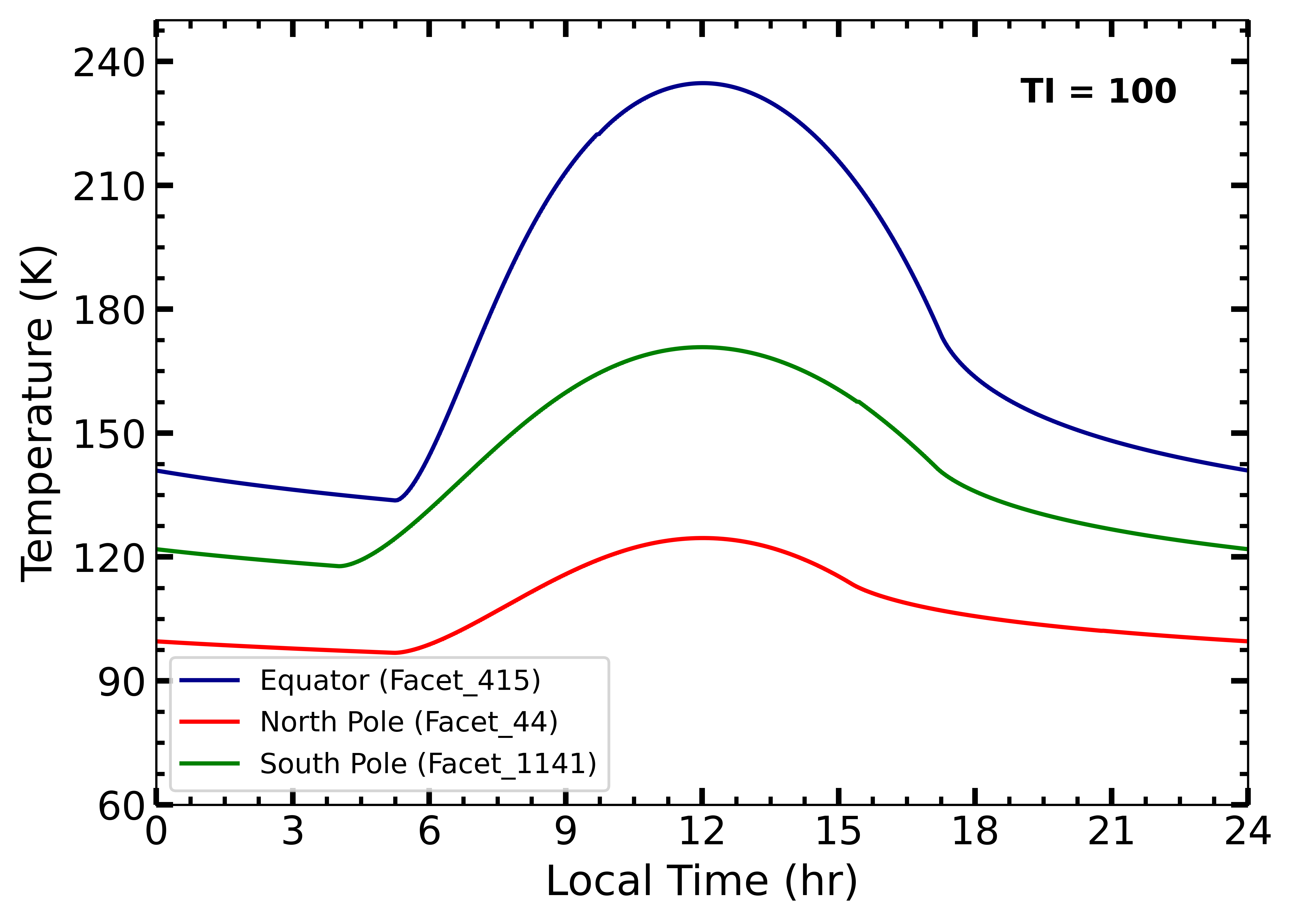}

    \vspace{0.2em} 
    \includegraphics[width=1.0\linewidth]{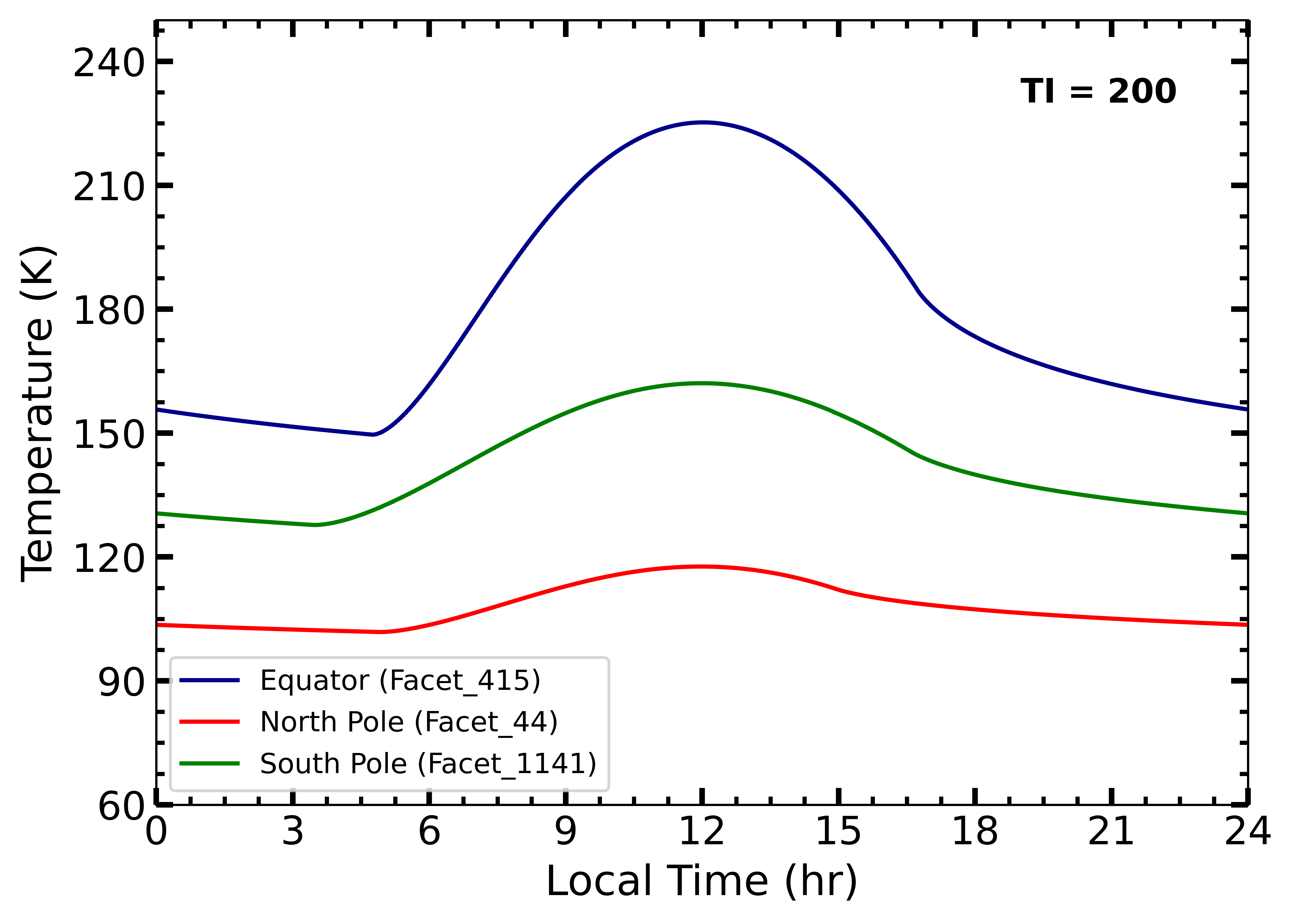}

    \caption{Predicted diurnal surface temperatures at three representative surface points corresponding to those shown in Fig. xx. Each plot is derived based on thermal inertia values of 40, 100, and 200~J~m\(^{-2}\)~K\(^{-1}\)~s\(^{-1/2}\).}
    \label{temperatures2}
\end{figure}

For each thermal inertia scenario, we calculated the global distribution of surface temperatures, the diurnal amplitude at the equator, and the subsolar temperature as a function of local time. In Figure \ref{temperatures1}, we report the surface temperatures of asteroid Justitia, whose variations show strong dependence on both latitude and longitude, owing to Justitia’s irregular shape and pole orientation/obliquity. For a nominal thermal inertia of 40~J~m$^{-2}$~K$^{-1}$~s$^{-1/2}$, equatorial temperatures range from approximately 240~K near local noon to about 112~K on the night side during one rotation. The north pole reaches a maximum of $\sim$136~K and a minimum of $\sim$87~K, while the south pole reaches $\sim$180~K with a pre-dawn temperature of $\sim$102~K. Considering a thermal inertia of 100~J~m$^{-2}$~K$^{-1}$~s$^{-1/2}$, equatorial temperatures range from $\sim$235~K to $\sim$134~K over one orbit, with the north pole varying between $\sim$125~K and $\sim$97~K, and the south pole between $\sim$170~K and $\sim$118~K. Increasing thermal inertia to 200~J~m$^{-2}$~K$^{-1}$~s$^{-1/2}$, equatorial temperatures vary from $\sim$225~K to $\sim$150~K, with the north pole ranging between $\sim$118~K and $\sim$100~K, and the south pole between $\sim$160~K and $\sim$128~K, reducing the amplitude by roughly 50\% on the equator in comparison to $\Gamma = 40$.

Furthermore, regarding the overall temperature variation, our results show a difference of $54\,K$ in relation to the minimum temperature and $10\,K$ in relation to the maximum temperature
when compared with the results of \citet{hayne2024AGU}

Figure \ref{temperatures2} presents the predicted diurnal surface temperatures for different thermal inertia values at three representative surface regions (equator, north pole, and south pole), corresponding to those shown in Figure \ref{temperatures1}. Table \ref{tab:facet_temp} reports the surface temperatures for different heliocentric distances, considering perihelion and aphelion values. In the perihelion, we find day-side temperatures reaching $\sim$275~K near local noon and night-side minima of $\sim$118~K, corresponding to a median diurnal amplitude of $\sim$157~K at the equator. On the other hand, at aphelion, temperatures reach $\sim$217~K and drop to $\sim$108~K on the night side. All simulations indicate that the South Pole receives more insolation than the North Pole, as evidenced by its higher surface temperatures. Therefore, biogenic/primitive materials could be found better preserved at the North Pole. Thermal emission derived from these models can be used to generate synthetic thermal phase curves based on possible thermal inertia scenarios, which may serve in future work for direct comparison with IR-cam and other instruments on the EMA payload.


\begin{table}[h!]
\centering
\caption{Physical and Thermal Properties of the Object}

{
\begin{tabular}{l c c}
\hline
Parameter & Value & Units \\
\hline
$\lambda$ & $73 \pm 11^{a}$ & deg \\
$\beta$ & $-81 \pm 15^{a}$ & deg \\
Albedo & 0.06$^{a}$ & -- \\
Emissivity & 0.9$^{a}$ & -- \\
Thermal inertia & 41$^{a}$ & J\,m$^{-2}$\,K$^{-1}$\,s$^{-1/2}$ \\
Solar distance (Main-Belt) & 2.06--3.17$^{d}$ & au \\
Rotational period & 33.12962$^{a}$ & h \\
Density & 1000$^{b}$ & kg\,m$^{-3}$ \\
Specific heat capacity & 750$^{c}$ & J\,kg$^{-1}$\,K$^{-1}$ \\
\hline
\end{tabular}
}

\vspace{2mm}
\raggedright
$^{a}$\cite{marciniak2025} \\
$^{b}$\citep{buie2025}\\
$^{c}$\citep{2021JGRE..12607003P}\\
$^{d}$ JPL Small-Body Database: \url{https://ssd.jpl.nasa.gov/}.
\end{table}


\section{Conclusion}\label{sec13}

In this work, we present for the first time a detailed analysis of the surface and orbital dynamics of asteroid (269) Justitia, a member of the main asteroid belt with an unusual characteristic: it is an ultra-red (Z-type) asteroid. We use its 3-D polyhedral shape model and its equivalent ellipsoidal model to analyze the surface dynamics and planar symmetric periodic orbits around it. 

The analysis of the geopotential and surface acceleration showed that the gravitational potential dominates the dynamical environment around asteroid Justitia, a consequence of its slow rotation. We also show that the global behavior of the geopotential and surface acceleration of Justitia remains unchanged for densities from 1.0\,g\,cm$^{-3}$ to 2.0\,g\,cm$^{-3}$.

We also identified that the variation in its tilts does not exceed 50$^{\circ}$. We found that 97.5\% of Justitia's surface has slopes below 40$^{\circ}$. Regarding its escape speed, it exhibits typical behavior for slowly rotating small bodies: the regions of maximum escape speed correspond to the regions with the minimum geopotential.

Furthermore, we analyzed the gravitational effects of Solar Radiation Pressure at the apocenter and pericenter of its current orbit, and also considering it in the past in the Kuiper belt at $\sim40$\,au. The data obtained indicate that particles on the surface of Justitia larger than $1\,\mu m$ at the pericentre, $3\times 10^{-1}\,\mu m$ at the apocentre, and $5.0\times 10^{-3}$ in the past, at $\sim 40$\, au, are negligibly affected by solar radiation pressure. We also analyzed the effects of third-body perturbation on Justitia's slopes, and our results indicate that this perturbation has a negligible influence on the overall behavior of the slopes.

Investigation of the zero-velocity curves in projection planes xOy, xOz, and yOz confirms the existence of the five equilibrium points about asteroid Justitia, regarding the densities from 1.0\,g\,cm$^{-3}$ to 2.0\,g\,cm$^{-3}$ and a rotational period of and a rotational period of 33.12962\,h. Regarding the topological structure of equilibrium points at the xOy plane, our results show that equilibrium points E$_1$ and E$_3$ are unstable. Equilibrium points E$_2$, E$_4$, and E$_5$ are linearly stable, with E$_5$ located internally to the asteroid. We also observed that as the density increases, the gravitational pull becomes stronger. This result is also consistent, as we analyzed the evolution of equilibrium points, assuming that Justitia had a faster spin in the past. Consequently, the external equilibrium points shift outward, while the internal equilibrium point shifts inward to the Justitia center of mass. The equilibrium point E$_3$ has the minimum geopotential.

The results about return speed show that the regions of maximum and minimum return speed correspond to the areas of maximum and minimum escape speed, respectively.

We also identified 28 new families of planar-symmetric periodic orbits around Justitia, using an equivalent ellipsoidal shape model for this purpose, expanded in a series of spherical harmonics up to fourth order. We analyzed the orbits of these families and their orbital stability. For a space mission, the ideal scenario is one of orbital stability (K < 2) or low values of K > 2, as the orbit can still be utilized in specific, controlled
orbital strategies.

Finally, we believe that our results could guide future investigations of the Emirates Mission to the Asteroid Belt (EMA) about the dynamic environment of the ultra-red asteroid (269) Justitia. Our analyses could also be applied to other similar small bodies in the Solar System that rotate slowly.

Our results have broader implications: the reliable constraints on the surface dynamics and orbital environment of Justitia, together with the knowledge of possible trajectories, will facilitate mission planning for the MBR Explorer, including in situ operations and landing strategies. In addition, we present thermal modeling results for Justitia at different thermal inertia values and heliocentric distances, along with detailed temperature maps of the asteroid. These results provide information on the surface temperatures and highlight which regions receive more insolation, thereby indicating the colder areas of the object where materials may be better preserved. We found that the south pole receives more insolation than the north pole, with minimum temperatures of 102 K and 87 K, respectively. Consequently, the quality of data obtained during the mission, when compared with pre-mission predictions, will provide crucial ground truth for asteroid studies based on remote sensing and ground-based methods.

\begin{acknowledgments}
A. Amarante and A. Ferreira thank the financial support of the São Paulo Research Foundation (FAPESP) [grants \#2023/11781-5 \& \#2025/15438-9].
FM thanks the financial support from the São Paulo Research Foundation (FAPESP), Brazil, under Process Number 2024/16260-6.
The authors also acknowledge the financial support of the Coordination for the Improvement of Higher Education Personnel (CAPES) - Brazil (finance code 001). This research was also supported by computational resources supplied by the Center for Scientific Computing (NCC/GridUNESP) of the São Paulo State University (UNESP) and the Center for Mathematical Sciences Applied to Industry (CeMEAI), funded by FAPESP [grant \#2013/07375-0].
\end{acknowledgments}

\section*{Data Availability}

\textbf{Data availability} The dataset generated in this work is publicly available on Zenodo at \url{https://doi.org/10.5281/zenodo.18327490}. \\
\textbf{Code availability} The numerical codes used in this work can be found at GitHub: \url{https://github.com/a-amarante}\\
\textbf{Conflict of interest} The authors declare no conflict of interest.

\begin{contribution}
LB, AA, AF, FM, and MM equally contributed to this work. LB, AA, AF, FM, and MM performed numerical simulations, figures, and wrote the manuscript. AA, AF, and FM revised the manuscript.
\end{contribution}

\appendix
\section{COMPUTATION OF GRAVITATIONAL FIELD}
\label{appendixA}
The analytical expressions for the gravitational potential, the gravitational attraction vector, and the gravitational gradient matrix, with singularity corrections, can be written as follows for a homogeneous asteroid of constant density \( \rho \) \citep{tsoulis2001}.
\begin{equation}
\begin{split}
U(x_1, x_2, x_3) = -\frac{G \rho}{2} \sum_{p=1}^{n} \sigma_p h_p \biggl[ 
& \sum_{q=1}^{m} \sigma_{pq} h_{pq} L_{N_{pq}} + \\ 
& + h_p \sum_{q=1}^{m} \sigma_{pq} A_{N_{pq}} + \sin(gA_p) \biggr],
\end{split}
\label{eq:potencial_split}
\end{equation}
\begin{equation}
\begin{split}
- \frac{\partial U(x_1, x_2, x_3)}{\partial x_i} = -G \rho \sum_{p=1}^{n} \cos(\mathbf{N}_p, \mathbf{e}_i) \biggl[ 
& \sum_{q=1}^{m} \sigma_{pq} h_{pq} L_{N_{pq}} + \\ 
& + h_p \sum_{q=1}^{m} \sigma_{pq} A_{N_{pq}} + \sin(gA_p) \biggr], \\&\quad (i = 1, 2, 3),
\end{split}
\label{eq:atracao_split}
\end{equation}
\begin{equation}
\begin{split}
- \frac{\partial^2 U(x_1, x_2, x_3)}{\partial x_i \partial x_j} = G \rho \sum_{p=1}^{n} \cos(\mathbf{N}_p, \mathbf{e}_i) \biggl[ 
& \sum_{q=1}^{m} \cos(\mathbf{n}_{pq}, \mathbf{e}_j) L_{N_{pq}} \\ 
& + \sigma_p \cos(\mathbf{N}_p, \mathbf{e}_j) \sum_{q=1}^{m} \sigma_{pq} A_{N_{pq}}+ \\ 
& + \sin(gB_{pj}) \biggr], \quad (i, j = 1, 2, 3).
\end{split}
\label{eq:gradiente_split}
\end{equation}
Auxiliary functions are defined by:
\begin{equation}
L_{N_{pq}} = \ln\left( \frac{s_{2pq} + l_{2pq}}{s_{1pq} + l_{1pq}} \right),
\label{eq:lnpq}
\end{equation}
\begin{equation}
A_{N_{pq}} = \arctan\left( \frac{h_p s_{2pq}}{h_{pq} l_{2pq}} \right) - \arctan\left( \frac{h_p s_{1pq}}{h_{pq} l_{1pq}} \right).
\label{eq:anpq}
\end{equation}
\FloatBarrier
The variables \( x_1, x_2, x_3 \) represent the coordinates of the test point in the reference system. The polyhedral model is homogeneous, with constant density \( \rho = M/V \), and its shape is composed of \( n \) faces, each containing \( m \) edges. The gravitational constant is \( G = 6{,}67408 \times 10^{-20} \ \text{km}^3\,\text{kg}^{-1}\,\text{s}^{-2} \). Each face defines a polygonal plane \( S_p \), whose unit normal is the vector \( \mathbf{N}_p \). The orthogonal projection of the point \( P(x_1, x_2, x_3) \) onto the plane of the face is denoted by \( P' \), and the projection of \( P' \) onto the edge \( G_{pq} \), belonging to the face \( S_p \), is called \( P'' \). The distance between \( P \) and the plane of the face is \( h_p \), while the distance between \( P' \) and \( P'' \) is \( h_{pq} \). The vectors \( \mathbf{e}_i \) and \( \mathbf{e}_j \) are the unit vectors of the geographic missions. The vector \( \mathbf{n}_{pq} \) is a unit vector contained in the plane of the face, perpendicular to the edge \( G_{pq} \), and is oriented outwards from the polyhedron. The sign \( \sigma_{pq} \) assumes the value \( -1 \) when \( \mathbf{n}_{pq} \) points to the half-plane containing \( P' \), and \( +1 \) otherwise.

The distances \( l_{1pq} \) and \( l_{2pq} \) are the distances between the point \( P \) and the two endpoints of the edge \( G_{pq} \), while \( s_{1pq} \) and \( s_{2pq} \) are the distances between \( P'' \) and these same endpoints. The terms \( \cos(\mathbf{N}_p, \mathbf{e}_i) \) and \( \cos(\mathbf{n}_{pq}, \mathbf{e}_j) \) are the directed cosines between the normal and common vectors. Finally, the terms \( \sin(gA_p) \) and \( \sin(gB_{pj}) \) involve the singularity corrections that arise when the point \( P \) is exactly on the face, on an edge, or at a vertex of the polyhedron. These terms are zero when the point is outside the projection of the corresponding face.
\FloatBarrier

\section{EIGENVALUES}
\label{appendixB}
\begin{table}
    \centering
    \footnotesize
    \setlength{\tabcolsep}{4pt}
    \caption{Eigenvalues ($\lambda_n \times 10^{-5}$, $n = 1, 2, \ldots, 6$) corresponding to the equilibrium points in the Justitia gravitational field, calculated for different density values and a rotational period of 33.12962\,h; utilizing \textsc{Minor-Equilibria} package \citep{amaranteandwinter2020}.}
    \label{tab:2}
\begin{tabular}{@{}l@{\hspace{2pt}}c@{\hspace{4pt}}ccccr@{}}
    \hline
    Point & \multicolumn{1}{c}{Density (g\,cm$^{-3}$)} & $\lambda_{1,2}$ & $\lambda_{3,4}$ & $\lambda_{5,6}$ & Topological Structure\\
    \hline
    \multirow{6}{*}{E$_1$} 
      & 1.0 & $\pm 0.863$ & $\pm 5.325 i$ & $\pm 5.282 i$ & \multirow{6}{*}{saddle-centre-centre} \\
      & 1.2 & $\pm 0.809$ & $\pm 5.318 i$ & $\pm 5.280 i$ & \\
      & 1.4 & $\pm 0.813$ & $\pm 5.313 i$ & $\pm 5.279 i$ & \\
      & 1.6 & $\pm 0.832$ & $\pm 5.309 i$ & $\pm 5.278 i$ & \\
      & 1.8 & $\pm 0.831$ & $\pm 5.306 i$ & $\pm 5.278 i$ & \\
      & 2.0 & $\pm 0.831$ & $\pm 5.304 i$ & $\pm 5.277 i$ & \\
    \hline
    \multirow{6}{*}{E$_2$} 
      & 1.0 & $\pm 0.831 i$ & $\pm 5.165 i$ & $\pm 5.305 i$ & \multirow{6}{*}{centre-centre-centre} \\
      & 1.2 & $\pm 0.783 i$ & $\pm 5.177 i$ & $\pm 5.300 i$ & \\
      & 1.4 & $\pm 0.793 i$ & $\pm 5.186 i$ & $\pm 5.292 i$ & \\
      & 1.6 & $\pm 0.713 i$ & $\pm 5.193 i$ & $\pm 5.295 i$ & \\
      & 1.8 & $\pm 0.686 i$ & $\pm 5.199 i$ & $\pm 5.198 i$ & \\
      & 2.0 & $\pm 0.678 i$ & $\pm 5.205 i$ & $\pm 5.277 i$ & \\
    \hline
    \multirow{6}{*}{E$_3$} 
      & 1.0 & $\pm 0.892$ & $\pm 5.328 i$ & $\pm 5.305 i$ & \multirow{6}{*}{saddle-centre-centre} \\
      & 1.2 & $\pm 0.835$ & $\pm 5.321 i$ & $\pm 5.281 i$ & \\
      & 1.4 & $\pm 0.831$ & $\pm 5.316 i$ & $\pm 5.280 i$ & \\
      & 1.6 & $\pm 0.832$ & $\pm 5.312 i$ & $\pm 5.278 i$ & \\
      & 1.8 & $\pm 0.823$ & $\pm 5.309 i$ & $\pm 5.278 i$ & \\
      & 2.0 & $\pm 0.831$ & $\pm 5.306 i$ & $\pm 5.277 i$ & \\
    \hline
    \multirow{6}{*}{E$_4$} 
      & 1.0 & $\pm 0.840 i$ & $\pm 5.301 i$ & $\pm 5.168 i$ & \multirow{6}{*}{centre-centre-centre} \\
      & 1.2 & $\pm 0.791 i$ & $\pm 5.297 i$ & $\pm 5.179 i$ & \\
      & 1.4 & $\pm 0.752 i$ & $\pm 5.188 i$ & $\pm 5.188 i$ & \\
      & 1.6 & $\pm 0.720 i$ & $\pm 5.195 i$ & $\pm 5.194 i$ & \\
      & 1.8 & $\pm 0.693 i$ & $\pm 5.200 i$ & $\pm 5.206 i$ & \\
      & 2.0 & $\pm 0.669 i$ & $\pm 5.205 i$ & $\pm 5.207 i$ & \\
    \hline
    \multirow{6}{*}{E$_5$} 
      & 1.0 & $\pm 59.278 i$ & $\pm 54.893 i$ & $\pm 43.767 i$ & \multirow{6}{*}{centre-centre-centre} \\
      & 1.2 & $\pm 64.933 i$ & $\pm 59.662 i$ & $\pm 48.417 i$ & \\
      & 1.4 & $\pm 70.134 i$ & $\pm 64.052 i$ & $\pm 52.689 i$ & \\
      & 1.6 & $\pm 74.975 i$ & $\pm 68.140 i$ & $\pm 56.663 i$ & \\
      & 1.8 & $\pm 79.521 i$ & $\pm 71.983 i$ & $\pm 60.392 i$ & \\
      & 2.0 & $\pm 83.822 i$ & $\pm 75.620 i$ & $\pm 63.916 i$ & \\
    \hline
\end{tabular}
\end{table}

\FloatBarrier

\section{Temperatures in different regions}
\begin{table}[ht]

\centering
\caption{Maximum and minimum temperatures of the three different regions/facets for different thermal inertia values and solar distances.}

{
\begin{tabular}{l c c c}
\hline
Facet & Distance 2.6 au (K) & Distance 2.05 au (K) & Distance 3.18 au (K) \\
\hline
\multicolumn{4}{c}{Thermal Inertia = 40~J~m\(^{-2}\)~K\(^{-1}\)~s\(^{-1/2}\)} \\
\hline
Facet\_415 (Equator)  & \makecell{Max: 241.85 \\ Min: 112.34} & \makecell{Max: 274.09 \\ Min: 117.59} & \makecell{Max: 217.11 \\ Min: 107.85} \\
Facet\_44 (North Pole)   & \makecell{Max: 135.55 \\ Min: 86.52}  & \makecell{Max: 156.78 \\ Min: 92.09}  & \makecell{Max: 119.37 \\ Min: 81.63} \\
Facet\_1141 (South Pole) & \makecell{Max: 179.71 \\ Min: 101.67} & \makecell{Max: 204.87 \\ Min: 107.20} & \makecell{Max: 160.31 \\ Min: 96.89} \\
\hline
\multicolumn{4}{c}{Thermal Inertia = 100~J~m\(^{-2}\)~K\(^{-1}\)~s\(^{-1/2}\)} \\
\hline
Facet\_415 (Equator)  & \makecell{Max: 234.72 \\ Min: 133.68} & \makecell{Max: 268.18 \\ Min: 141.01} & \makecell{Max: 208.87 \\ Min: 127.27} \\
Facet\_44 (North Pole)  & \makecell{Max: 124.57 \\ Min: 96.78}  & \makecell{Max: 145.12 \\ Min: 104.98} & \makecell{Max: 109.65 \\ Min: 89.76} \\
Facet\_1141 (South Pole) & \makecell{Max: 170.79 \\ Min: 117.75} & \makecell{Max: 196.89 \\ Min: 125.82} & \makecell{Max: 151.01 \\ Min: 110.72} \\
\hline
\multicolumn{4}{c}{Thermal Inertia = 200~J~m\(^{-2}\)~K\(^{-1}\)~s\(^{-1/2}\)} \\
\hline
Facet\_415 (Equator)  & \makecell{Max: 225.22 \\ Min: 149.59} & \makecell{Max: 259.71 \\ Min: 159.38} & \makecell{Max: 198.94 \\ Min: 140.98} \\
Facet\_44 (North Pole)  & \makecell{Max: 117.67 \\ Min: 101.81} & \makecell{Max: 136.16 \\ Min: 112.05} & \makecell{Max: 104.42 \\ Min: 93.44} \\
Facet\_1141 (South Pole) & \makecell{Max: 162.06 \\ Min: 127.73} & \makecell{Max: 187.61 \\ Min: 138.50} & \makecell{Max: 143.32 \\ Min: 118.58} \\
\hline
\end{tabular}
}

\label{appendixC}
\label{tab:facet_temp}
\end{table}
\FloatBarrier

\bibliography{references}
\bibliographystyle{aasjournalv7}

\end{document}